\documentclass[amssymb,aps,prd,twocolumn,footinbib]{revtex4-2}%longbibliography

\usepackage{subfigure}
\usepackage{enumitem}
\usepackage[colorlinks=true,linkcolor=blue,urlcolor=blue,citecolor=blue]{hyperref}
\usepackage{graphicx}% Include figure files
\usepackage{dcolumn}% Align table columns on decimal point
\usepackage{bm}% bold math
\usepackage{bbold}
\usepackage{amsmath}
\usepackage{amsfonts}
\usepackage{amssymb}

\usepackage{tabularx}
\usepackage[overload]{empheq}
\usepackage{eufrak}
\usepackage{siunitx}
\usepackage{slashed}
\usepackage{tikz}

\usepackage{comment}

\DeclareMathOperator{\sgn}{sgn}

\newcommand{\nn}{\nonumber}
\newcommand{\unitmatrix}{\mathbb{1}}

\tikzset{every picture/.style={line width=0.75pt}} %set default line width to 0.75pt

\begin{document}

\title{Boundary critical behavior of the Gross-Neveu-Yukawa model}

\author{Andrei A. Fedorenko$^1$\ and Ilya A. Gruzberg$^2$ }

\affiliation{\mbox{$^1$
Laboratoire de Physique, UMR CNRS 5672, ENS de Lyon, Universit\'{e} de Lyon, F-69342 Lyon, France}\\
\mbox{$^2$ Ohio State University, Department of Physics, 191 West Woodruff Avenue, Columbus, Ohio 43210, USA}}
\date{July 23, 2026}

%%%%%%%%%%%%%%%%%%%%%%%%%%%%%%%%%%%%%%%
%%%%%%%%%%%%%%%%%%%%%%%%%%%%%%%%%%%%%%%
\begin{abstract}
We study the critical behavior of the semi-infinite Gross-Neveu-Yukawa model, a quantum field theory describing Dirac fermions interacting with bosonic fields via a Yukawa coupling. We consider Neumann and Dirichlet boundary conditions for the bosonic fields, and the most general boundary conditions for the fermions that preserve unitarity, conformal invariance, and charge conjugation symmetry. We analyze the phase diagram and identify distinct fixed points corresponding to different universality classes of boundary critical behavior. The associated boundary critical exponents, which govern the scaling behavior and crossover phenomena, are computed to one-loop order. We also discuss the relevance of our results to the semi-infinite pseudoscalar Yukawa model.
\end{abstract}

\maketitle

\section{Introduction} \label{sec:intro}

In recent years, there has been a revival of interest in field theories with boundaries, driven both by advances in conformal field theory (CFT), such as the conformal bootstrap~\cite{PolandRychkovVichi:2019}, and by the discovery of new materials, including Dirac materials such as graphene~\cite{Neto:2009,Cayssol:2013} and nodal semimetals~\cite{Armitage2018}, which make it possible to test boundary field theory predictions in the laboratory.

Seminal studies of boundary critical behavior in spin systems with discrete symmetries have revealed a remarkably rich phase structure, encompassing the \emph{ordinary}, \emph{special}, and \emph{extraordinary} transitions~\cite{Diehl-book:1986}. These correspond, respectively, to situations in which the bulk orders in the presence of a disordered, critically enhanced, or ordered boundary.
For 3D systems with continuous symmetries, the extraordinary transition is precluded by the Mermin--Wagner theorem, which forbids spontaneous symmetry breaking of continuous symmetries in two or fewer dimensions at finite temperature. Since the boundary of a 3D system is effectively two-dimensional, it cannot sustain true long-range order, and thus an ordered surface coexisting with a disordered bulk is not possible.

However, it was recently discovered that the 3D $O(N)$ $\phi^4$ model exhibits a novel \emph{extraordinary-log} boundary universality class for a finite range of $2 \le N \le N_c \approx 5$~\cite{Metlitski:2022,PadayasiKrishnanMetlitskiGruzbergMeineri:2022}. The universal properties of this class appear to be related to those of the \emph{normal} universality class, which corresponds to the case where an explicit symmetry-breaking field is applied at the boundary~\cite{ParisenToldin:2021,ParisenToldinMetlitski:2022}.Related phenomena have been identified in several quantum systems, including the quantum edge criticality of the two-dimensional Bose--Hubbard model~\cite{SunLv2022} and the boundary criticality arising from fractional exciton condensation in quantum Hall bilayers~\cite{ZhangZhuVishwanath:2023}.

Boundary criticality has also been explored in fer\-mi\-onic systems. The earliest examples involved noninteracting fermions in the presence of quenched disorder, leading to Anderson localization~\cite{Abrahams:2010} and continuous Anderson transitions, which include both metal--insulator and topological transitions such as the integer quantum Hall plateau transitions~\cite{Evers:2008}. Anderson transitions are characterized by the absence of a conventional order parameter. Instead, their criticality is encoded in transport coefficients and in the full distribution of the local density of states (LDOS). Different moments of the LDOS exhibit independent scaling behavior, reflecting the multifractal nature of the critical wave functions described by a continuum of exponents -- the multifractal spectrum.

Anderson transitions in the presence of boundaries have been investigated in Refs.~\cite{Subramaniam:2006, Mildenberger-Boundary-2007, Obuse-Multifractality-2007, Obuse-Corner-2008, Obuse-Boundary-2008, Subramaniam-Boundary-2008, Babkin-Generalized-2023, Babkin-Boundary-2023}, with particular focus on the multifractality of critical wave functions near the boundary. These studies have shown that the multifractal spectrum is modified near boundaries, but only at the \emph{ordinary} boundary critical point. The possibility of \emph{extraordinary} or \emph{special} boundary Anderson transitions remains an open question of significant interest.

A distinct class of disorder-driven quantum phase transitions arises in \emph{nodal semimetals}, where sufficiently strong disorder drives a transition from a clean (ballistic) regime to a diffusive metallic phase~\cite{Syzranov:2018}. This transition is described by the $O(N)$ Gross--Neveu model in the limit of vanishing number of fermion flavors (replicas) $N \to 0$~\cite{Roy:2014,Louvet:2016} or its supersymmetric extension~\cite{Syzranov:2015b}. Here, the average LDOS at the nodal point serves as an effective order parameter, becoming finite beyond a critical disorder strength~\cite{Sbierski:2014, Sbierski:2016, Fradkin:1986, Roy:2016b, Goswami:2011, Hosur:2012, Ominato:2014, Chen:2015, Altland:2015:2016, SzaboRoy:2020}. This transition has been extensively studied through both numerical simulations~\cite{Kobayashi:2014, Sbierski:2015, Liu:2016, Bera:2016, Fu:2017, Sbierski:2017} and analytical approaches~\cite{Roy2016,Balog:2018, Louvet:2017, Sbierski:2019, Klier:2019}. Although the critical wave functions are again multifractal, the fluctuations of non-self-averaging quantities such as the LDOS are significantly narrower than at Anderson transitions~\cite{Brillaux:2020}. The boundary manifestations of such \emph{non-Anderson} disorder-driven transitions in semi-infinite Dirac semimetals have recently been investigated in Refs.~\cite{Brillaux:2021, Brillaux:2024},
where only the \emph{extraordinary} and \emph{special} transitions were identified.

Recent breakthroughs in CFT methods, such as conformal bootstrap, applied to higher-dimensional systems, together with the continuing strong interest in the AdS/CFT correspondence have also sparked significant activity within the high-energy theory community in the study of boundary conformal field theories~\cite{MiaoChuGuo:2017,HerzogHuang:2017,Shpot:2021,HerzogSchaub:2024}. For instance, interacting fermionic systems are often described by the Gross-Neveu-Yukawa (GNY) model, in which a scalar bosonic field mediates the interaction between fermions. In the high-energy context, this model provides a renormalizable framework for studying dynamical mass generation and spontaneous symmetry breaking~\cite{Zinn-Justin:1986}. The GNY model exhibits a quantum critical point separating a symmetric (massless) phase from a phase with nonzero fermion condensate, and plays a central role in understanding chiral phase transitions in relativistic field theories.

The critical properties of the GNY model have been investigated extensively using a variety of complementary approaches. Perturbative renormalization group (RG) analyses have been carried out up to two-loop~\cite{Rosenstein:1993, Fei:2016}, three-loop~\cite{Mihaila:2017}, and four-loop~\cite{Zerf:2017,Ihrig:2018} order, while nonperturbative methods such as the conformal bootstrap~\cite{Erramilli:2023}, nonperturbative RG~\cite{RosaVitaleWetterich:2001,Knorr:2016} and quantum Monte Carlo simulations~\cite{Chandrasekharan:2013} have provided valuable insights into its critical behavior. More recently, the boundary critical behavior of the semi-infinite GNY model has been investigated using conformal field theory techniques~\cite{Giombi:2022, Herzog:2023}.

In addition, the \emph{boundary} pseudoscalar Yukawa (pY) model has been studied in~\cite{Jiang:2025} in the context of interacting Dirac fermions on a honeycomb lattice with armchair boundaries, revealing results that differ quantitatively  from those of Refs.~\cite{Giombi:2022, Herzog:2023}. The GNY type theories emerging from the interplay between edge fermions and bulk bosons were also studied in the context of topological insulators~\cite{Shen:2024} and superconductors~\cite{GeJiangYaoJian:2025}.

In the present paper, we study the GNY model in a semi-infinite space with general boundary conditions (BCs) using perturbative RG in \(D = 4-\varepsilon\) dimensions. We consider Neumann and Dirichlet BCs for the bosonic fields, and the most general BCs for the fermions that preserve unitarity, conformal invariance, and charge conjugation symmetry. This allows us to identify all possible universality classes and compute the corresponding dimensions of scaling operators and critical exponents. Mapping our model onto the pY model, we reconcile the discrepancy between the results obtained in~\cite{Giombi:2022, Herzog:2023} and in~\cite{Jiang:2025}.

The paper is organized as follows. In Sec.~\ref{sec:model}, we introduce the semi-infinite GNY semimetal, discuss the BCs compatible with various symmetries, analyze the existence of boundary states, and compute the corresponding Green's functions. Section~\ref{sec:phase-diagram} summarizes the phase diagram and identifies the universality classes of boundary phase transitions. In Sec.~\ref{sec:Renormalization}, we discuss the field-theoretic renormalization procedure and define the boundary critical exponents. Section~\ref{sec:bulk} reviews the known results for bulk critical behavior, while Sec.~\ref{sec:boundary} is devoted to the analysis of boundary transitions, including the computation of critical exponents and the stability of different fixed points (FPs). In Sec.~\ref{sec:Pseudoscalar-Yukawa}, we examine the boundary criticality of the pY model.
Finally, Sec.~\ref{sec:conclusions} presents our conclusions and outlook.

\section{Semi-infinite Gross-Neveu-Yukawa model}
\label{sec:model}

We define the GNY model in the semi-infinite $D=d+1$-dimensional Euclidean spacetime $\bm{x}=(x_0,...,x_d)\equiv (\vec{r},z)$  where $x_0$ is the imaginary (Euclidean) time coordinate,  $z \equiv x_d \ge 0$ and $\vec{r}=(x_0,...,x_{d-1})$ is a $d$-dimensional vector in the spacetime plane $z=0$.

The Euclidean action of the semi-infinite GNY model can be written as a sum
\begin{align}
S &= S_\mathrm{GNY}+ S_\mathrm{B},
\label{eq:action-GNY}
\end{align}
where the bulk part $S_\mathrm{GNY}$ is the standard action for the GNY model in $D=d+1$ dimensions~\cite{Zinn-Justin:1986} restricted to the half space $z \equiv x_d  \ge 0$,
%\begin{eqnarray}
\begin{align}
    S_\mathrm{GNY} & = \int d^{d} r  \int\limits_0^{\infty}d z \,  \Big\{ \sum\limits_{a=1}^{\tilde{N}} \bar{\psi}_a(\bm{x})  \left[ \slashed{\partial}  + g \varphi(\bm{x}) \right] \psi_a (\bm{x}) \nonumber \\
 &\quad + \frac12 (\nabla \varphi(\bm{x}))^2+\frac12 m^2 \varphi^2(\bm{x}) +\frac{\lambda}{4!}\varphi^4(\bm{x}) \Big\},
\label{eq:action-GNY-bulk}
\end{align}
%\end{eqnarray}
where $\psi$ is a $\tilde{N}$-component spinor fermionic (Grassmann) field,  $\varphi$ is a scalar bosonic field, and the conjugate Dirac field is $\bar{\psi}=\psi^{\dagger} \gamma_0$. We define $\slashed{\partial} \equiv \sum_{i=0}^{d} \gamma_i \partial_{i}$ where the Euclidean Dirac gamma matrices $\gamma_i$ are generators of the Clifford algebra $Cl_{D,0}(\mathbb{R})$ in a certain representation.

Usually, when using dimensional regularization to carry out perturbative calculations within the $D = 4 - \varepsilon$ expansion, one treats the $\gamma$ matrices as formal generators of a Clifford algebra in $D$ dimensions, without specifying a particular finite-dimensional representation. Unlike the Gross-Neveu model, the action of the GNY model is quadratic in fermions, and does not generate products of many $\gamma$ matrices which become independent in $4 - \varepsilon$ dimensions and require special treatment~\cite{Louvet:2016}.
As a consequence, all results depend on the rank of the $\gamma$ matrices only through the parameter $N = \tilde{N} \, \mathrm{tr}\, \mathbb{1}$, which gives us the freedom to account for additional internal degrees of freedom, such as spin, valley, or flavor, either by enlarging $N$ or by increasing the matrix rank of the $\gamma$'s~\cite{Zinn-Justin:1986}.
However, when studying boundary phenomena, the representation independence of the $\gamma$ matrices becomes more subtle, since one must specify BCs whose general form depends on both the spacetime dimension and the rank of the $\gamma$ matrices.

It is known that in strictly $2+1$-dimensional theories formulated in terms of a single irreducible two-component Dirac fermion, the parity anomaly leads to the loss of time-reversal invariance (TRI) upon regularization~\cite{Redlich:1984}. However, our model employs a reducible four-component Dirac representation, as specified in Appendix~\ref{sec:Appendix-A}. In this doubled formulation, the contributions to the parity anomaly from the two irreducible components can cancel, so that both TRI and non-TRI BCs can be consistently defined and analyzed on the same footing within the dimensional regularization scheme.

The boundary (surface) part $S_\mathrm{B}$ including only relevant in RG sense terms reads
\begin{align}
S_\mathrm{B} &= \int d^{d} r \Big\{  \sum\limits_{a=1}^{\tilde{N}} \bar{\psi}_{a}(\vec{r}) \check{M} \psi_a(\vec{r}) + \frac{c}2 \varphi^2(\vec{r})   \Big\},
\label{eq:action-GNY-surface}
\end{align}
where $\check{M}$ is a matrix in spinor space, i.e. restricted  to U($\tilde{N}$)-invariant BCs, and $c$ is related to the boundary enhancement of the bosonic coupling constant in the corresponding lattice model.
From the boundary terms of the classical equations of motion $\delta S = 0$, we derive the Robin BC for the bosonic field
\begin{align}
\partial_z \varphi |_{z=0} = c\, \varphi |_{z=0},
\label{eq:BC-Robin-1}
\end{align}
and the BC for the fermionic field
\begin{align}
M \psi |_{z=0} = \psi |_{z=0},
\label{eq-BC-fermions}
\end{align}
where we define matrix $M$ related to $\check{M}$ in Eq.~\eqref{eq:action-GNY-surface} by
\begin{align}
  M = - \gamma_d \check{M}.
  \label{eq-MM-relation}
\end{align}
The general form of the matrix $M$ for $(2+1)$-dimensional Dirac fermions has been discussed in the context of high-energy physics in Ref.~\cite{Biswas:2022}, and in condensed-matter systems such as carbon nanotubes and graphene in Refs.~\cite{McCann-Falco:2004, Akhmerov-Boundary-2008, ShtankoLevitov2018}, where it was shown that without loss of generality the matrix \(M\) can be chosen unitary and Hermitian, that is \(M^2 = 1\). The corresponding analysis for $(3+1)$-dimensional fermions has also been carried out in the context of Weyl semimetals~\cite{hashimoto_boundary_2017,faraei_greens_2018} and Dirac semimetals~\cite{Brillaux:2024}. Since our goal is to study the semi-infinite GNY model in $D = 4 - \varepsilon$ dimensions, we follow the approach and conventions of the latter reference.

The Hamiltonian of the free Dirac fermions corresponding to the fermionic part of the action~\eqref{eq:action-GNY} with $g=0$ can be written as
\begin{align}
\hat{H}_0 = - i \sum\limits_{j=1}^{d} \alpha_j \partial_j + m_D \beta + \tilde{M} \delta(z),
\label{eq-Hamiltonian-1}
\end{align}
where for the sake of generality we introduced a nonzero Dirac mass $m_D$. The standard Dirac matrices $\alpha_j$ and $\beta$ related to the Euclidean Dirac matrices $\gamma_j$ in the action~\eqref{eq:action-GNY-bulk} by $\alpha_j= i \gamma_0 \gamma_j $,  $j=1,..,d$ and $\beta = \gamma_0$, see Appendix~\ref{sec:Appendix-A}. Also, the boundary is represented in Eq.~\eqref{eq-Hamiltonian-1} as an extended ``impurity'' or defect involving the matrix
\begin{align}
    \tilde{M} &= i \alpha_d M = - \gamma_0 \gamma_d M = \gamma_0 \check{M}.
\end{align}
Integrating the Schr\"odinger equation with the Hamiltonian~\eqref{eq-Hamiltonian-1} across the boundary at \(z = 0\), it is easy to see that the term \(\tilde{M} \delta(z)\) leads to the BC~\eqref{eq-BC-fermions}.

The Hamiltonian~\eqref{eq-Hamiltonian-1} supplemented by the BC \eqref{eq-BC-fermions} is Hermitian in the semi-infinite space $z\ge 0$ if
\begin{align}
    \langle \psi_1 | \hat{H}_0 \psi_2 \rangle = \langle \hat{H}_0 \psi_1 | \psi_2 \rangle
\end{align}
for arbitrary $\psi_1$ and $\psi_2$ satisfying the BC~\eqref{eq-BC-fermions}. This implies  that $\alpha_d$ anticommutes with $M$, and thus, the $z$ component of the current $J_i = \psi^\dagger \alpha_i \psi$ (normal to the boundary) vanishes at $z=0$. The Hermiticity of the Hamiltonian~\eqref{eq-Hamiltonian-1} ensures that the field theory~\eqref{eq:action-GNY} is unitary.

We impose additional symmetries on our problem and require it to be invariant under the conformal group preserving the boundary \(z=0\) and under the charge conjugation. The BC~\eqref{eq-BC-fermions} preserves conformal invariance in the \(z=0\) plane
if the matrix $M$ commutes with  translations $P^\mu$, rotations $M^{\mu\nu}$, dilatation $D$, and special conformal generators $K^{\mu}$, where the indices \(\mu\) and \(\nu\) are restricted to the range \((0,\ldots, d-1)\). The Hamiltonian~\eqref{eq-Hamiltonian-1}
is invariant under the charge conjugation if matrix $M$ satisfies $\hat{\mathcal{C}} M \hat{\mathcal{C}}^{-1} = M$ (which is equivalent to \(\hat{\mathcal{C}} \tilde{M} \hat{\mathcal{C}}^{-1} = - \tilde{M}\)), where the charge-conjugation operator $\hat{\mathcal{C}}$ is defined in Appendix~\ref{sec:Appendix-B}.

A one-parameter family of matrices that satisfies all these symmetry conditions (conformal symmetry and charge conjugation) is
\begin{align}
M &= -(\gamma_d \sin \phi + i \gamma_d\gamma_S \cos\phi),
\nonumber \\
\check{M} &= \sin \phi + i \gamma_S \cos\phi,
\nonumber \\
\tilde{M} &= \gamma_0 \sin \phi + i \gamma_0 \gamma_S \cos\phi,
\label{eq:M-phi}
\end{align}
with $\phi \in (-\pi,\pi]$, where the ``fifth'' matrix $\gamma_S$  anticommuting with other Dirac matrices is defined for arbitrary $D$ in Appendix~\ref{sec:Appendix-A}. As shown in~\cite{Brillaux:2024}, for $D=3$ and $D=4$, where the size of the gamma matrices is $4 \times 4$, this family, known as the chiral MIT bag BCs~\cite{Rohim:2021}, exhausts all $M$ matrices satisfying the symmetries above.
In these dimensions our specific choice of the gamma matrices, Eqs.~\eqref{eq:gamma4-1} and~\eqref{eq:gamma4-2}, gives
\begin{align}
    M_{D=3} &= (\cos\phi \, \tau_1 - \sin \phi \, \tau_2 )\otimes \sigma_2,
    \label{eq:M-phi-D=3}
    \\
    M_{D=4} &= (\cos\phi \, \tau_1 - \sin \phi \, \tau_2 )\otimes \sigma_3,
    \label{eq:M-phi-D=4}
    \\
    \check{M} &= (\sin \phi \, \tau_0 + i \cos\phi \, \tau_3 ) \otimes \sigma_0,
    \\
    \tilde{M} &= - (\sin \phi \, \tau_1 + \cos\phi \, \tau_2) \otimes \sigma_0.
    \label{eq:tilde-M-D=4}
\end{align}
(Eq.~\eqref{eq:M-phi-D=4} is essentially the same as the second line of Eq.~(5) in Ref.~\cite{Brillaux:2024} with \(\theta = 0\) and the replacement \(\phi \to \pi -\phi\).)
In higher dimensions there may be other such \(M\) matrices characterized by more parameters. Since we work in \(D = 4-\varepsilon\), the family~\eqref{eq:M-phi} is all we need.

If, in addition, we impose time-reversal symmetry, the matrix \(M\) should satisfy
$\hat{\mathcal{T}} M \hat{\mathcal{T}}^{-1} = M$ (equivalently, \(\hat{\mathcal{T}} \tilde{M} \hat{\mathcal{T}}^{-1} = \tilde{M}\)), where the time-reversal operator $\hat{\mathcal{T}}$ is defined in Appendix~\ref{sec:Appendix-B}, Eqs.~\eqref{eq:T-sym-1} and~\eqref{eq:UT-1}. It follows that the only TRI BCs among the family~\eqref{eq:M-phi} are those with $\phi = \pm \tfrac{\pi}{2}$.

The meaning of the operator \(\hat{\mathcal{T}}\) depends on the physical context.
For example, in the case of graphene, \(\hat{\mathcal{T}}\) does not correspond to the physical time-reversal operator, because the Pauli matrices \(\sigma\) act in the sublattice (pseudospin) space rather than on the real spin of the electrons. Instead, the physical time-reversal operator \(\hat{\mathcal{T}}^{(g)}\) is given by Eq.~\eqref{eq:UT-g}, and it exchanges the valleys. This symmetry is preserved by the BCs~\eqref{eq:M-phi} for all values of the angle \(\phi\). We can then call the operator \(\hat{\mathcal{T}}\) the {\it intravalley\,} time reversal, and it is preserved only for \(\phi = \pm \tfrac{\pi}{2}\), when reflection from the boundary exchanges the two valleys without introducing a phase shift.%\(\pi/2\)-phase shift.

We note in passing that the above discussion of symmetries of the boundary matrix \(M\) has a complete analogy with the symmetry classification of disordered single-particle Dirac Hamiltonians in the context of Anderson localization, see Sec.~VI.G in Ref.~\cite{Evers:2008}. This should not come as a surprise since disordered Dirac fermions are described by a Hamiltonian of the form of Eq.~\eqref{eq-Hamiltonian-1} but with the boundary term replaced by a random disorder ``potential'' \(\tilde{M}(x_1,...,x_d)\).

The interpretation of specific symmetries of disordered Hamiltonians again depends on particular systems of interest. For example, in the study of disordered graphene, the charge-conjugation symmetry \(\hat{\mathcal{C}}\) becomes one of the particle-hole or ``Bogoliubov-de Gennes'' type symmetries, while the Dirac time-reversal \(\hat{\mathcal{T}}\) is one of the four symmetries of time-reversal type. For reference, our operators \(\hat{\mathcal{C}}\), \(\hat{\mathcal{T}}\), and \(\hat{\mathcal{T}}^{(g)}\) are denoted by \(CT_z\), \(T_x\), and \(T_0\) in Ref.~\cite{Evers:2008}. In this context, our boundary matrix \(\tilde{M}\) contains the matrix structures \(\sigma_0 \tau_{1,2} \) listed in the fourth line of the table V in Ref.~\cite{Evers:2008}, and these types of disorder are seen to preserve \(T_0\) and \(CT_z\). At the same time, only the real structure \(\sigma_0 \tau_{1}  \) preserves \(T_x\). This symmetry is not seen in the table V since there the authors assumed statistically isotropic disorder, which requires the two structures to come with the same coefficient. On the other hand, our boundary certainly breaks the isotropy in the \(x\)-\(y\) plane, and the two structures \(\sigma_0 \tau_{1,2} \) can appear with different coefficients.

As a preliminary step, before addressing the complete theory, we analyze the impact of the BCs on free fermions at $g = \lambda = 0$. We employ Fourier transforms in the subspace  $\vec{r}\,'=(x_1,...,x_{d-1})$, while retaining the coordinate dependence on $z=x_d$. In this setting, an eigenstate of the Hamiltonian~\eqref{eq-Hamiltonian-1}, localized at the boundary $z=0$ with $m_D=0$ and eigenvalue $\epsilon$, obeys the equation
\begin{align}
& \left [ \vec{\alpha} \vec{q}\,'  - i \alpha_d \frac{\partial }{\partial z}\right ] \psi(\vec{q}\,', z) = \epsilon  \psi(\vec{q}\,', z),  \label{eq:psi-1}
\end{align}
where $\vec{q}\,'$ is a $d-1$ dimensional wave vector in the plane $z=0$, and the BCs are
\begin{align}
& M \psi(\vec{q}\,', z=0) = \psi(\vec{q}\,', z=0)  \label{eq:psi-2}
\end{align}
with matrix $M$ given by \eqref{eq:M-phi}. The solution of Eqs.~\eqref{eq:psi-1}-\eqref{eq:psi-2} localized at the boundary can be written as
\begin{align}
%\psi = (\psi_1,\psi_2,\psi_3,\psi_4)^\mathrm{T} e^{-\mu z },
\psi(\vec{q}\,', z)  = \psi(\vec{q}\,',0)  e^{-\mu z },
\label{eq:psi-3}
\end{align}
where $\mu$ is the inverse penetration length.  We find that the only existing solutions of the form \eqref{eq:psi-3} correspond to $\mu = 0$, indicating the absence of boundary states for the BCs~\eqref{eq:M-phi}. For more general BCs, there exist solutions with nonzero $\mu$ and energies $-q' < \epsilon < q'$, i.e., within the gap between the valence and conduction bands.  However, as the matrix $M$ approaches the form given by \eqref{eq:M-phi}, $\mu$ vanishes while $\epsilon$ approaches $q'$ or $-q'$, and thus the boundary state merges into the bulk states. The presence of boundary states, which would appear, for example, if the charge-conjugation symmetry were broken, complicates the description of boundary critical behavior, in a way similar to the situation encountered at the extraordinary transition of the pure bosonic $\varphi^4$ theory.

Let us now consider the Gaussian approximation. The bosonic and fermionic propagators satisfy the following BCs:
\begin{align} \label{eq:BC-boson-general}
\left. \partial_z {G}_\mathrm{B}(q,z,z') \right|_{z=0} &= c\, {G}_\mathrm{B}(q,0,z'), \\
M  {G}_\mathrm{F}(q,0,z')  &=  {G}_\mathrm{F}(q,0,z'),
\end{align}
where $\vec{q}$ is a $d$-dimensional wave vector lying in the $z=0$ plane of the $(D=d+1)$-dimensional spacetime. The bare propagators follow from the Gaussian part of the action~\eqref{eq:action-GNY}–\eqref{eq:action-GNY-surface} and take the form
\begin{align}
{G}_\mathrm{B}(\vec{q},z,z') &= \frac1{2\kappa} \left[ e^{-\kappa|z-z'|} - \frac{c-\kappa}{c+\kappa} e^{-\kappa(z+z')} \right],
\label{eq:boson-propagator-general} \\
{G}_\mathrm{F}(\vec{q},z,z') &=  \frac12 \left[-\frac{i \slashed{q}}{q} +\gamma_d \sgn(z-z') \right] e^{-q |z-z'|} \nn \\
& \hspace{-10mm} - \frac{1}2  M \left[\frac{i \slashed{q}}{q} + \gamma_d \sgn(z+z') \right] e^{-q (z+z')}, \label{eq:fermion-propagator-general}
\end{align}
where we redefined $\slashed{q} \equiv \sum_{i=0}^{d-1} \gamma_i q_{i}$ and $\kappa(\vec{q}) = \sqrt{q^2+m^2}$.
It is important to note that Eq.~\eqref{eq:fermion-propagator-general} is valid only for $M$ as defined in Eq.~\eqref{eq:M-phi}, and does not hold for arbitrary matrices~$M$.

%%%%%%%%%%%%%%%%%%%  Fig. 1 %%%%%%%%%%%%%%%%%%%%%%%
\begin{figure}[t!]
\includegraphics[scale=0.65]{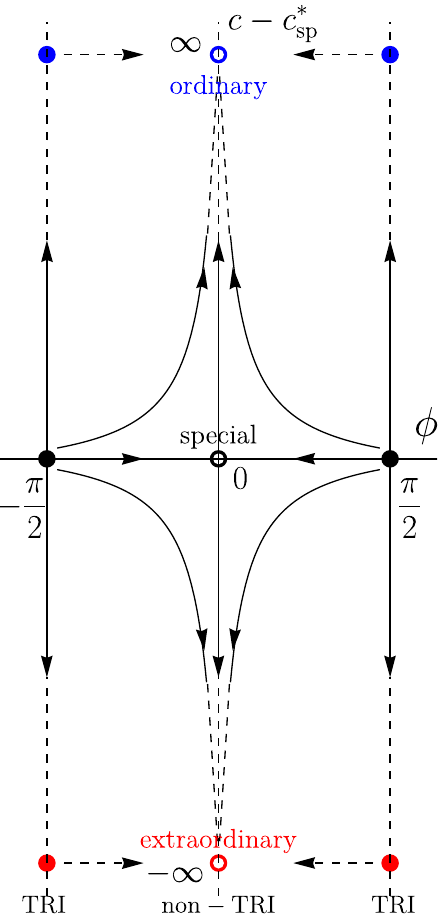}
\caption{RG flow and FPs in the $(\phi,c)$ plane.
Although the matrix $M$ given by Eq.~\eqref{eq:M-phi} itself is $2\pi$-periodic in $\phi$, the RG flows  exhibit an effective periodicity of $\pi$.
There are six independent FPs, corresponding to ordinary ($c^*=\infty$), special ($c=c^*_{\rm sp}$), extraordinary ($c^*=-\infty$) transitions with TRI ($\phi^*=\pm\frac{\pi}2$) and without TRI ($\phi^*=0$). The TRI FPs are unstable with respect to non-TRI perturbations.   \label{fig:1-RGFlow} }
\end{figure}
%%%%%%%%%%%%%%%%%%%%%%%%%%%%%%%%%%%%%%%%%%%%%%%%%%%

\section{Phase diagram and boundary universality classes} \label{sec:phase-diagram}

The RG flows of the field theory~\eqref{eq:action-GNY}-\eqref{eq:action-GNY-surface} with matrix $M$ given by \eqref{eq:M-phi}, drawn in the $(c,\phi)$ plane,  are shown schematically in Fig.~\ref{fig:1-RGFlow}. This diagram illustrates how different initial values of $(c, \phi)$ evolve under RG, indicating the stability domains and the trajectories connecting ordinary, special, and extraordinary FPs both with and without TRI.

Before delving into details, let us outline the methods used to derive the RG flow structure and to identify the FPs.
Within the RG description of boundary critical phenomena, correlation functions evaluated using dimensional regularization contain additional poles in $\varepsilon$ that cannot be removed by bulk renormalization alone. The theory therefore requires an additional renormalization of boundary couplings or boundary fields. Technically, however, such a renormalization procedure is feasible, i.e., the theory remains renormalizable, only if the loop corrections possess the same functional and spinor structure as the corresponding tree-level expressions. Only in this case can the additional divergences be absorbed into a finite number of new $Z$-factors.
For example, in the purely bosonic case, i.e. the $\phi^4$ theory describing boundary criticality in the Ising universality class, the loop corrections retain the same structure as the tree-level functions only for $c=0$ and $c=\pm\infty$, corresponding to RG FPs. For generic values of $c$, the corrections acquire a different structure, leading to an RG flow of $c$ toward the FPs.
In the present case, the BCs depend not only on $c$ but also on the continuous parameter $\phi$. We have computed the one-loop corrections to the boundary two-point functions for arbitrary $\phi$, and we find that the loop corrections preserve the tree-level structure only for particular discrete values of $\phi$, namely for \(\phi = 0\) and \(\phi = \pm \pi/2\). We therefore identify these special values with distinct RG FPs corresponding to different universality classes of boundary critical behavior.

We do not report the expressions obtained for arbitrary values of $\phi$, since they are extremely cumbersome and, more importantly, cannot be consistently renormalized. Indeed, if the theory were renormalizable for generic $\phi$, one would obtain not isolated FPs but rather a continuous line of FPs parametrized by $\phi$.
In principle, the general expressions could be used to derive the RG flow of the boundary parameter $\phi$, whose FPs we have identified from renormalizability. In practice, however, the expressions obtained within dimensional regularization are reliable only in the vicinity of these FPs and cannot be extrapolated far away from them in order to reconstruct the global flow connecting different FPs.

One might then worry that the RG flow from an unstable to a stable FP could leave the restricted space of BCs parametrized solely by $\phi$. However, this cannot occur, since the imposed charge-conjugation symmetry forbids the generation of more general boundary terms.

Let us now analyze the RG flow structure shown in Fig.~\ref{fig:1-RGFlow}.
In the bosonic sector there are three types of FPs: the ordinary FPs corresponding to $c=c^*_{\textrm{ord}} = \infty$, the special FPs at $c=c^*_{\textrm{sp}}$, and the extraordinary FPs at $c=c^*_{\textrm{ext}}=-\infty$ which are generalizations of that for the pure bosonic systems~\cite{Diehl-book:1986}. The ordinary FPs are attractive for $c>c^*_{\textrm{sp}}$, the extraordinary FP is attractive for
$c<c^*_{\textrm{sp}}$, while the special FP with  $c^*_{\textrm{sp}}\le 0$ is always unstable. Within the mean-field approximation and the MS scheme, the special transition occurs at $c^*_{\textrm{sp}}=0$.
In the fermionic sector there are two types of FPs: the TRI FPs corresponding  to $\phi = \pm \frac{\pi}2$, which are unstable with respect to  TRI breaking perturbations, and the non-TRI FPs  corresponding to $\phi = 0;\pi  $, which are stable.
The matrices $M$ corresponding to the two classes of FPs are labeled  by $\sigma = +1$ for the TRI case and $\sigma = -1$ for the non-TRI case, and are given by
\begin{align}
\check{M}_\sigma =
\begin{dcases}
\pm \mathbb{1}    & \text{\ for \ } \sigma=+1 \text{\ (TRI),\ } \\
\pm i \gamma_S   & \text{\ for \ } \sigma=-1  \text{\ (non-TRI)\ }.
\end{dcases}
\label{eq:M-sigma}
\end{align}
To establish the RG flow from the TRI FPs with $\phi=\pm \frac{\pi}{2}$ ($\sigma=+1$) toward the non-TRI FP with $\phi=0$ ($\sigma=-1$), we analyze the relevance of boundary composite operators (see Sec.~\ref{sec:composite-operators}). We show that the TRI FPs, characterized by a boundary action~\eqref{eq:action-GNY-surface} containing only the operator $(\bar\psi \psi)_s$, are destabilized by the addition of the operator $(\bar\psi \gamma_S \psi)_s$. In contrast, the non-TRI FPs, where only $(\bar\psi \gamma_S \psi)_s$ is present in the boundary action, remain stable with respect to the perturbation $(\bar\psi \psi)_s$, which is found to be irrelevant.
This analysis establishes the RG flow shown in Fig.~\ref{fig:1-RGFlow}, at least locally in the vicinity of the FPs, while the global structure of the RG flow is constrained by symmetry considerations.

\begin{figure}[t!]
\includegraphics[scale=0.38]{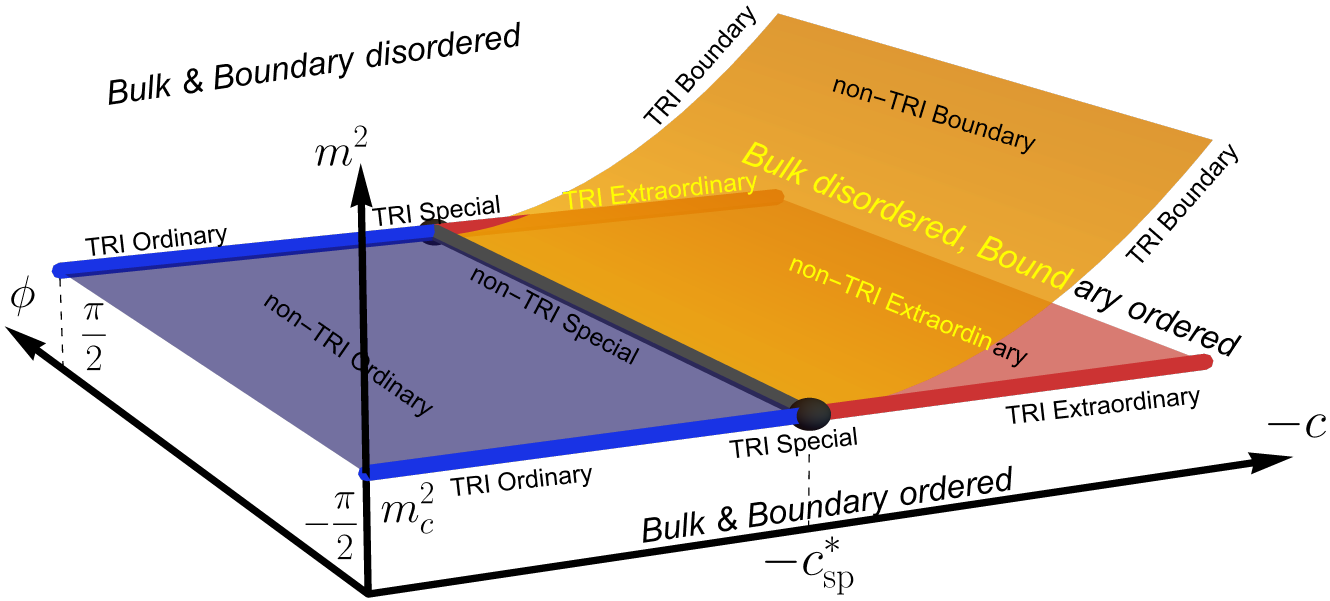}
\caption{Phase diagram exhibits three distinct phases: (i) both bulk and boundary ordered, (ii) bulk disordered but boundary ordered, and (iii) both bulk and boundary disordered. These phases are separated by the ordinary, extraordinary, and surface transition lines, which meet at the line of special transitions. Each of these transitions belongs to a non-TRI universality class for $|\phi| < \frac{\pi}{2}$, and to a TRI universality class for $\phi = \pm \frac{\pi}{2}$.
The corresponding critical exponents for the TRI and non-TRI ordinary and special universality classes are summarized in Table~\ref{tab:1-dimensions}
 \label{fig:2-phase-diagram} }
\end{figure}

The bulk phase transition is controlled by the boson mass $m^2$, such that $\delta m^2 = m^2 - m_c^2$ plays the role of the ``reduced temperature'' in the language of statistical mechanics of second-order phase transitions.
Note that within the MS scheme the critical mass vanishes, i.e., $m_c^2 = 0$. Indeed, the mass counterterm is chosen to cancel only the divergent part, so that the finite shift of the critical point due to interactions is not included, and the critical point is effectively located at zero renormalized mass similar to $c^*_{\textrm{sp}}=0$ within the MS scheme.
The resulting phase diagram, indicating both bulk and boundary phases, is shown in Fig.~\ref{fig:2-phase-diagram}.

In the present paper, we consider only the special and ordinary transitions.
Note that the TRI boundary universality classes have been studied using conformal field theory methods in Refs.~\cite{Giombi:2022, Herzog:2023}. Our results, obtained using a direct RG approach for the special and ordinary TRI transitions, are in full agreement with those found for the B2 and B2$'$ universality classes in Ref.~\cite{Giombi:2022}. In addition, we compute the scaling dimensions of boundary composite operators, which were not previously known even in the TRI cases. It is also interesting to note that in Ref.~\cite{Giombi:2022} the B1 universality class was argued to describe the TRI extraordinary transition, which lies beyond the scope of the methods used in the present work.

At the special transition, within the MS scheme, $c_{\textrm{sp}}^* = 0$, so that the boson field satisfies the Neumann BCs:
\begin{align} \label{eq:BC-boson-Neumann}
\partial_z {G}_\mathrm{B}(q,z,z') |_{z=0} = 0.
\end{align}
At the ordinary transition  $c^*_{\textrm{ord}}=\infty$ so that the boson field satisfies the Dirichlet BC
\begin{align} \label{eq:BC-boson-Dirichlet}
{G}_\mathrm{B}(q,z,z') |_{z=0} = 0.
\end{align}
We denote the corresponding  limiting propagators by
\begin{align}
{G}_\mathrm{B}^\mathrm{\chi}(\vec{q},z,z') = \frac1{2\kappa} \left[ e^{-\kappa|z-z'|} + \chi e^{-\kappa(z+z')} \right],
\label{eq:boson-propagator-chi}
\end{align}
where $\chi = +1$ corresponds to the propagator satisfying the Neumann BCs at the special transition, and $\chi = -1$ corresponds to the propagator satisfying the Dirichlet BCs at the ordinary transition.
%%%%%%%%%%%%%%%%%%%  Fig. 2 %%%%%%%%%%%%%%%%%%%%%%%
%%%%%%%%%%%%%%%%%%%%%%%%%%%%%%%%%%%%%%%%%%%%%%%%%%%%%%%
%%%%%%%%%%%%%%%%%%%%   Tab 1   %%%%%%%%%%%%%%%%%%%%%%%%%
\begin{table}
\begin{tabular}{c|l}
\hline \\[-2mm]
Operator & \mbox{}\hspace{15mm}  Scaling dimension \\[1mm]
\hline \\[-2mm]
$\psi_s$  & \mbox{}\hspace{2mm} $\Delta_{\psi_s} = \dfrac{3}{2}-\dfrac{(\sigma+3)N+2(2+\chi)\sigma+12 }{4(N+6)}\varepsilon$  \\[3mm]
$\varphi_s$   & \mbox{}\hspace{2mm} $\Delta_{\varphi_s} = \dfrac{3-\chi}2-\dfrac{\left(Y(N)+(6\sigma \chi+5)N+42\right)  \varepsilon }{12 (N+6)}$  \\[3mm]
$(\bar{\psi}\psi)_s$  & \mbox{}\hspace{2mm} $\Delta_{(\bar{\psi}\psi)_s}^{\textrm{non-TRI}} = 3 + \dfrac{N+3\chi}{N+6}\varepsilon$, \ \ \ $\sigma=-1$  \\[3mm]
$(\bar{\psi}\gamma_S\psi)_s$  & \mbox{}\hspace{2mm} $\Delta_{(\bar{\psi}\gamma_S\psi)_s}^{\textrm{TRI}} = 3 - \dfrac{2(N+6)+3\chi}{N+6}\varepsilon$, \ \ \ $\sigma=+1$  \\[3mm]
$\varphi^2_s$ (\(\chi = 1\))   & \mbox{}\hspace{2mm} $\Delta_{\varphi^2_s}^{\textrm{Sp}} = 2 + \dfrac{\left(Y(N)-(6\sigma+7) N-30\right) \varepsilon  }{6 (N+6)}$ \\[3mm]
\hline
\end{tabular}
\caption{
Dimensions of the boundary operators corresponding to the ordinary ($\chi=-1$) and special ($\chi=+1$) transitions, both with TRI ($\sigma=+1$) and without TRI ($\sigma=-1$). Here, $\varepsilon = 4 - D$, the function $Y(N) = \sqrt{N^2 + 132 N + 36}$, see Eq.~\eqref{Y-N}, and, for comparison with \cite{Giombi:2022,Herzog:2023,Jiang:2025}, one should substitute $N = \tilde{N}\,\mathrm{tr} \, \mathbb{1} = 4 \tilde{N}$.
\label{tab:1-dimensions} }
\end{table}
%%%%%%%%%%%%%%%%%%%%%%%%%%%%%%%%%%%%%%%

\section{Renormalization and critical exponents} \label{sec:Renormalization}

The bare action is defined by Eqs.~\eqref{eq:action-GNY}--\eqref{eq:action-GNY-surface}, upon replacing all fields and couplings with their bare counterparts, denoted by circles, e.g., $\psi \to \mathring{\psi}$. We also define the bare Green's function $\mathring{G}^{(n_1,n_2,n_3,n_4;l)}(\{\vec{k}\},\{z\})$ as the expectation value of the product of $n_1$ bulk fermion fields $\bar{\psi}$ or $\psi$, $n_2$ boundary fermion fields $\bar{\psi}_s$ or $\psi_s$, $n_3$ bulk boson fields $\varphi$, $n_4$ boundary boson fields $\varphi_s$, and, for $l>0$, one composite operator $O_l$. Here, $\{\vec{k}\}$ denotes the set of all momenta of the fields, and $\{z\}$ denotes the set of all coordinates along the $z$-direction, and we employ the following shorthand notations for brevity:
\begin{align}
O_1 &\equiv \varphi^2_s,
&
O_2 &\equiv \bar\psi \psi,
\nonumber \\
O_3 &\equiv (\bar\psi \psi)_s,
&
O_4 &\equiv (\bar\psi \gamma_S\psi)_s.
\end{align}
The Green's functions computed perturbatively in the couplings $\mathring{g}$ and $\mathring{\lambda}$ are UV divergent in $D=4$, which is the upper critical dimension of the model. To regularize the theory, we employ dimensional regularization by computing the Green's functions in $D = 4 - \varepsilon$ dimensions, thereby turning the UV divergences into poles at $\varepsilon = 0$. We renormalize the theory using the MS scheme and absorb these poles into the redefinition of the fields and couplings~\cite{Zinn-Justin:1986}. In the case of the special transition, we impose
\begin{align}
\mathring{\psi} &= Z_{\psi}^{1/2} \psi,
&
\mathring{\psi_s} &= Z_{\psi_s}^{1/2} {\psi_s},
\label{eq:Z-factors1} \\
%
%&\mathring{\bar{\psi}}{\psi} = Z_{\bar{\psi} \psi} \bar{\psi} \psi,
%&(\mathring{\bar{\psi}}\psi)_s = Z_{(\bar{\psi}\psi)_s} (\bar{\psi}\psi)_s, %\label{eq:Z-factors12} \\
%&\mathring{\bar{\psi}}_s{\psi_s} = Z_{\bar{\psi}_s \psi_s} \bar{\psi}_s \psi_s, \label{eq:Z-factors12} \\
%
\mathring{\varphi} &= Z_{\varphi}^{1/2} \varphi,
&
\mathring{\varphi_s} &= Z_{\varphi_s}^{1/2} {\varphi_s},
\label{eq:Z-factors2} \\
\mathring{m}^2 &= Z_{m}\, m^2,
%&\mathring{c} = m Z_{c}\, c, \label{eq:Z-factors3} \\
&
\mathring{O_l} &= Z_{O_l} O_l,
\label{eq:Z-factors3} \\
%
%&\mathring{\varphi_s^2} = Z_{\varphi_s^2} Z_{\varphi_s}^{-1} {\varphi_s^2},
%&\mathring{m}^2 = Z_{\varphi^2} Z_{\varphi}^{-1} m^2  \label{eq:Z-factors3} \\
%
\mathring{g}^2 &= \frac{2 m^{\varepsilon }}{K_D} \frac{Z_{g^2}}{Z_{\psi}^2 Z_{\varphi}} g^2,
&
\mathring{\lambda} &= \frac{2 m^{\varepsilon }}{K_D} \frac{Z_{\lambda}}{Z_{\varphi}^{2}} \lambda,
\label{eq:Z-factors4}
\end{align}
such that the Green's functions, expressed in terms of the renormalized couplings, remain finite in $D=4$. Here, we have introduced the notation $K_D = S_D / (2\pi)^D$, where $S_D = 2 \pi^{D/2} / \Gamma(D/2)$ is the area of the unit sphere in $D$ dimensions. The ordinary transition requires a slightly different renormalization scheme. Indeed, the boson field on the boundary, $\varphi_s$, vanishes identically due to the Dirichlet BCs. Thus, instead of the condition~\eqref{eq:Z-factors2} on $\varphi_s$ at the boundary, we impose a renormalization condition on its derivative at the boundary as
\begin{align} \label{eq:Z-partial-phi-s}
& \partial_z \mathring{\varphi} \big|_{z=0^+} = Z^{1/2}_{\partial \varphi_s } \partial_z \varphi \big|_{z=0^+},
\end{align}
defining the renormalization constant $Z_{\partial \varphi_s }$.
The $\beta$-functions, which encode the dependence of the renormalized Green's functions on the renormalization scale and describe how the effective interactions in the theory evolve under renormalization, are given by
\begin{align}
&\beta_{g^2} (g^2,\lambda) = m\frac{\partial g^2}{\partial m} \bigg|_{\mathring{g}^2,\mathring{\lambda}}, \ \ \
 \beta_{\lambda} (g^2,\lambda) = m\frac{\partial \lambda}{\partial m} \bigg|_{\mathring{g}^2,\mathring{\lambda}}.
 \label{eq:beta-functions-gen}
\end{align}
The FPs of the $\beta$-functions \eqref{eq:beta-functions-gen}, defined by
\begin{align}
\beta_{g^2}(g^{*2},\lambda^*) &= 0,
&
\beta_{\lambda}(g^{*2},\lambda^*) &= 0,
\label{eq:fixpoint0}
\end{align}
correspond to scale-invariant or critical behavior when they are stable, i.e., when the eigenvalues of the matrix
\begin{align}
\begin{pmatrix}
    \dfrac{\partial \beta_{g^2}}{\partial g^2} &  \dfrac{\partial \beta_{g^2}}{\partial \lambda} \\[1em]
   \dfrac{\partial \beta_{\lambda}}{\partial g^2} &  \dfrac{\partial \beta_{\lambda}}{\partial \lambda}
\end{pmatrix}
_{g^2=g^{2*},\lambda=\lambda^*}
\label{eq:stability0}
\end{align}
have positive real parts. The critical exponents describing the scale-invariant behavior can be computed by evaluating the scaling functions
%\ig{for \(i=\psi,\psi_s,\varphi,\varphi_s, \partial\varphi_s, m, O_l \)}
\begin{align}
\eta_i  &= \frac{\partial \ln Z_i}{\partial \ln m}\bigg|_{\mathring{g}^2,\mathring{\lambda}},
&
(i=\psi,\psi_s,\varphi,\varphi_s, \partial\varphi_s, m, O_l ),
\label{eq:critical-exponents-gen}
\end{align}
at the corresponding FP \eqref{eq:fixpoint0}.
For instance, the correlation length diverges at the transition as $\xi \sim |t|^{-\nu}$, where the critical exponent $\nu$ characterizes the divergence of the correlation length near the critical point, and $t$ denotes the reduced temperature or distance from the critical point. The correlation length exponent reads
\begin{align}
 \frac1{\nu} = 2+ \eta_m(g^{*2},\lambda^*).
\end{align}
The fermion and boson two-point functions exhibit characteristic scaling behavior in the bulk  at the transition, reflecting the underlying scale invariance of the critical point,
\begin{align}
G^{(2,0,0,0;0)}(\bm{x}) &\sim \frac{\gamma_\mu x_\mu}{|\bm{x}|^{D+\eta_{\psi}}},
&
\eta_{\psi} &= \eta_{\psi}(g^{*2},\lambda^*),
\\
G^{(0,0,2,0;0)}(\bm{x}) &\sim \frac1{|\bm{x}|^{D-2+\eta_{\varphi}}},
&
\eta_{\varphi} &= \eta_{\varphi}(g^{*2},\lambda^*),
\end{align}
where the critical exponents $\eta_\psi$ and $\eta_\varphi$ characterize the anomalous scaling of the bulk fermion and boson fields, and are directly related to their scaling dimensions
\begin{align}
\Delta_{\psi}  &= \frac12(D-1+\eta_{\psi}),
&
\Delta_{\varphi} &= \frac12(D-2+\eta_{\varphi})
\end{align}
at the critical point.

The critical fermion and boson two-point functions at the boundary exhibit anomalous scaling behavior, which depends on the surface universality class, and are given by
\begin{align}
 G^{(0,2,0,0;0)}(\vec{r}) &\sim \frac{(\unitmatrix + M)\, \vec{\gamma}\cdot \vec{r}}{|\vec{r}|^{D+\eta_{\psi_\parallel}}}, \\
 G^{(0,0,0,2;0)}(\vec{r}) &\sim \frac{1}{|\vec{r}|^{D-2+\eta_{\varphi_\parallel}}},
\end{align}
with the surface critical exponents
\begin{align}
\eta_{\psi_\parallel} = \eta_{\psi_s}(g^{*2},\lambda^*), \qquad
\eta_{\varphi_\parallel}^{\mathrm{(sp)}} = \eta_{\varphi_s}(g^{*2},\lambda^*).
\label{eq:phi-parallel-def}
\end{align}
The latter expression is valid only at the special transition, whereas at the ordinary transition it should be replaced by
\begin{align} \label{eq:eta-phi-parallel-D0}
\eta_{\varphi_\parallel}^{\mathrm{(ord)}} = 2+\eta_{\partial\varphi_s}(\lambda^*,g^*),
\end{align}
where $\eta_{\partial\varphi_s}$ is computed using Eq.~\eqref{eq:critical-exponents-gen}.

The critical boundary/bulk two-point functions exhibit power-law decay along the direction perpendicular to the boundary and are given by
\begin{align}
 G^{(1,1,0,0;0)}(z) &\sim \frac{(\unitmatrix + M)}{z^{D-1+\eta_{\psi_\perp}}},
\quad
\eta_{\psi_\perp} = \frac{1}{2} (\eta_{\psi} + \eta_{\psi_\parallel}),
\\
 G^{(0,0,1,1;0)}(z) &\sim \frac{1}{z^{D-2+\eta_{\varphi_\perp}}},
\quad
\eta_{\varphi_\perp} = \frac{1}{2} (\eta_{\varphi} + \eta_{\varphi_\parallel}).
\end{align}
The scaling dimensions of the boundary fermion and scalar operators are then expressed as
\begin{align}
\Delta_{\psi_s} &= \frac{1}{2} (D-1 + \eta_{\psi_s}),
&
\Delta_{\varphi_s} &= \frac{1}{2} (D-2 + \eta_{\varphi_s}).
\end{align}

At the special transition, the surface coupling $c$ is a relevant parameter.
Physically, this means that a small deviation of $c$ from its fixed-point value
($c = 0$  within the MS scheme) grows under the RG flow.
This relevance corresponds to the surface composite operator $\varphi_s^2$ being relevant, i.e. its scaling dimension satisfies
$\Delta_{\varphi_s^2} < D - 1$.
The flow of $c$ defines  the crossover exponent $\Phi$, which characterizes how the system crosses over
from the special transition to the ordinary surface transition.
Near the special transition point $m=c=0$, all observables-such as the free
energy, correlation functions, or response functions depend on the scaling combination
$c\, m^{-2\Phi}$.
In terms of the anomalous dimension given by Eq.~\eqref{eq:critical-exponents-gen} the dimension of the surface composite operator $\varphi_s^2$ and the crossover exponent can be written as
\begin{align}
\Delta_{\varphi_s^2} &= D - 2 + \eta_{\varphi_s^2},
&
\Phi &= \nu \left[1-\eta_{\varphi_s^2}(\lambda^*,g^*)\right].
\label{eq:Phi-def}
\end{align}
The scaling dimensions of the fermion composite operators in the bulk and on the boundary are given by
\begin{align}
& \Delta_{O_l} = D-1+\eta_{O_l}, \ \ \ (l\ge2). \label{eq:scaling-dimension-O-l}
\end{align}
In the bulk, the composite operator $O_2=\bar{\psi}\psi$ is a descendant of the operator $\varphi$, so that their scaling dimensions are related by~\cite{Fei:2016}
\begin{align}
& \Delta_{\bar{\psi}\psi} = 2 + \Delta_{\varphi}.  \label{eq:Delta-psi-Delta-phi-bulk}
\end{align}
As we will see below this is not the case for the composite operator $(\bar{\psi}\psi)_s$ on the boundary, its scaling dimension cannot be related to that of $\varphi_s$.
We will show that the boundary operator $O_3=(\bar{\psi}\psi)_s$ is irrelevant at the non-TRI FPs, i.e. $\Delta_{(\bar{\psi}\psi)_s}>D-1$, while
the boundary operator $O_4=(\bar\psi \gamma_S\psi)_s$ is relevant at the TRI FPs, i.e. $\Delta_{(\bar\psi \gamma_S\psi)_s}<D-1$. This is consistent
with the RG flows depicted in Fig.~\ref{fig:1-RGFlow}.

\section{Renormalization in the bulk to one-loop order} \label{sec:bulk}

In this Section, we outline the known results on the renormalization of the GNY model in the bulk to one-loop order with proper normalization, which we need to study the boundary criticality.
The bulk renormalization constants to one-loop order are given by~\cite{Zinn-Justin:1986}
\begin{align}
Z_{\psi}  &= 1 -   g^2 \frac1{\varepsilon} ,
\label{eq:psi-bulk-renormalization} \\
Z_{\varphi}  &= 1 -   N g^2 \frac1{\varepsilon} ,
\label{eq:phi-bulk-renormalization} \\
\mathring{m}^2  &= m^{2}  \Big[1+ (N g^2 + \lambda) \frac1{\varepsilon} \Big],
\label{eq:mass-renormalization} \\
\mathring{g}^2  &= \frac{2 m^{\varepsilon }}{K_D} g^2 \Big[1+ ({N+6}) g^2 \frac1{\varepsilon} \Big],
\label{eq:g-renormalization} \\
\mathring{\lambda} &= \frac{2 m^{\varepsilon }}{K_D} \Big[\lambda + (3\lambda^2 + 2 N \lambda g^2 - 12 N g^4 ) \frac1{\varepsilon} \Big],
\label{eq:lambda-renormalization}
\end{align}
where we define $N = \tilde{N} \, \mathrm{tr}\,\mathbb{1}$ and $D = 4 - \varepsilon$. Using Eqs.~\eqref{eq:g-renormalization}-\eqref{eq:lambda-renormalization} we derive the corresponding $\beta$-functions defined by Eqs.~\eqref{eq:beta-functions-gen}, which read to one loop order
\begin{align}
\beta_{g^2} (g^2,\lambda)  &= -\varepsilon g^2 + (N+6) g^4,
\label{eq:beta-g2} \\
\beta_{\lambda} (g^2,\lambda)  &= - \varepsilon \lambda + 3\lambda^2 + 2 N \lambda g^2 - 12 N g^4.
\label{eq:beta-lambda}
\end{align}
These $\beta$-functions have three FPs: (i) the Gaussian FP ($g^* = \lambda^* = 0$) with eigenvalues $(-\varepsilon, -\varepsilon)$ of the stability matrix~\eqref{eq:stability0}, indicating instability in both directions; (ii) the Ising FP ($g^* = 0, \lambda^* = \frac{\varepsilon}{3}$) with eigenvalues $(-\varepsilon, \varepsilon)$, which is unstable along one direction;  (iii) the GNY FP with
\begin{align}
&g^{*2} = \frac{\varepsilon}{N+6},
&
\lambda^{*}  &=  \frac{ Y(N) - N + 6 }{6(N+6)}\varepsilon,
\label{eq:fp-lambda}
\end{align}
where we have introduced the shorthand notation
\begin{align}
Y(N)= \sqrt{N^2+132N+36}.
\label{Y-N}
\end{align}
The corresponding eigenvalues are $\left( \varepsilon, \frac{Y(N)}{N+6}\varepsilon \right)$
and thus, the GNY FP is IR stable for $D<4$.
Using Eqs.~\eqref{eq:psi-bulk-renormalization}--\eqref{eq:mass-renormalization}, we calculate the bulk scaling functions~\eqref{eq:critical-exponents-gen} as
\begin{align}
\eta_{\psi}(g^2,\lambda) &= g^2,
\qquad
\eta_{\varphi}(g^2,\lambda) = N g^2, \label{eq:eta-bulk-1loop} \\
\eta_{m}(g^2,\lambda) &= - N g^2 - \lambda, \label{eq:eta-m-bulk-1loop}
\end{align}
which yield the bulk critical exponents at the GNY FP to one-loop order:
\begin{align}
\eta_{\psi}  &= \frac{\varepsilon}{N+6},
\qquad
\eta_{\varphi} = \frac{N\varepsilon}{N+6},
\nonumber \\
\frac1{\nu} &= 2 - \frac{ Y(N) +5 N + 6 }{6(N+6)}\varepsilon.
\label{eq:eta-exp-bulk-1loop}
\end{align}
In the following Section, we turn to the renormalization of the fermion and boson operators at the boundary for both the special and ordinary transitions.

\section{Special and ordinary transitions to one-loop order} \label{sec:boundary}

While the renormalization of the boundary fermion can be treated on the same footing for both the special and ordinary transitions, the renormalization of the boundary boson requires distinct treatments in each case.

\subsection{Boundary fermion renormalization  at the special and ordinary transitions}
The renormalization of the boundary fermion field can be carried out in the same manner for both the special and ordinary transitions. To determine the renormalization constant $Z_{\psi_s}$ for the boundary fermions, we consider the two-point Green function $\mathring{G}^{(0,2,0,0;0)}(\vec{p})$ evaluated at the boundary ($z=0$). At one-loop order, its bare value can be represented diagrammatically as
\begin{equation}
\begin{tikzpicture}[baseline={([yshift=-2.5ex]current bounding box.center)},x=0.75pt,y=0.75pt,yscale=-0.7,xscale=0.7]
%uncomment if require: \path (0,76); %set diagram left start at 0, and has height of 76
%Straight Lines
\draw    (143,51) -- (219.5,51) ;
\draw [shift={(186.25,51)}, rotate = 180] [fill={rgb, 255:red, 0; green, 0; blue, 0 }  ][line width=0.1]  [draw opacity=0] (8.93,-4.29) -- (0,0) -- (8.93,4.29) -- cycle    ;
%Curve Lines
\draw  [dash pattern={on 5pt off 5pt}]  (285,51) .. controls (296.65,31.81) and (310.3,22.08) .. (324.51,22.79) .. controls (327.43,22.93) and (330.36,23.52) .. (333.31,24.54) .. controls (343.76,28.17) and (354.34,37.33) .. (364.5,51) ;
%Straight Lines
\draw    (268,51) -- (380.5,51) ;
\draw [shift={(329.25,51)}, rotate = 180] [fill={rgb, 255:red, 0; green, 0; blue, 0 }  ][line width=0.08]  [draw opacity=0] (8.93,-4.29) -- (0,0) -- (8.93,4.29) -- cycle    ;
%Straight Lines
\draw    (259.3,50.9) -- (285,51) ;
\draw [shift={(277.15,51)}, rotate = 180.22] [fill={rgb, 255:red, 0; green, 0; blue, 0 }  ][line width=0.08]  [draw opacity=0] (8.93,-4.29) -- (0,0) -- (8.93,4.29) -- cycle    ;
%Straight Lines
\draw    (363.3,51) -- (389,51) ;
\draw [shift={(381.15,51)}, rotate = 180.22] [fill={rgb, 255:red, 0; green, 0; blue, 0 }  ][line width=0.08]  [draw opacity=0] (8.93,-4.29) -- (0,0) -- (8.93,4.29) -- cycle    ;
% Text Node
\draw (235,41.4) node [anchor=north west][inner sep=0.75pt]    {$+$};
\draw (176,26) node [anchor=north west][inner sep=0.75pt]    {\scalebox{0.9}{$\vec{p}$}};
\draw (267,26) node [anchor=north west][inner sep=0.75pt]    {\scalebox{0.9}{$\vec{p}$}};
\draw (369,26) node [anchor=north west][inner sep=0.75pt]    {\scalebox{0.9}{$\vec{p}$}};
\draw (317,26) node [anchor=north west][inner sep=0.75pt]    {\scalebox{0.9}{$\vec{q}$}};
\draw (302,0) node [anchor=north west][inner sep=0.75pt]   {\scalebox{0.9}{$\vec{p}-\vec{q}$}};
\draw (139,54) node [anchor=north west][inner sep=0.75pt]    {\scalebox{0.8}{$0$}};
\draw (383,54) node [anchor=north west][inner sep=0.75pt]    {\scalebox{0.8}{$0$}};
\draw (212,54) node [anchor=north west][inner sep=0.75pt]    {\scalebox{0.8}{$0$}};
\draw (258,54) node [anchor=north west][inner sep=0.75pt]    {\scalebox{0.8}{$0$}};
\draw (280,54) node [anchor=north west][inner sep=0.75pt]    {$z_{1}$};
\draw (355,54) node [anchor=north west][inner sep=0.75pt]    {$z_{2}$};
%Straight Lines [id:da27375291005843794]
\draw    (448,51) -- (530.5,51) ;
%Straight Lines [id:da6782358690128715]
\draw    (439.3,51) -- (465,51) ;
\draw [shift={(457.15,51)}, rotate = 180.22] [fill={rgb, 255:red, 0; green, 0; blue, 0 }  ][line width=0.08]  [draw opacity=0] (8.93,-4.29) -- (0,0) -- (8.93,4.29) -- cycle    ;
%Straight Lines [id:da5460049952348773]
\draw    (523.3,51) -- (529,51) ;
\draw [shift={(521.15,51)}, rotate = 180.22] [fill={rgb, 255:red, 0; green, 0; blue, 0 }  ][line width=0.08]  [draw opacity=0] (8.93,-4.29) -- (0,0) -- (8.93,4.29) -- cycle    ;
%Shape: Circle [id:dp21572831570197504]
\draw   (470.89,14.7) .. controls (470.97,6.94) and (477.32,0.71) .. (485.08,0.78) .. controls (492.85,0.86) and (499.08,7.22) .. (499,14.98) .. controls (498.92,22.74) and (492.57,28.97) .. (484.81,28.89) .. controls (477.04,28.81) and (470.82,22.46) .. (470.89,14.7) -- cycle ;
%Straight Lines [id:da5620708730925607]
%\draw    (498,10) -- (498,20.75) ;
\draw [shift={(498,8.88)}, rotate = 91.95] [fill={rgb, 255:red, 0; green, 0; blue, 0 }  ][line width=0.08]  [draw opacity=0] (8.93,-4.29) -- (0,0) -- (8.93,4.29) -- cycle    ;
%Straight Lines [id:da02037541292362055]
\draw  [dash pattern={on 4.5pt off 4.5pt}]  (485,28.89) -- (485,51.8) ;
%%%%%%%%%%%%%%%%%%%%%%%
% Text Node
\draw (410,41.4) node [anchor=north west][inner sep=0.75pt]    {$+$};
\draw (445,26) node [anchor=north west][inner sep=0.75pt]    {\scalebox{0.9}{$\vec{p}$}};
\draw (517,26) node [anchor=north west][inner sep=0.75pt]    {\scalebox{0.9}{$\vec{p}$}};
% Text Node
\draw (520,54) node [anchor=north west][inner sep=0.75pt]    {\scalebox{0.8}{$0$}};
% Text Node
\draw (439,54) node [anchor=north west][inner sep=0.75pt]    {\scalebox{0.8}{$0$}};
% Text Node
\draw (464,27) node [anchor=north west][inner sep=0.75pt]    {\scalebox{1}{$z_{1}$}};
% Text Node
\draw (480.25,53.2) node [anchor=north west][inner sep=0.75pt]    {\scalebox{1}{$z_{2}$}};
% Text Node
\draw (502,2.2) node [anchor=north west][inner sep=0.75pt]    {\scalebox{0.9}{$\vec{q}$}};
\end{tikzpicture}, \label{eq:G-02000-diag}
\end{equation}
where the solid lines with arrows represent the fermion propagator~\eqref{eq:fermion-propagator-general}, with $\sigma = -1$ corresponding to the non-TRI BCs and $\sigma = +1$ to the TRI BCs, while the dashed lines denote the boson propagator~\eqref{eq:boson-propagator-chi}, which satisfies either the Neumann BC ($\chi = +1$) at the special transition or the Dirichlet BC ($\chi = -1$) at the ordinary transition. Evaluating the diagrams in Eq.~\eqref{eq:G-02000-diag} using dimensional regularization and retaining only the pole contributions in $\varepsilon = 4 - D$, we arrive at
\begin{align}
\mathring{G}^{(0,2,0,0;0)}(\vec{p}) &= (\unitmatrix - \gamma_d \check{M}_\sigma)  \frac{-i \slashed{p}}{2 p} \Big[1 + \nn \\
&+ \frac{K_{d}}2 m^{-\varepsilon} \frac{(2+\chi)\sigma}{4 \varepsilon} \mathring{g}^2 \nn \\
&+ \frac{K_{d}}2 m^{-\varepsilon} \frac{(1+\sigma)}{8 \varepsilon} N \mathring{g}^2
 \Big]. \label{eq:G-02000-expression}
\end{align}
Note that the last diagram in Eq.~\eqref{eq:G-02000-diag}, which corresponds to the last line in Eq.~\eqref{eq:G-02000-expression}, vanishes at the non-TRI FP with $\sigma = -1$. Within the MS scheme, the renormalization condition reads
\begin{align}
Z^{-1}_{\psi_s} \, \mathring{G}^{(0,2,0,0;0)}(\vec{p};\mathring{m},\mathring{g},\mathring{\lambda}) = \mathrm{finite},
\end{align}
where the bare quantities $\mathring{m}$, $\mathring{g}$, and $\mathring{\lambda}$ are expressed in terms of the renormalized parameters $m$, $g$, and $\lambda$ according to Eqs.~(\ref{eq:mass-renormalization})--(\ref{eq:lambda-renormalization}). Taking into account that, to lowest order in $\varepsilon$, $K_d = 4 K_D + O(\varepsilon)$, we obtain
\begin{align} \label{eq:psi-surface-renormalization}
Z_{\psi_s} = 1 + \left[ (2+\chi)\sigma + N \frac{\sigma+1}{2} \right] \frac{g^2}{\varepsilon} + O(g^4, \lambda^2, g^2 \lambda).
\end{align}
Substituting Eq.~\eqref{eq:psi-surface-renormalization} into Eq.~\eqref{eq:critical-exponents-gen} yields the scaling function
\begin{align} \label{eq:eta-psi-s}
\eta_{\psi_s}(g^2,\lambda) = - \left[ (2+\chi)\sigma + N \frac{\sigma+1}{2} \right] g^2
\end{align}
to one-loop order. Evaluating Eq.~\eqref{eq:eta-psi-s} at the FP~\eqref{eq:fp-lambda} gives the surface critical exponent and the scaling dimension of the boundary fermion:
\begin{align} \label{eq:eta-psi-parallel}
\eta_{\psi_\parallel} &= - \left[ (2+\chi)\sigma + N \frac{\sigma+1}{2} \right] \frac{\varepsilon}{N+6}, \\
\Delta_{\psi_s} &= \frac{3}{2} - \frac{(\sigma+3)N + 2(2+\chi)\sigma + 12}{4 (N+6)} \varepsilon,
\end{align}
for the TRI ($\sigma=+1$) and non-TRI ($\sigma=-1$) special ($\chi=+1$) and ordinary ($\chi=-1$) transitions. Note that the scaling dimensions of the boundary fermion for the TRI BC ($\sigma=+1$) are consistent with those reported in Ref.~\cite{Giombi:2022}.

\subsection{Boundary boson renormalization at the special transition}
In this Section, we discuss the renormalization of the boundary boson field at the special transition. To determine the renormalization constant $Z_{\phi_s}$ for the boundary scalar field, we consider the two-point Green's function $\mathring{G}^{(0,0,0,2;0)}(\vec{p})$, which satisfies the Neumann BC at $z=0$. At one-loop order, its bare value can be represented diagrammatically as
\begin{eqnarray}
%\!\!\!\!&&\mathring{G}^{(0,0,0,2,0)}_{\mathrm{+}}(\vec{p})= \nonumber \\
 \begin{tikzpicture}[baseline={([yshift=-2.2ex]current bounding box.center)},x=0.75pt,y=0.75pt,yscale=-0.75,xscale=0.75]
%Shape: Ellipse
\draw   (274.99,57) .. controls (274.98,42) and (290.63,29.85) .. (309.96,29.83) .. controls (329.29,29.81) and (344.98,41.92) .. (344.99,57) .. controls (345.01,71.84) and (329.35,83.98) .. (310.02,84) .. controls (290.69,84.02) and (275.01,71.91) .. (274.99,57) -- cycle ;
%Straight Lines
\draw  [dash pattern={on 4pt off 4pt}]  (244.33,57) -- (274.99,57) ;
%Straight Lines
\draw  [dash pattern={on 4pt off 4pt}]  (346.33,57) -- (380,57) ;
%Arrows
\draw [shift={(315.13,29.67)}, rotate = 179.08] [fill={rgb, 255:red, 0; green, 0; blue, 0 }  ][line width=0.08]  [draw opacity=0] (8.93,-4.29) -- (0,0) -- (8.93,4.29) -- cycle    ;
\draw [shift={(305.02,84)}, rotate = 360] [fill={rgb, 255:red, 0; green, 0; blue, 0 }  ][line width=0.08]  [draw opacity=0] (8.93,-4.29) -- (0,0) -- (8.93,4.29) -- cycle    ;
%Straight Lines
\draw  [dash pattern={on 4.5pt off 4.5pt}]  (143.33,58.83) -- (210.33,57.83) ;
%Straight Lines
\draw  [dash pattern={on 4.5pt off 4.5pt}]  (403.15,57.59) -- (485.74,57.7) ;
%Shape: Tear Drop
\draw  [dash pattern={on 4.5pt off 4.5pt}] (429.3,43.41) .. controls (422.04,35.75) and (422.56,23.46) .. (430.46,15.96) .. controls (438.37,8.46) and (450.67,8.59) .. (457.93,16.24) .. controls (465.2,23.9) and (464.68,36.19) .. (456.77,43.69) .. controls (447.24,52.75) and (442.45,57.39) .. (442.45,57.64) .. controls (442.45,57.39) and (438.08,52.65) .. (429.3,43.41) -- cycle ;
% Text Node
\draw (171,33.4) node [anchor=north west][inner sep=0.75pt]    {$\vec{p}$};
\draw (250,33.4) node [anchor=north west][inner sep=0.75pt]    {$\vec{p}$};
\draw (364,33.4) node [anchor=north west][inner sep=0.75pt]    {$\vec{p}$};
\draw (305,59) node [anchor=north west][inner sep=0.75pt]    {$\vec{q}$};
\draw (290,7) node [anchor=north west][inner sep=0.75pt]    {$\vec{p}+\vec{q}$};
\draw (140,61.4) node [anchor=north west][inner sep=0.75pt]    {$0$};
\draw (371,61.4) node [anchor=north west][inner sep=0.75pt]    {$0$};
\draw (205,61.4) node [anchor=north west][inner sep=0.75pt]    {$0$};
\draw (241,61.4) node [anchor=north west][inner sep=0.75pt]    {$0$};
\draw (258,61.4) node [anchor=north west][inner sep=0.75pt]    {$z_{1}$};
\draw (345,61.4) node [anchor=north west][inner sep=0.75pt]    {$z_{2}$};
\draw (222,50) node [anchor=north west][inner sep=0.75pt]    {$+$};
\draw (406,33.4) node [anchor=north west][inner sep=0.75pt]    {$\vec{p}$};
\draw (405,61.4) node [anchor=north west][inner sep=0.75pt]    {$0$};
\draw (472,61.4) node [anchor=north west][inner sep=0.75pt]    {$0$};
\draw (472,33.4) node [anchor=north west][inner sep=0.75pt]    {$\vec{p}$};
\draw (438,61.4) node [anchor=north west][inner sep=0.75pt]    {$z$};
\draw (438,14.4) node [anchor=north west][inner sep=0.75pt]    {$\vec{q}$};
\draw (384,48.4) node [anchor=north west][inner sep=0.75pt]    {$+$};
\end{tikzpicture},  \ \ \  \label{eq:G-00020-diag}
\end{eqnarray}
where the solid lines with arrows represent the fermion propagator~\eqref{eq:fermion-propagator-general}, and the dashed lines denote the boson propagator~\eqref{eq:boson-propagator-chi}, which satisfies the Neumann BC ($\chi = +1$) at the special transition. The diagrams in Eq.~\eqref{eq:G-00020-diag} correspond to the one-loop corrections to the boundary boson two-point function at zero momentum, $\vec{p}=0$. Evaluating these diagrams within dimensional regularization and retaining only the pole contributions in $\varepsilon = 4-D$, we obtain
\begin{align}
\mathring{G}^{(0,0,0,2;0)}(\vec{p}=0) & = \frac{1}{\mathring{m}} \Bigg[ 1 +
\frac{K_{d} m^{-\varepsilon}}{16\varepsilon} \nn \\
& \quad \times \left( 3 \mathring{\lambda} + N (2\sigma+1) \mathring{g}^2 \right) \Bigg],
\label{eq:G00020-0-final}
\end{align}
where the first term $1/\mathring{m}$ corresponds to the tree-level diagram, and the second term represents the leading-order quantum corrections due to both the quartic boson interaction $\mathring{\lambda}$ and the Yukawa coupling $\mathring{g}$.
The renormalization  condition for $\mathring{G}^{(0,0,0,2;0)}$ within the MS scheme reads
\begin{align}
Z^{-1}_{\varphi_s} \mathring{G}^{(0,0,0,2;0)}(\vec{p}=0;\mathring{m},\mathring{g},\mathring{\lambda}) = \mathrm{finite},
\end{align}
where all bare parameters must be expressed in terms of the renormalized ones according to
Eqs.~(\ref{eq:mass-renormalization})--(\ref{eq:lambda-renormalization}). This procedure leads to
\begin{align} \label{eq:phi-surface-renormalization}
Z_{\varphi_s}= 1+\frac{\lambda}{\varepsilon} +\frac{N}{\varepsilon} \sigma  g^2 + O(g^4,\lambda^2, g^2\lambda).
\end{align}
The scaling function $\eta_{\varphi_s}$ defined in Eq.~\eqref{eq:critical-exponents-gen}  is given to  one-loop order by
\begin{align} \label{eq:eta-phi-s}
\eta_{\varphi_s}= -\lambda-N \sigma g^2.
\end{align}
Thus, the boundary critical exponent for the scalar field~\eqref{eq:phi-parallel-def} at the special transition is given by
\begin{align} \label{eq:eta-phi-parallel-N}
\eta_{\varphi_\parallel} = -\frac{\left(Y(N)+ (6\sigma-1) N + 6\right) \varepsilon}{6 (N+6)},
\end{align}
and the boundary boson dimension is
\begin{align} \label{eq:Delta-phi-N}
\Delta_{\varphi_s} = 1-\frac{\left(Y(N)+(6\sigma+5)N+42\right)  \varepsilon }{12 (N+6)},
\end{align}
whose value in the TRI BC case ($\sigma=+1$) is in agreement with the results of Ref.~\cite{Giombi:2022}.

\subsection{Boundary boson renormalization at the ordinary transition}

The ordinary transition is described by the scalar boson field, which satisfies the Dirichlet BC~(\ref{eq:BC-boson-Dirichlet}), obtained from Eq.~\eqref{eq:BC-boson-general} in the limit $c=c_{\mathrm{ord}}^*=\infty$.
In this case, the boson field vanishes at the boundary. To extract the scaling behavior near the boundary, we therefore consider the renormalization of the operator  $\partial_z \mathring{\phi}(\vec{r},z)\big|_{z=0^+}$ following  Eq.~\eqref{eq:Z-partial-phi-s}.  For this purpose, we define the  function
\begin{align}
\mathring{G}_2(\vec{p}) = \frac{\partial^2}{\partial z \partial z'} \mathring{G}^{(0,0,0,2;0)}(p,z,z')\Big|_{z=0,z'=0}.
\label{eq:Green-Dirichlet-1}
\end{align}
At one-loop order, its bare value can be expressed diagrammatically as
\begin{eqnarray}
%\!\!\!\!&&\mathring{G}^{(0,0,0,2,0)}_{\mathrm{+}}(\vec{p})= \nonumber \\
 \begin{tikzpicture}[baseline={([yshift=-2.2ex]current bounding box.center)},x=0.75pt,y=0.75pt,yscale=-0.75,xscale=0.75]
%Shape: Ellipse
\draw   (274.99,57) .. controls (274.98,42) and (290.63,29.85) .. (309.96,29.83) .. controls (329.29,29.81) and (344.98,41.92) .. (344.99,57) .. controls (345.01,71.84) and (329.35,83.98) .. (310.02,84) .. controls (290.69,84.02) and (275.01,71.91) .. (274.99,57) -- cycle ;
%Straight Lines
\draw  [dash pattern={on 4pt off 4pt}]  (244.33,57) -- (274.99,57) ;
%Straight Lines
\draw  [dash pattern={on 4pt off 4pt}]  (346.33,57) -- (380,57) ;
%Arrows
\draw [shift={(315.13,29.67)}, rotate = 179.08] [fill={rgb, 255:red, 0; green, 0; blue, 0 }  ][line width=0.08]  [draw opacity=0] (8.93,-4.29) -- (0,0) -- (8.93,4.29) -- cycle    ;
\draw [shift={(305.02,84)}, rotate = 360] [fill={rgb, 255:red, 0; green, 0; blue, 0 }  ][line width=0.08]  [draw opacity=0] (8.93,-4.29) -- (0,0) -- (8.93,4.29) -- cycle    ;
%Straight Lines
\draw  [dash pattern={on 4.5pt off 4.5pt}]  (143.33,58.83) -- (210.33,57.83) ;
%Straight Lines
\draw  [dash pattern={on 4.5pt off 4.5pt}]  (403.15,57.59) -- (485.74,57.7) ;
%Shape: Tear Drop
\draw  [dash pattern={on 4.5pt off 4.5pt}] (429.3,43.41) .. controls (422.04,35.75) and (422.56,23.46) .. (430.46,15.96) .. controls (438.37,8.46) and (450.67,8.59) .. (457.93,16.24) .. controls (465.2,23.9) and (464.68,36.19) .. (456.77,43.69) .. controls (447.24,52.75) and (442.45,57.39) .. (442.45,57.64) .. controls (442.45,57.39) and (438.08,52.65) .. (429.3,43.41) -- cycle ;
%Straight small Lines
\draw [line width=1.5]    (153,52) -- (153,62) ;
\draw [line width=1.5]    (203,52) -- (203,62) ;
\draw [line width=1.5]    (253,52) -- (253,62) ;
\draw [line width=1.5]    (370,52) -- (370,62) ;
\draw [line width=1.5]    (410,52) -- (410,62) ;
\draw [line width=1.5]    (477,52) -- (477,62) ;
% Text Node
\draw (171,33.4) node [anchor=north west][inner sep=0.75pt]    {$\vec{p}$};
\draw (250,33.4) node [anchor=north west][inner sep=0.75pt]    {$\vec{p}$};
\draw (364,33.4) node [anchor=north west][inner sep=0.75pt]    {$\vec{p}$};
\draw (305,59) node [anchor=north west][inner sep=0.75pt]    {$\vec{q}$};
\draw (290,7) node [anchor=north west][inner sep=0.75pt]    {$\vec{p}+\vec{q}$};
\draw (140,61.4) node [anchor=north west][inner sep=0.75pt]    {$0$};
\draw (371,61.4) node [anchor=north west][inner sep=0.75pt]    {$0$};
\draw (205,61.4) node [anchor=north west][inner sep=0.75pt]    {$0$};
\draw (241,61.4) node [anchor=north west][inner sep=0.75pt]    {$0$};
\draw (258,61.4) node [anchor=north west][inner sep=0.75pt]    {$z_{1}$};
\draw (345,61.4) node [anchor=north west][inner sep=0.75pt]    {$z_{2}$};
\draw (222,50) node [anchor=north west][inner sep=0.75pt]    {$+$};
\draw (406,33.4) node [anchor=north west][inner sep=0.75pt]    {$\vec{p}$};
\draw (405,61.4) node [anchor=north west][inner sep=0.75pt]    {$0$};
\draw (472,61.4) node [anchor=north west][inner sep=0.75pt]    {$0$};
\draw (472,33.4) node [anchor=north west][inner sep=0.75pt]    {$\vec{p}$};
\draw (438,61.4) node [anchor=north west][inner sep=0.75pt]    {$z$};
\draw (438,14.4) node [anchor=north west][inner sep=0.75pt]    {$\vec{q}$};
\draw (384,48.4) node [anchor=north west][inner sep=0.75pt]    {$+$};
\end{tikzpicture},  \ \ \  \label{eq:G-00020-diag-Dirichlet}
\end{eqnarray}
where the solid lines with arrows stand for the fermion propagator \eqref{eq:fermion-propagator-general},
the dashed lines represent the boson propagator \eqref{eq:boson-propagator-chi} with $\chi=-1$
satisfying the Dirichlet BC, and the bold ticks denote derivatives with respect to $z$.
The diagrams in Eq.~\eqref{eq:G-00020-diag-Dirichlet} yield
\begin{align}
\mathring{G}_2(\vec{p}=0) = - \mathring{m}\left[ 1 +
\frac{K_{d} m^{-\varepsilon}}{16\varepsilon}(2\mathring{\lambda}-N (2\sigma+1) \mathring{g}^2)\right].
\label{eq:G2-0-final}
\end{align}
The renormalization condition~\eqref{eq:Z-partial-phi-s} in the MS scheme can be rewritten in terms of the two-point function~\eqref{eq:Green-Dirichlet-1} as
\begin{align} \label{eq:renorm-2}
Z^{-1}_{\partial\varphi_s} \mathring{G}_2(\vec{p}=0;\mathring{m},\mathring{g},\mathring{\lambda}) = \mathrm{finite},
\end{align}
where the bare parameters $\mathring{m}$, $\mathring{g}$, and $\mathring{\lambda}$ have been expressed
in terms of the renormalized parameters $m$, $g$, and $\lambda$
according to Eqs.~(\ref{eq:mass-renormalization})-(\ref{eq:lambda-renormalization}).
We derive the renormalization constant from Eq.~\eqref{eq:renorm-2},
\begin{align} \label{eq:partial-phi-surface-renormalization}
Z_{\partial\varphi_s}= 1+\frac{\lambda}{\varepsilon} -\frac{N}{\varepsilon} \sigma  g^2 + O(g^4,\lambda^2, g^2\lambda).
\end{align}
The corresponding scaling function~\eqref{eq:critical-exponents-gen} to one loop order is
\begin{align} \label{eq:eta-phi-s-ord}
\eta_{\partial\varphi_s}(\lambda,g)= -\lambda+N \sigma g^2.
\end{align}
The boundary critical exponent for the boson field at the ordinary transition defined by~\eqref{eq:eta-phi-parallel-D0}
%\begin{align} \label{eq:eta-phi-parallel-D}
%\eta_{\phi \parallel}^{\mathrm{(ord)}} = 2+\eta_{\partial\phi_s}(\lambda^*,g^*)
%\end{align}
then reads
\begin{align} \label{eq:eta-phi-parallel-norm-1loop}
\eta_{\varphi_\parallel} = 2-\frac{\left(Y(N)-(6\sigma+1) N+6\right) \varepsilon  }{6 (N+6)},
\end{align}
and the scaling dimension of the boundary boson  at the ordinary transition is
\begin{align} \label{eq:Delta-phi-D}
\Delta_{\varphi_s} = 2-\frac{\left(Y(N)-(6\sigma-5)N+42\right) \varepsilon  }{12 (N+6)},
\end{align}
whose value for the TRI BC ($\sigma=+1$) is in agreement with Ref.~\cite{Giombi:2022}.

\section{Renormalization of the composite operators} \label{sec:composite-operators}

\subsection{Renormalization of composite operator $(\bar\psi \psi)_s$ at non-TRI surface transitions}

In the bulk, the composite operator $(\bar{\psi}\psi)$ is a descendant of the bosonic operator $\varphi$, so their scaling dimensions are related by Eq.~\eqref{eq:Delta-psi-Delta-phi-bulk}. We now show that this relation does not hold for the corresponding boundary operators $(\bar{\psi}\psi)_s$ and $\varphi_s$. It is straightforward to verify that inserting the boundary composite operator $(\bar{\psi}\psi)_s$ into Green's functions leads to their vanishing in the case of  TRI BCs for fermions, similar to the case where the insertion of the boundary operator $\varphi_s$ into Green's functions vanishes under Dirichlet BCs for bosons. Here, we restrict ourselves to the case of non-TRI
BCs~\eqref{eq:M-sigma} with $\sigma=-1$.

To find the renormalization constant $Z_{(\bar{\psi}\psi)_s}$, we consider the two-point Green's function $\mathring{G}^{(0,0,0,2;3)}(\vec{p})$ with an insertion of one boundary composite operator $(\bar{\psi}\psi)_s$. To one-loop order, its bare value can be written diagrammatically as:
\begin{eqnarray}
\tikzset{every picture/.style={line width=0.75pt}} %set default line width to 0.75pt
\begin{tikzpicture}[baseline={([yshift=-2.2ex]current bounding box.center)},x=0.75pt,y=0.75pt,yscale=-0.7,xscale=0.7]
%uncomment if require: \path (0,335); %set diagram left start at 0, and has height of 335
%
%Straight Lines [id:da524471051925741]
\draw    (48.63,99) -- (88,99) ;
\draw [shift={(73.32,99)}, rotate = 180.44] [fill={rgb, 255:red, 0; green, 0; blue, 0 }  ][line width=0.08]  [draw opacity=0] (8.93,-4.29) -- (0,0) -- (8.93,4.29) -- cycle    ;
%Curve Lines [id:da15236466891029288]
\draw  [dash pattern={on 4.5pt off 4.5pt}]  (169.63,99) .. controls (175.63,80.75) and (175.3,70.08) .. (189.51,70.79) .. controls (203.72,71.5) and (195.36,71.52) .. (198.31,72.54) .. controls (201.26,73.57) and (203.63,82.75) .. (209.63,99) ;
%Straight Lines [id:da6913231230142012]
\draw    (150,99) -- (227.63,99) ;
\draw [shift={(193.82,99)}, rotate = 179.82] [fill={rgb, 255:red, 0; green, 0; blue, 0 }  ][line width=0.08]  [draw opacity=0] (8.93,-4.29) -- (0,0) -- (8.93,4.29) -- cycle    ;
%Straight Lines [id:da003613862166628179]
\draw    (137.63,99) -- (149,99) ;
\draw [shift={(148.31,99)}, rotate = 180] [fill={rgb, 255:red, 0; green, 0; blue, 0 }  ][line width=0.08]  [draw opacity=0] (8.93,-4.29) -- (0,0) -- (8.93,4.29) -- cycle    ;
%Straight Lines [id:da47575836933187177]
\draw    (205.63,99) -- (234,99) ;
\draw [shift={(224.82,99)}, rotate = 180] [fill={rgb, 255:red, 0; green, 0; blue, 0 }  ][line width=0.08]  [draw opacity=0] (8.93,-4.29) -- (0,0) -- (8.93,4.29) -- cycle    ;
%Straight Lines [id:da16332510916447107]
\draw    (182,188) -- (238.63,188) ;
%Straight Lines [id:da07030418937610294]
\draw    (174,188) -- (199.7,188) ;
\draw [shift={(191.85,188)}, rotate = 180] [fill={rgb, 255:red, 0; green, 0; blue, 0 }  ][line width=0.08]  [draw opacity=0] (8.93,-4.29) -- (0,0) -- (8.93,4.29) -- cycle    ;
%Straight Lines [id:da34221715633825445]
\draw    (213.75,188) -- (237.63,188) ;
\draw [shift={(230.69,188)}, rotate = 180] [fill={rgb, 255:red, 0; green, 0; blue, 0 }  ][line width=0.08]  [draw opacity=0] (8.93,-4.29) -- (0,0) -- (8.93,4.29) -- cycle    ;
%Shape: Circle [id:dp6664629901104352]
\draw   (195.89,147.7) .. controls (195.97,139.94) and (202.32,133.71) .. (210.08,133.78) .. controls (217.85,133.86) and (224.08,140.22) .. (224,147.98) .. controls (223.92,155.74) and (217.57,161.97) .. (209.81,161.89) .. controls (202.04,161.81) and (195.82,155.46) .. (195.89,147.7) -- cycle ;
%Straight Lines [id:da906421075499679]
\draw    (223,143) -- (223,153.75) ;
\draw [shift={(223,141.88)}, rotate = 90] [fill={rgb, 255:red, 0; green, 0; blue, 0 }  ][line width=0.08]  [draw opacity=0] (8.93,-4.29) -- (0,0) -- (8.93,4.29) -- cycle    ;
%Straight Lines [id:da5193870757810901]
\draw  [dash pattern={on 4.5pt off 4.5pt}]  (210,161.89) -- (210,188.2) ;
%Straight Lines [id:da6733512306314186]
\draw    (68,99) -- (107.63,98.75) ;
\draw [shift={(92.82,99)}, rotate = 180] [fill={rgb, 255:red, 0; green, 0; blue, 0 }  ][line width=0.08]  [draw opacity=0] (8.93,-4.29) -- (0,0) -- (8.93,4.29) -- cycle    ;
%Curve Lines [id:da16334916659149756]
\draw [line width=2.25]    (78.28,76.25) .. controls (86.43,81.77) and (70.13,81.77) .. (78.08,87.88) ;
%Curve Lines [id:da8626273372179926]
\draw [line width=2.25]    (78.08,87.88) .. controls (86.23,93.4) and (69.94,93.4) .. (77.89,99.5) ;
%
%Curve Lines [id:da08589197801772619]
\draw [line width=2.25]    (155.28,76.25) .. controls (163.43,81.77) and (147.13,81.77) .. (155.08,87.88) ;
%Curve Lines [id:da3870385192731519]
\draw [line width=2.25]    (155.08,87.88) .. controls (163.23,93.4) and (146.94,93.4) .. (154.89,99.5) ;
%
%Straight Lines [id:da140066799990613]
\draw    (157.63,99) -- (169,99) ;
\draw [shift={(168.31,99)}, rotate = 180] [fill={rgb, 255:red, 0; green, 0; blue, 0 }  ][line width=0.08]  [draw opacity=0] (8.93,-4.29) -- (0,0) -- (8.93,4.29) -- cycle    ;
%Curve Lines [id:da5772664456268147]
\draw  [dash pattern={on 4.5pt off 4.5pt}]  (280.63,99) .. controls (286.63,81.75) and (286.3,71.08) .. (300.51,71.79) .. controls (314.72,72.5) and (306.36,72.52) .. (309.31,73.54) .. controls (312.26,74.57) and (314.63,83.75) .. (320.63,99) ;
%Straight Lines [id:da4590539432746792]
\draw    (278,99) -- (343.13,99) ;
\draw [shift={(315.57,99)}, rotate = 180] [fill={rgb, 255:red, 0; green, 0; blue, 0 }  ][line width=0.08]  [draw opacity=0] (8.93,-4.29) -- (0,0) -- (8.93,4.29) -- cycle    ;
%Straight Lines [id:da05785783289240343]
\draw    (263.63,99) -- (278.63,99) ;
\draw [shift={(276.12,99)}, rotate = 180] [fill={rgb, 255:red, 0; green, 0; blue, 0 }  ][line width=0.08]  [draw opacity=0] (8.93,-4.29) -- (0,0) -- (8.93,4.29) -- cycle    ;
%Straight Lines [id:da1561083583941536]
\draw    (319.63,99) -- (340.63,99) ;
\draw [shift={(335.13,99)}, rotate = 180] [fill={rgb, 255:red, 0; green, 0; blue, 0 }  ][line width=0.08]  [draw opacity=0] (8.93,-4.29) -- (0,0) -- (8.93,4.29) -- cycle    ;
%Curve Lines [id:da13612305544560177]
\draw [line width=2.25]    (302.28,77.25) .. controls (310.43,82.77) and (294.13,82.77) .. (302.08,88.88) ;
%Curve Lines [id:da39653160022940337]
\draw [line width=2.25]    (302.08,88.88) .. controls (310.23,94.4) and (293.94,94.4) .. (301.89,100.5) ;
%
%Straight Lines [id:da18868731776671988]
\draw    (284.63,99) -- (296,99) ;
\draw [shift={(295.31,99)}, rotate = 180] [fill={rgb, 255:red, 0; green, 0; blue, 0 }  ][line width=0.08]  [draw opacity=0] (8.93,-4.29) -- (0,0) -- (8.93,4.29) -- cycle    ;
%Curve Lines [id:da5685034201101412]
\draw [line width=2.25]    (194.28,166.25) .. controls (202.43,171.77) and (186.13,171.77) .. (194.08,177.88) ;
%Curve Lines [id:da8435793772995859]
\draw [line width=2.25]    (194.08,177.88) .. controls (202.23,183.4) and (185.94,183.4) .. (193.89,189.5) ;
%
%Straight Lines [id:da7759340463822407]
\draw    (189.63,188) -- (218,188) ;
\draw [shift={(208.82,188)}, rotate = 180] [fill={rgb, 255:red, 0; green, 0; blue, 0 }  ][line width=0.08]  [draw opacity=0] (8.93,-4.29) -- (0,0) -- (8.93,4.29) -- cycle    ;
%Curve Lines [id:da7476518372560567]
\draw  [dash pattern={on 4.5pt off 4.5pt}]  (75.63,188) .. controls (81.63,169.75) and (81.3,159.08) .. (95.51,159.79) .. controls (109.72,160.5) and (101.36,160.52) .. (104.31,161.54) .. controls (107.26,162.57) and (109.63,171.75) .. (115.63,188) ;
%Straight Lines [id:da9052089818176963]
\draw    (64.32,188) -- (129.45,188) ;
\draw [shift={(101.88,188)}, rotate = 180] [fill={rgb, 255:red, 0; green, 0; blue, 0 }  ][line width=0.08]  [draw opacity=0] (8.93,-4.29) -- (0,0) -- (8.93,4.29) -- cycle    ;
%Straight Lines [id:da05742632200035769]
\draw    (58.63,188) -- (73.63,188) ;
\draw [shift={(71.12,188)}, rotate = 180] [fill={rgb, 255:red, 0; green, 0; blue, 0 }  ][line width=0.08]  [draw opacity=0] (8.93,-4.29) -- (0,0) -- (8.93,4.29) -- cycle    ;
%Straight Lines [id:da3795045866791317]
\draw    (130.45,188) -- (151.45,188) ;
\draw [shift={(145.95,188)}, rotate = 180] [fill={rgb, 255:red, 0; green, 0; blue, 0 }  ][line width=0.08]  [draw opacity=0] (8.93,-4.29) -- (0,0) -- (8.93,4.29) -- cycle    ;
%Curve Lines [id:da019214548833602962]
\draw [line width=2.25]    (133.28,165.25) .. controls (141.43,170.77) and (125.13,170.77) .. (133.08,176.88) ;
%Curve Lines [id:da11452331967455398]
\draw [line width=2.25]    (133.08,176.88) .. controls (141.23,182.4) and (124.94,182.4) .. (132.89,188.5) ;
%
%Straight Lines [id:da5385910724513064]
\draw    (118.63,188) -- (130,188) ;
\draw [shift={(129.31,188)}, rotate = 180] [fill={rgb, 255:red, 0; green, 0; blue, 0 }  ][line width=0.08]  [draw opacity=0] (8.93,-4.29) -- (0,0) -- (8.93,4.29) -- cycle    ;
%Straight Lines [id:da8764566232343617]
\draw    (271,188) -- (327.63,188) ;
%Straight Lines [id:da5652466474275477]
\draw    (302.75,188) -- (326.63,188) ;
\draw [shift={(319.69,188)}, rotate = 180] [fill={rgb, 255:red, 0; green, 0; blue, 0 }  ][line width=0.08]  [draw opacity=0] (8.93,-4.29) -- (0,0) -- (8.93,4.29) -- cycle    ;
%Shape: Circle [id:dp4710649654008622]
\draw   (284.89,148.7) .. controls (284.97,140.94) and (291.32,134.71) .. (299.08,134.78) .. controls (306.85,134.86) and (313.08,141.22) .. (313,148.98) .. controls (312.92,156.74) and (306.57,162.97) .. (298.81,162.89) .. controls (291.04,162.81) and (284.82,156.46) .. (284.89,148.7) -- cycle ;
%Straight Lines [id:da47662056421162413]
\draw    (312,144) -- (312,154.75) ;
\draw [shift={(312,142.88)}, rotate = 90] [fill={rgb, 255:red, 0; green, 0; blue, 0 }  ][line width=0.08]  [draw opacity=0] (8.93,-4.29) -- (0,0) -- (8.93,4.29) -- cycle    ;
%Straight Lines [id:da8543493370601879]
\draw  [dash pattern={on 4.5pt off 4.5pt}]  (298.81,162.89) -- (299.25,189.2) ;
%Curve Lines [id:da2567103524540332]
\draw [line width=2.25]    (285.76,148) .. controls (280.24,156) and (280.24,140) .. (274.13,148) ;
%Curve Lines [id:da18833039598556178]
\draw [line width=2.25]    (274.13,148) .. controls (268.61,156) and (268.61,140) .. (262.51,148) ;
%
%Straight Lines [id:da8419616480539469]
\draw    (270.63,188) -- (299,188) ;
\draw [shift={(289.82,188.92)}, rotate = 180.5] [fill={rgb, 255:red, 0; green, 0; blue, 0 }  ][line width=0.08]  [draw opacity=0] (8.93,-4.29) -- (0,0) -- (8.93,4.29) -- cycle    ;
%Straight Lines [id:da2905995591849262]
\draw    (357,188) -- (413.63,188) ;
%Straight Lines [id:da2344901310839248]
\draw    (358,188) -- (383.7,188) ;
\draw [shift={(375.85,188)}, rotate = 180] [fill={rgb, 255:red, 0; green, 0; blue, 0 }  ][line width=0.08]  [draw opacity=0] (8.93,-4.29) -- (0,0) -- (8.93,4.29) -- cycle    ;
%Straight Lines [id:da6871601488216461]
\draw    (395.75,188) -- (419.63,188) ;
\draw [shift={(412.69,188)}, rotate = 180] [fill={rgb, 255:red, 0; green, 0; blue, 0 }  ][line width=0.08]  [draw opacity=0] (8.93,-4.29) -- (0,0) -- (8.93,4.29) -- cycle    ;
%Shape: Circle [id:dp16244424646751232]
\draw   (370.89,147.7) .. controls (370.97,139.94) and (377.32,133.71) .. (385.08,133.78) .. controls (392.85,133.86) and (399.08,140.22) .. (399,147.98) .. controls (398.92,155.74) and (392.57,161.97) .. (384.81,161.89) .. controls (377.04,161.81) and (370.82,155.46) .. (370.89,147.7) -- cycle ;
%Straight Lines [id:da35036586126339797]
\draw    (398,143) -- (398,153.75) ;
\draw [shift={(398,141.88)}, rotate = 90] [fill={rgb, 255:red, 0; green, 0; blue, 0 }  ][line width=0.08]  [draw opacity=0] (8.93,-4.29) -- (0,0) -- (8.93,4.29) -- cycle    ;
%Straight Lines [id:da3248854717710107]
\draw  [dash pattern={on 4.5pt off 4.5pt}]  (384.81,161.89) -- (385.25,188) ;
%Curve Lines [id:da4086331496856235]
\draw [line width=2.25]    (400.28,165.25) .. controls (408.43,170.77) and (392.13,170.77) .. (400.08,176.88) ;
%Curve Lines [id:da7752730138430365]
\draw [line width=2.25]    (400.08,176.88) .. controls (408.23,182.4) and (391.94,182.4) .. (399.89,188) ;
%
%Straight Lines [id:da5513460315390474]
\draw    (377.63,188) -- (406,188) ;
\draw [shift={(396.82,187.92)}, rotate = 180.5] [fill={rgb, 255:red, 0; green, 0; blue, 0 }  ][line width=0.08]  [draw opacity=0] (8.93,-4.29) -- (0,0) -- (8.93,4.29) -- cycle    ;
%
% Text Node
\draw (116,89.4) node [anchor=north west][inner sep=0.75pt]    {$+$};
\draw (44,102) node [anchor=north west][inner sep=0.75pt]    {$0$};
\draw (227,102) node [anchor=north west][inner sep=0.75pt]    {$0$};
\draw (102,102) node [anchor=north west][inner sep=0.75pt]    {$0$};
\draw (134,102) node [anchor=north west][inner sep=0.75pt]    {$0$};
\draw (227.69,191) node [anchor=north west][inner sep=0.75pt]    {$0$};
\draw (171,191) node [anchor=north west][inner sep=0.75pt]    {$0$};
\draw (242,89.4) node [anchor=north west][inner sep=0.75pt]    {$+$};
\draw (331,102) node [anchor=north west][inner sep=0.75pt]    {$0$};
\draw (255,102) node [anchor=north west][inner sep=0.75pt]    {$0$};
\draw (154,179) node [anchor=north west][inner sep=0.75pt]    {$+$};
\draw (37,179) node [anchor=north west][inner sep=0.75pt]    {$+$};
\draw (132,191) node [anchor=north west][inner sep=0.75pt]    {$0$};
\draw (56,191) node [anchor=north west][inner sep=0.75pt]    {$0$};
\draw (316.69,191) node [anchor=north west][inner sep=0.75pt]    {$0$};
\draw (267,191) node [anchor=north west][inner sep=0.75pt]    {$0$};
\draw (246,179) node [anchor=north west][inner sep=0.75pt]    {$+$};
\draw (408.69,191) node [anchor=north west][inner sep=0.75pt]    {$0$};
\draw (354,191) node [anchor=north west][inner sep=0.75pt]    {$0$};
\draw (334,179) node [anchor=north west][inner sep=0.75pt]    {$+$};
\draw (420,175) node [anchor=north west][inner sep=0.75pt]    {,};
\end{tikzpicture}
\label{eq:diagrams-psi-composite}
\end{eqnarray}
where the wavy lines denote the $(\bar\psi \psi)_s$-insertions.
Evaluation of the diagrams in Eq.~\eqref{eq:diagrams-psi-composite} yields
\begin{align}
\delta^{(0)} \mathring{G}^{(0,0,0,2;3)}(\vec{p}=0) & = -\frac12 (\unitmatrix - \gamma_d \check{M}_\sigma) \Big[ 1 -  \frac{K_{d}}{2}   \nn \\
 & \times \frac{m^{-\varepsilon}}{\varepsilon}\mathring{g}^2 \left( 2+\chi + \frac{N}2  \right)\Big].
\label{eq:G00021xxx-0-final}
\end{align}
The renormalization  condition within the MS scheme reads
\begin{align}
Z^{-1}_{\psi_s} Z^{-1}_{(\bar{\psi}\psi)_s}  \mathring{G}^{(0,0,0,2;3)}(\vec{p}=0;\mathring{m},\mathring{g},\mathring{\lambda}) = \mathrm{finite},
\end{align}
where all bare parameters have to be expressed in terms of the renormalized ones according to
Eqs.~(\ref{eq:mass-renormalization})-(\ref{eq:lambda-renormalization}). This leads to
\begin{align} \label{eq:Z-c-1}
Z_{(\bar{\psi}\psi)_s} = 1-\left(3(2+\chi) +2 N \right)  \frac{g^2}{\varepsilon} + O(g^4,\lambda^2, g^2\lambda).
\end{align}
Using Eq.~\eqref{eq:scaling-dimension-O-l} we obtain the scaling dimension of the boundary fermion composite operator
\begin{align} \label{eq:Z-c-2}
\Delta_{(\bar{\psi}\psi)_s} = 3+\frac{N+3\chi}{N+6}\varepsilon,
\end{align}
which cannot be expressed as a simple linear function of the scaling dimension $\Delta_{\varphi_s}$  given by Eqs.~\eqref{eq:Delta-phi-N}  and  \eqref{eq:Delta-phi-D} for $\chi=\pm1$,  similarly to Eq.~\eqref{eq:Delta-psi-Delta-phi-bulk}.
Moreover, since $\Delta_{(\bar{\psi}\psi)_s}>D-1$, the boundary operator $(\bar{\psi}\psi)_s$ is irrelevant at both the non-TRI special and ordinary FPs, implying that these FPs are stable.  This conclusion is consistent with the RG flows shown in Fig.~\ref{fig:1-RGFlow}.

\subsection{Renormalization of composite operator $(\bar\psi \gamma_S \psi)_s$ at TRI surface transitions}

It is straightforward to verify that inserting the boundary composite operator $(\bar\psi \gamma_S \psi)_s$ into Green's functions leads to their vanishing in the case of non-TRI BC for fermions. Here we consider only TRI BC~\eqref{eq:M-sigma} with $\sigma=+1$.
To find the renormalization constant $Z_{(\bar\psi \gamma_S \psi)_s}$ we consider the two-point Green's function $\mathring{G}^{(0,0,0,2;4)}(\vec{p})$ with an insertion of one boundary composite operator $(\bar\psi \gamma_S \psi)_s$. To one-loop order its bare value can be written diagrammatically as in Eq.~\eqref{eq:diagrams-psi-composite} with the difference that the wavy lines denote the $(\bar\psi \gamma_S \psi)_s$ insertions.
Evaluation of the diagrams in Eq.~\eqref{eq:diagrams-psi-composite} yields
\begin{align}
\delta^{(0)} \mathring{G}^{(0,0,0,2;4)}(\vec{p}=0) & = \frac12 (\unitmatrix - \gamma_d \check{M}_\sigma) \gamma_S \Big[ 1 + \frac{K_{d}}{2}   \nn \\
 &  \times \frac{m^{-\varepsilon}}{\varepsilon} \mathring{g}^2 \left( 2+\chi + \frac{N}2  \right)\Big].
\label{eq:G00021xxxcc-0-final}
\end{align}
The renormalization  condition within the MS  scheme reads
\begin{align} \label{eq:Z-cond-2}
Z^{-1}_{\psi_s} Z^{-1}_{(\bar{\psi}\gamma_S\psi)_s}  \mathring{G}^{(0,0,0,2;4)}(\vec{p}=0;\mathring{m},\mathring{g},\mathring{\lambda}) = \mathrm{finite},
\end{align}
where all bare parameters are expressed in terms of the corresponding renormalized ones, as prescribed by Eqs.~(\ref{eq:mass-renormalization})-(\ref{eq:lambda-renormalization}). Then the condition~\eqref{eq:Z-cond-2} gives
\begin{align} \label{eq:Z-c-3}
Z_{(\bar{\psi}\gamma_S\psi)_s} = 1+\left(3(2+\chi) +N \right)  \frac{g^2}{\varepsilon} + O(g^4,\lambda^2, g^2\lambda).
\end{align}
The scaling dimension of the boundary composite operator $(\bar\psi \gamma_S \psi)_s$, as given by Eq.~\eqref{eq:scaling-dimension-O-l}, reads at one-loop order
\begin{align} \label{eq:Z-c-4}
\Delta_{(\bar{\psi}\gamma_S\psi)_s} = 3-\frac{2(N+6)+3\chi}{N+6}\varepsilon.
\end{align}
Since $\Delta_{(\bar\psi \gamma_S\psi)_s} < D-1$, the boundary operator $(\bar\psi \gamma_S\psi)_s$ is relevant at the TRI ordinary and special FPs, in agreement with the RG flows shown in Fig.~\ref{fig:1-RGFlow}.

\subsection{Bosonic crossover exponent at the special transitions}

In this Section, we consider the renormalization of the boundary parameter $c$ at the special transition. The parameter $c$ appears in the boundary part of the action as a coupling to the operator $\varphi_s^2$, and hence its renormalization is directly related to the renormalization of the composite boundary operator $\varphi_s^2$.

To determine the corresponding renormalization constant $Z_{\varphi^2_s}$, we consider the two-point Green's function $\mathring{G}^{(0,0,0,2;2)}(\vec{p})$ with an insertion of one boundary composite operator $\varphi_s^2$. To one-loop order, its bare value can be written diagrammatically as:
\begin{eqnarray}
 \begin{tikzpicture}[baseline={([yshift=-2.2ex]current bounding box.center)},x=0.75pt,y=0.75pt,yscale=-0.7,xscale=0.7]
%uncomment if require: \path (0,157); %set diagram left start at 0, and has height of 157
%Shape: Ellipse [id:dp4468363650881282]
\draw   (267.99,31.21) .. controls (267.97,19.98) and (278.63,10.86) .. (291.78,10.85) .. controls (304.93,10.83) and (315.6,19.93) .. (315.61,31.16) .. controls (315.62,42.39) and (304.97,51.51) .. (291.82,51.52) .. controls (278.67,51.54) and (268,42.44) .. (267.99,31.21) -- cycle ;
%Straight Lines [id:da37525237454727645]
\draw  [dash pattern={on 4.5pt off 4.5pt}]  (237.33,31.09) -- (267.99,31.21) ;
%Straight Lines [id:da4629311921035586]
\draw    (285.96,10.83) -- (296.3,10.67) ;
\draw [shift={(296.13,10.67)}, rotate = 179.08] [fill={rgb, 255:red, 0; green, 0; blue, 0 }  ][line width=0.08]  [draw opacity=0] (8.93,-4.29) -- (0,0) -- (8.93,4.29) -- cycle    ;
%Straight Lines [id:da8828904567558743]
\draw    (295.02,52) -- (289.02,52) ;
\draw [shift={(287.02,52)}, rotate = 360] [fill={rgb, 255:red, 0; green, 0; blue, 0 }  ][line width=0.08]  [draw opacity=0] (8.93,-4.29) -- (0,0) -- (8.93,4.29) -- cycle    ;
%Straight Lines [id:da9930710621796206]
\draw  [dash pattern={on 4.5pt off 4.5pt}]  (143.33,31.83) -- (210.33,30.83) ;
%Straight Lines [id:da6088190375113437]
\draw  [dash pattern={on 4.5pt off 4.5pt}]  (270.15,130.59) -- (352.74,130.7) ;
%Shape: Tear Drop [id:dp6575751854512973]
\draw  [dash pattern={on 4.5pt off 4.5pt}] (296.3,116.41) .. controls (296.3,116.41) and (296.3,116.41) .. (296.3,116.41) .. controls (296.3,116.41) and (296.3,116.41) .. (296.3,116.41) .. controls (289.04,108.75) and (289.56,96.46) .. (297.46,88.96) .. controls (305.37,81.46) and (317.67,81.59) .. (324.93,89.24) .. controls (332.2,96.9) and (331.68,109.19) .. (323.77,116.69) .. controls (314.24,125.75) and (309.45,130.39) .. (309.45,130.64) .. controls (309.45,130.39) and (305.08,125.65) .. (296.3,116.41) -- cycle ;
%Straight Lines [id:da8442821541689595]
\draw  [dash pattern={on 4.5pt off 4.5pt}]  (157.15,130.59) -- (239.74,130.7) ;
%Shape: Tear Drop [id:dp6857267676271542]
\draw  [dash pattern={on 4.5pt off 4.5pt}] (183.3,116.41) .. controls (183.3,116.41) and (183.3,116.41) .. (183.3,116.41) .. controls (183.3,116.41) and (183.3,116.41) .. (183.3,116.41) .. controls (176.04,108.75) and (176.56,96.46) .. (184.46,88.96) .. controls (192.37,81.46) and (204.67,81.59) .. (211.93,89.24) .. controls (219.2,96.9) and (218.68,109.19) .. (210.77,116.69) .. controls (201.24,125.75) and (196.45,130.39) .. (196.45,130.64) .. controls (196.45,130.39) and (192.08,125.65) .. (183.3,116.41) -- cycle ;
%Straight Lines [id:da7839482074178296]
\draw  [dash pattern={on 4.5pt off 4.5pt}]  (315.61,31.16) -- (346.27,31.28) ;
%Shape: Ellipse [id:dp9450362344897855]
\draw   (405.99,32.21) .. controls (405.97,20.98) and (416.63,11.86) .. (429.78,11.85) .. controls (442.93,11.83) and (453.6,20.93) .. (453.61,32.16) .. controls (453.62,43.39) and (442.97,52.51) .. (429.82,52.52) .. controls (416.67,52.54) and (406,43.44) .. (405.99,32.21) -- cycle ;
%Straight Lines [id:da9733576347680692]
\draw  [dash pattern={on 4.5pt off 4.5pt}]  (375.33,32.09) -- (405.99,32.21) ;
%Straight Lines [id:da513926902072152]
\draw    (423.96,11.83) -- (434.3,11.67) ;
\draw [shift={(434.13,11.67)}, rotate = 179.08] [fill={rgb, 255:red, 0; green, 0; blue, 0 }  ][line width=0.08]  [draw opacity=0] (8.93,-4.29) -- (0,0) -- (8.93,4.29) -- cycle    ;
%Straight Lines [id:da9372459535683161]
\draw    (433.02,53) -- (427.02,53) ;
\draw [shift={(425.02,53)}, rotate = 360] [fill={rgb, 255:red, 0; green, 0; blue, 0 }  ][line width=0.08]  [draw opacity=0] (8.93,-4.29) -- (0,0) -- (8.93,4.29) -- cycle    ;
%Straight Lines [id:da2193727255310598]
\draw  [dash pattern={on 4.5pt off 4.5pt}]  (453.61,32.16) -- (484.27,32.28) ;
%Straight Lines [id:da08634920598596874]
\draw  [dash pattern={on 4.5pt off 4.5pt}]  (403.15,130.09) -- (485.74,130.2) ;
%Shape: Tear Drop [id:dp2674937628181281]
\draw  [dash pattern={on 4.5pt off 4.5pt}] (429.3,115.91) .. controls (429.3,115.91) and (429.3,115.91) .. (429.3,115.91) .. controls (429.3,115.91) and (429.3,115.91) .. (429.3,115.91) .. controls (422.04,108.25) and (422.56,95.96) .. (430.46,88.46) .. controls (438.37,80.96) and (450.67,81.09) .. (457.93,88.74) .. controls (465.2,96.4) and (464.68,108.69) .. (456.77,116.19) .. controls (447.24,125.25) and (442.45,129.89) .. (442.45,130.14) .. controls (442.45,129.89) and (438.08,125.15) .. (429.3,115.91) -- cycle ;
%Curve Lines [id:da5181998884369777]
\draw [line width=2.25]    (444.28,61.25) .. controls (452.43,66.77) and (436.13,66.77) .. (444.08,72.88) ;
%Curve Lines [id:da7789153228615646]
\draw [line width=2.25]    (444.08,72.88) .. controls (452.23,78.4) and (435.94,78.4) .. (443.89,84.5) ;
%Curve Lines [id:da49016806414459]
\draw [line width=2.25]    (177.28,8.25) .. controls (185.43,13.77) and (169.13,13.77) .. (177.08,19.88) ;
%Curve Lines [id:da44152921493553676]
\draw [line width=2.25]    (177.08,19.88) .. controls (185.23,25.4) and (168.94,25.4) .. (176.89,31.5) ;
%Curve Lines [id:da900601133465001]
\draw [line width=2.25]    (252.28,9.25) .. controls (260.43,14.77) and (244.13,14.77) .. (252.08,20.88) ;
%Curve Lines [id:da1713624230074915]
\draw [line width=2.25]    (252.08,20.88) .. controls (260.23,26.4) and (243.94,26.4) .. (251.89,32.5) ;
%Curve Lines [id:da6029985421312357]
\draw [line width=2.25]    (470.28,10.25) .. controls (478.43,15.77) and (462.13,15.77) .. (470.08,21.88) ;
%Curve Lines [id:da4326664351544358]
\draw [line width=2.25]    (470.08,21.88) .. controls (478.23,27.4) and (461.94,27.4) .. (469.89,33.5) ;
%Curve Lines [id:da8312461933564729]
\draw [line width=2.25]    (168.28,107.25) .. controls (176.43,112.77) and (160.13,112.77) .. (168.08,118.88) ;
%Curve Lines [id:da5533079861302032]
\draw [line width=2.25]    (168.08,118.88) .. controls (176.23,124.4) and (159.94,124.4) .. (167.89,130.5) ;
%Curve Lines [id:da07297807103032095]
\draw [line width=2.25]    (340.28,107.25) .. controls (348.43,112.77) and (332.13,112.77) .. (340.08,118.88) ;
%Curve Lines [id:da22425820128796492]
\draw [line width=2.25]    (340.08,118.88) .. controls (348.23,124.4) and (331.94,124.4) .. (339.89,130.5) ;
% Text Node
\draw (140,34.4) node [anchor=north west][inner sep=0.75pt]    {$0$};
\draw (337,35.4) node [anchor=north west][inner sep=0.75pt]    {$0$};
\draw (205,34.4) node [anchor=north west][inner sep=0.75pt]    {$0$};
\draw (241,34.4) node [anchor=north west][inner sep=0.75pt]    {$0$};
\draw (218,20.4) node [anchor=north west][inner sep=0.75pt]    {$+$};
\draw (272.15,133.99) node [anchor=north west][inner sep=0.75pt]    {$0$};
\draw (339,134.4) node [anchor=north west][inner sep=0.75pt]    {$0$};
\draw (356,21.4) node [anchor=north west][inner sep=0.75pt]    {$+$};
\draw (159.15,133.99) node [anchor=north west][inner sep=0.75pt]    {$0$};
\draw (226,134.4) node [anchor=north west][inner sep=0.75pt]    {$0$};
\draw (244,119.4) node [anchor=north west][inner sep=0.75pt]    {$+$};
\draw (475,36.4) node [anchor=north west][inner sep=0.75pt]    {$0$};
\draw (379,35.4) node [anchor=north west][inner sep=0.75pt]    {$0$};
\draw (134,121.4) node [anchor=north west][inner sep=0.75pt]    {$+$};
\draw (405.15,133.49) node [anchor=north west][inner sep=0.75pt]    {$0$};
\draw (472,133.9) node [anchor=north west][inner sep=0.75pt]    {$0$};
\draw (372,118.9) node [anchor=north west][inner sep=0.75pt]    {$+$};
\draw (490,125) node [anchor=north west][inner sep=0.75pt]    {,};
\end{tikzpicture} \label{eq:G-00021-diag} \ \ \
\end{eqnarray}
where the solid lines with arrows represent the fermion propagator \eqref{eq:fermion-propagator-general},
the dashed lines correspond to the boson propagator \eqref{eq:boson-propagator-chi} satisfying at the special transition the Neumann BC ($\chi = +1$), and
the wavy lines indicate the insertions of the composite operator $\varphi_s^2$.
The diagrams in Eq.~\eqref{eq:G-00021-diag} then yield:
\begin{align}
\delta^{(0)} \mathring{G}^{(0,0,0,2;2)}(\vec{p}=0) & = \frac{1}{\mathring{m}^2}\left[ 1 +
\frac{K_{d} m^{-\varepsilon}}{8\varepsilon} \right.  \nn \\
 & \times  \left. \left(\mathring{\lambda}+  N (2\sigma+1) \mathring{g}^2 \right)\right].
\label{eq:G00021-0-final}
\end{align}
The renormalization  condition within the MS scheme reads
\begin{align}
Z^{-1}_{\varphi_s} Z^{-1}_{\varphi^2_s} \mathring{G}^{(0,0,0,2;2)}(\vec{p}=0;\mathring{m},\mathring{g},\mathring{\lambda}) = \mathrm{finite},
\end{align}
where all bare parameters are expressed in terms of the renormalized ones according to
Eqs.~(\ref{eq:mass-renormalization})-(\ref{eq:lambda-renormalization}). This leads to
\begin{align} \label{eq:Z-c-5}
Z_{\varphi^2_s}= 1-\frac{\lambda}{\varepsilon} + \frac{N}{\varepsilon} \sigma g^2 + O(g^4,\lambda^2, g^2\lambda).
%Z_{c}= 1+\frac{\lambda}{\varepsilon} - \frac{N}{\varepsilon} \sigma g^2 + O(g^4,\lambda^2, g^2\lambda).
\end{align}
The scaling function~\eqref{eq:critical-exponents-gen}  to one loop order is
\begin{align} \label{eq:eta-phi-s-1}
\eta_{\varphi^2_s}= \lambda - N \sigma g^2.
%\eta_{c}= -\lambda+N \sigma g^2.
\end{align}
Evaluating the scaling function~\eqref{eq:eta-phi-s-1} at the GNY FP~\eqref{eq:fp-lambda} we get the critical exponent
\begin{align} \label{eq:eta-c-1}
\eta_{\varphi^2_s} = \frac{\left(Y(N)-(6\sigma+1) N+6\right) \varepsilon  }{6 (N+6)}.
\end{align}
The crossover exponent is related to $\eta_{\varphi^2_s}$ by \eqref{eq:Phi-def} and reads
\begin{align} \label{eq:crossover-Phi}
\Phi = \frac12 -\frac{\left(Y(N)-(12\sigma+7) N+6\right) \varepsilon  }{24 (N+6)}.
\end{align}
The scaling dimension of the boundary boson composite operator $\varphi_s^2$ is then given by
\begin{align} \label{eq:eta-c-2}
\Delta_{\varphi^2_s} = 2+ \frac{\left(Y(N)-(6\sigma+7) N-30\right) \varepsilon  }{6 (N+6)}.
\end{align}

\section{Pseudoscalar Yukawa model} \label{sec:Pseudoscalar-Yukawa}

A model closely related to the GNY model is the pseudoscalar Yukawa (pY) model, in which fermions $\psi$ interact with a pseudoscalar boson $\phi$ through a Yukawa-type coupling of the form $i g \varphi \bar\psi  \gamma_S \psi$.
The Euclidean action of the  pY model can be  written as
\begin{eqnarray}
S_\mathrm{pY} &= & \int d^{d} r  \int\limits_0^{\infty}d z \,  \Big\{ \sum\limits_{a=1}^{\tilde{N}} \bar{\psi}_a(\bm{x})  \left[ \slashed{\partial}  + i g \gamma_S \varphi(\bm{x}) \right] \psi_a (\bm{x}) \nonumber \\
 &+& \frac12 (\nabla \varphi(\bm{x}))^2+\frac12 m^2 \varphi^2(\bm{x}) +\frac{\lambda}{4!}\varphi^4(\bm{x}) \Big\}.
\label{eq:action-pseudoGNY-bulk}
\end{eqnarray}
In the absence of the mass term $\bar{\psi}\psi$ the  pY model~\eqref{eq:action-pseudoGNY-bulk}  has the same renormalization constants as the GNY model~\eqref{eq:action-GNY-bulk}.
Indeed after the chiral rotation
\begin{align}
\psi &\to  e^{-i \pi \gamma_S/4} \psi,
&
\bar{\psi} &\to  \bar{\psi} e^{-i \pi \gamma_S/4}
\label{eq:chiral-rotation}
\end{align}
the action~\eqref{eq:action-pseudoGNY-bulk}  becomes identical to the action~\eqref{eq:action-GNY-bulk}.
However, this does not hold for the pY model defined in semi-infinite space as
\begin{align}
S=S_\mathrm{pY}+S_\mathrm{B},
\end{align}
which was studied recently in the condensed matter context~\cite{Jiang:2025}.
Indeed, the boundary term $S_\mathrm{B}$ given by Eq.~\eqref{eq:action-GNY-surface} is not invariant under the chiral rotation~\eqref{eq:chiral-rotation}, in a manner analogous to the fermion mass term.
Under this rotation, the matrix $\check{M}$ appearing in the boundary term~\eqref{eq:action-GNY-surface} transforms as
\begin{align}
\check{M} \to e^{-i \pi \gamma_S/4} \, \check{M} \, e^{-i \pi \gamma_S/4}.
\label{eq:M-chiral-rotation}
\end{align}
For example, the matrices $\check{M}$ corresponding to the TRI and non-TRI FPs exchange under transformation.~\eqref{eq:M-chiral-rotation} as follows:
\begin{align}
& \check{M} = \pm \mathbb{1}   \to  \check{M} = \mp i \gamma_S, \label{eq:M-chiral-rotation-1} \\
& \check{M} = \pm i \gamma_S \to  \check{M} = \pm \mathbb{1}. \label{eq:M-chiral-rotation-2}
\end{align}
This explains the observed differences in the boundary critical exponents between the models studied in \cite{Giombi:2022,Herzog:2023} and  in \cite{Jiang:2025}.

\section{Conclusions} \label{sec:conclusions}

In this work we studied the semi-infinite GNY model, describing Dirac fermions coupled to a real scalar bosonic field via a Yukawa interaction.
We considered general Robin BCs for the bosons, parametrized by a boundary enhancement parameter $c$, together with the most general fermionic BCs compatible with unitarity, conformal invariance, and charge-conjugation symmetry. The latter are parametrized by an angle $\phi$, with the special values $\phi = \pm \frac{\pi}{2}$ corresponding to TRI BCs.

We study the RG flows of the model in the $(c,\phi)$ plane, shown in Fig.~\ref{fig:1-RGFlow}, and identify six distinct universality classes of surface critical behavior, corresponding to the ordinary, special, and extraordinary transitions, each of which can be either TRI or non-TRI. The TRI FPs are unstable with respect to TRI-breaking perturbations in the BCs. The resulting phase diagram is presented in Fig.~\ref{fig:2-phase-diagram}. Note that in the case of graphene, TRI here does not refer to physical TRI that exchanges the valleys, but to the intravalley TRI, which is preserved only when the reflection off the boundary mixes the two valleys without introducing a phase shift.

Using a perturbative RG analysis in $D = 4 - \varepsilon$ dimensions, we computed the critical exponents and the scaling dimensions of bosonic and fermionic composite operators to one-loop order at the TRI and non-TRI ordinary and special transitions, corresponding to Dirichlet and Neumann BCs in the bosonic sector.

We further examined the transformation of the boundary theory under chiral rotations and clarified why the semi-infinite pY model is not equivalent to the GNY model at the boundary, despite their equivalence in the bulk. This distinction accounts for the differences between the boundary critical exponents obtained here and those reported in Refs.~\cite{Giombi:2022,Herzog:2023,Jiang:2025}.

Our results offer a systematic classification of surface criticality in fermion-boson theories and establish a basis for future higher-loop studies and applications to systems with emergent Dirac fermions.

\begin{acknowledgments}
We would like to thank David Carpentier and Ivan Balog for inspiring discussions. We acknowledge support from the CNRS International Research Project ``Disordered quantum matter'' (DisQ). The work of A.A.F. has been partially supported by the French National Agency for Research (ANR) within the project PROCURPHY ANR-23-CE30-0051-02. I.A.G. thanks the CNRS visiting researcher program and the ENS de Lyon, where a part of this work was done, for hospitality.
\end{acknowledgments}

\textit{Note added in proofs}. - Soon after this work appeared, the TRI universality classes of boundary criticality in the GNY model were studied beyond the leading-order approximation considered here in~\cite{Diatlyk:2026}.

\appendix

\section{Dirac matrices in \(D=d+1\) spacetime dimensions}\label{sec:Appendix-A}

We now construct inductively an explicit representation of the Euclidean Clifford algebra using  hermitian unitary matrices. In a space of even dimension $D$, the minimal rank of these matrices is $2^{D/2}$. In $D=2$ dimensions we define
\begin{align}
    \gamma_j^{(D=2)} &= \sigma_{j+1},
    &
    j &= 0,1,2,
\end{align}
where $\sigma_j$, $j=1,2,3$ are the Pauli matrices and $\gamma_2^{(D=2)} \equiv \gamma_S^{(D=2)}$ is the
additional  Dirac matrix which in $D=4$ is known as the ``fifth'' matrix  $\gamma_5$.
In even $D$ dimensions we define the gamma matrices recursively by
\begin{align}
    \gamma_0^{(D)} &= - \sigma_{1} \otimes \mathbb{1}_{2^{D/2-1}} , \\
    \gamma_j^{(D)} &=   \sigma_{2} \otimes \gamma_{j-1}^{(D-2)}, \qquad j=1,\cdots,D-1, \\
    \gamma_S^{(D)} &=  \gamma_D^{(D)} = - i^{-D/2} \gamma_0^{(D)}...\gamma_{D-1}^{(D)},
\end{align}
where  $\mathbb{1}_{D}$ is the unity matrix of rank $D$.

In the space of an odd dimension $D$, the minimal rank of the Dirac matrices is $2^{[D/2]}$. However, in order to fulfill certain discrete symmetries and to relate the model under consideration to condensed-matter systems such as graphene, instead of using the irreducible representation of the Clifford algebra, we construct a representation of higher rank, namely $2^{[D/2]+1}$, as
\begin{align}
    \gamma_j^{(D)} &= \gamma_j^{(D+1)},  \qquad j=0,..,D-1, \\
    \gamma_S^{(D)} &=  \gamma_D^{(D)} = \gamma_{D+1}^{(D+1)}.
\end{align}
In all dimensions the gamma matrices satisfy the anti-commutation relation
\begin{align}
    \{\gamma_i^{(D)}, \gamma_j^{(D)}\} &= 2\delta_{i j},
    &
    i,j = 0, \dots, D-1,
\end{align}
%$\{\gamma_i^{(D)}, \gamma_j^{(D)}\} = 2\delta_{i j}  $, with $i,j = 0, \dots, D-1 $
and $\gamma_S^{(D)}$ anti-commutes with all \(\gamma_i^{(D)}\).

The corresponding alpha and beta Dirac matrices are given by
\begin{align}
 \alpha^{(D)}_{j} &= i \gamma_0^{(D)} \gamma_j^{(D)},
 &
 j &= 1,\cdots,D-1, \\
\beta^{(D)} &=  \gamma_0^{(D)}.
\end{align}
In what follows, we will omit the superscript $(D)$.

In $D=4$ the above definition gives (the alternative form of) the chiral (Weyl) representation of the Dirac matrices
\begin{align}
%\gamma^0_{\textrm{M}} &=
\gamma_0 &= -\tau_1 \otimes \sigma_0,
&
%\gamma^j_{\textrm{M}} &=i
\gamma_j &=  \tau_2 \otimes \sigma_j, \label{eq:gamma4-1}
\end{align}
where $\tau_j$ and $\sigma_j$ are two sets of the usual Pauli matrices, $\tau_0$ and $\sigma_0$ are the $2$ by $2$ unity matrix.
The ``fifth'' Dirac matrix
\begin{align}
\gamma_S
%= i\gamma_\mathrm{M}^0 \gamma_\mathrm{M}^1 \gamma_\mathrm{M}^2\gamma_\mathrm{M}^3
&= \gamma_0\gamma_1\gamma_2\gamma_3 =  \tau_3 \otimes \sigma_0
\label{eq:gamma4-2}
\end{align}
anti-commutes with other Dirac $\gamma$ matrices. The corresponding alpha and beta matrices are
\begin{align}
    \alpha_j &= \tau_3 \otimes \sigma_j,
    &
    \beta &= -\tau_1 \otimes \sigma_0.
    \label{eq:alpha-beta-D=4}
\end{align}

In the $D=3$ case corresponding to graphene, the matrices $\gamma_0$, $\gamma_1$, $\gamma_2$, $\gamma_S$, \(\alpha_1\), \(\alpha_2\), and \(\beta\)  are still given by Eqs.~(\ref{eq:gamma4-1})-(\ref{eq:alpha-beta-D=4}), where $\tau_i$ act in the space of valleys $K$ and $K'$ and $\sigma_j$ act in the space of sublattices A and B.

\section{Discrete symmetries of the Dirac Hamiltonian with boundary conditions} \label{sec:Appendix-B}

\subsection{Charge-conjugation symmetry} In the condensed matter context the charge-con\-ju\-gation symmetry corresponds to the particle-hole symmetry. In terms of the Hamiltonian it can be written as an anti-unitary operator
\begin{align}
\hat{\mathcal{C}}=U_C \hat{\mathcal{K}},
\label{eq:C-sym-1}
\end{align}
where $\hat{\mathcal{K}}$  is the operator of complex conjugation and $U_C$ is a unitary matrix in the spinor space satisfying the following relations
\begin{align}
U_C \alpha_j^* U_C^{-1} &= \alpha_j,
&
U_C \beta^* U_C^{-1} &= -\beta,
\label{eq:C-sym-2}
\end{align}
and, in the presence of a boundary, also
\begin{align}
U_C M^* U_C^{-1} &= M,
&
U_C \tilde{M}^* U_C^{-1} &= -\tilde{M},
\label{eq:C-sym-2b}
\end{align}
so that
\begin{align}
\hat{\mathcal{C}} \hat{H}_0 \hat{\mathcal{C}}^{-1} = -\hat{H}_0.
\label{eq:C-sym-3}
\end{align}
In terms of the Euclidean action the charge conjugation symmetry is the invariance of the action with respect to the transformation
\begin{align}
\psi &\to  C \bar{\psi}^{T},
&
\bar{\psi} &\to  - {\psi}^{T} C^{-1},
\label{eq:C-sym-4}
\end{align}
where the matrix $C$ anti-commutes with $\gamma_0$ and satisfies
\begin{align}
C \gamma_j^T C^{-1} = - \gamma_j,
\label{eq:C-sym-5}
\end{align}
and related to the matrix $U_C$ by
\begin{align}
U_C = C \gamma_0.
\label{eq:C-sym-6}
\end{align}
For even $D$  the matrix $C$  can be chosen as
\begin{align}
 C^{(D)} = i \big(\gamma_S\big)^{D/2}\prod\limits_{\mathrm{even} \ j=0}^{D} \gamma_j,
\label{eq:C-sym-7}
\end{align}
while for odd $D$ it is given by  $C^{(D)}= C^{(D+1)}$. In the representation (\ref{eq:gamma4-1}) and (\ref{eq:gamma4-2}) it reads
\begin{align}
U_C &= -i  \tau_2 \otimes \sigma_2,
&
C &=  \tau_3 \otimes \sigma_2
\label{eq:UC-1}
\end{align}
in both $D=4$ and $D=3$~\cite{McCann-Falco:2004}.

\subsection{Time reversal invariance for Dirac spin}

In the Hamiltonian formalism, time-reversal is implemented by
an anti-unitary operator
\begin{align}
\hat{\mathcal{T}}=U_T \hat{\mathcal{K}},
\label{eq:T-sym-1}
\end{align}
where $\hat{\mathcal{K}}$  is the operator of complex conjugation and $U_T$ is a unitary matrix in the spinor space satisfying the following relations
\begin{align}
U_T \alpha_j^* U_T^{-1}& = -\alpha_j,
&
U_T \beta^* U_T^{-1} &= \beta.
\label{eq:T-sym-2}
\end{align}
For the system with a boundary to satisfy time-reversal invariance, the boundary matrix \(\tilde{M}\) should satisfy the same relation as the \(\beta\) matrix:
\begin{align}
U_T \tilde{M}^* U_T^{-1} = \tilde{M}, \ \ \ \ \ \ \mathrm{i.e.,} \ \ \ \ \
U_T M^* U_T^{-1}  = M,
\label{eq:T-sym-2b}
\end{align}
ensuring that the Hamiltonian~\eqref{eq-Hamiltonian-1} commutes with \(\hat{\mathcal{T}}\):
\begin{align}
\hat{\mathcal{T}} \hat{H}_0 \hat{\mathcal{T}}^{-1} = \hat{H}_0.
\label{eq:T-sym-3}
\end{align}
For even $D$  the matrix $U_T$  can be chosen as
\begin{align}
 U_T^{(D)} = i \big(\gamma_S\big)^{D/2}\prod\limits_{\mathrm{odd} \ j=1}^{D-1} \gamma_j,
\label{eq:T-sym-7}
\end{align}
while for odd $D$ it is given by  $U_T^{(D)}= U_T^{(D+1)}$. In the representation (\ref{eq:gamma4-1}) and (\ref{eq:gamma4-2}) it reads
\begin{align}
U_T = \tau_0 \otimes \sigma_2
\label{eq:UT-1}
\end{align}
in both $D=4$ and $D=3$. It is easy to see that only the first term in the matrix~\eqref{eq:tilde-M-D=4} satisfies the condition~\eqref{eq:T-sym-2b}, and we conclude that the TRI is preserved only when \(\phi = \pm \pi/2\).

\subsection{Time reversal invariance in disordered graphene}

In the case of graphene ($D=3$), $\sigma_j$ correspond not to the spin of the electrons but to the
A/B sublattices. The physical  time-reversal operator for graphene (the operator \(T_0\) in section~VI.G in Ref.~\cite{Evers:2008})
%Therefore, the true time-reversal operator must exchange the valleys~\cite{Cayssol:2013} and
is given by~\cite{Cayssol:2013}
\begin{align}
\hat{\mathcal{T}}^{(g)} &= U_T^{(g)} \hat{\mathcal{K}},
&
U_T^{(g)} &= \tau_1 \otimes \sigma_1,
\label{eq:UT-g}
\end{align}
and it involves the exchange of the valleys \(K\) and \(K'\). The matrix \(U_T^{(g)}\) satisfies

\begin{align}
U_T^{(g)} \alpha_{1,2}^* [U_T^{(g)}]^{-1} &= -\alpha_{1,2},
\nonumber \\
U_T^{(g)} \beta^* [U_T^{(g)}]^{-1}  &=  \beta,
\end{align}
so the Hamiltonian for clean graphene without a boundary (Eq.~\eqref{eq-Hamiltonian-1} with \(d=2\) and \(\tilde{M} = 0\)) is time-reversal invariant. For the kind of boundaries considered in this paper the matrix \(\tilde{M}\) in Eq.~\eqref{eq:tilde-M-D=4} satisfies the same relation  as the \(\beta\) matrix, \(U_T^{(g)} \tilde{M}^* [U_T^{(g)}]^{-1}  =  \tilde{M}\), for any value of the angle \(\phi\).

On the other hand, the operator $\hat{\mathcal{T}}$ in graphene represents an effective \textit{intravalley} time-reversal symmetry. This symmetry protects the structure of each Dirac cone rather than strictly preserving the valley index.
The intravalley time-reversal symmetry is satisfied either in the absence of intervalley scattering, corresponding to terms proportional to $\tau_0$ and $\tau_3$ (as in zigzaglike boundaries), or when intervalley scattering is present, as in armchairlike boundaries, but does not introduce a relative phase shift between the valleys. The latter case corresponds to a scattering matrix $\tilde{M}$ containing $\tau_1$ but not $\tau_2$. The $\tau_2$ term is imaginary and shifts the phase by $\pi/2$, while for an arbitrary $\tilde{M}$ with $\phi \ne \pm \pi/2$, the phase shift is proportional to $\phi$.
Note that zigzaglike boundaries host zero-energy edge modes, which complicate the low-energy field theory description. These modes are not considered in the present work, and we therefore restrict ourselves to armchair-like boundaries, where the effective value of $\phi$ can be controlled, for example by tuning the sample width~\cite{BreyFertig:2006}.


%%% BIBLIO %%%

%\bibliographystyle{apsrev4-2}
%\bibliography{biblioSurface}

\begin{thebibliography}{78}%
\makeatletter
\providecommand \@ifxundefined [1]{%
 \@ifx{#1\undefined}
}%
\providecommand \@ifnum [1]{%
 \ifnum #1\expandafter \@firstoftwo
 \else \expandafter \@secondoftwo
 \fi
}%
\providecommand \@ifx [1]{%
 \ifx #1\expandafter \@firstoftwo
 \else \expandafter \@secondoftwo
 \fi
}%
\providecommand \natexlab [1]{#1}%
\providecommand \enquote  [1]{``#1''}%
\providecommand \bibnamefont  [1]{#1}%
\providecommand \bibfnamefont [1]{#1}%
\providecommand \citenamefont [1]{#1}%
\providecommand \href@noop [0]{\@secondoftwo}%
\providecommand \href [0]{\begingroup \@sanitize@url \@href}%
\providecommand \@href[1]{\@@startlink{#1}\@@href}%
\providecommand \@@href[1]{\endgroup#1\@@endlink}%
\providecommand \@sanitize@url [0]{\catcode `\\12\catcode `\$12\catcode
  `\&12\catcode `\#12\catcode `\^12\catcode `\_12\catcode `\%12\relax}%
\providecommand \@@startlink[1]{}%
\providecommand \@@endlink[0]{}%
\providecommand \url  [0]{\begingroup\@sanitize@url \@url }%
\providecommand \@url [1]{\endgroup\@href {#1}{\urlprefix }}%
\providecommand \urlprefix  [0]{URL }%
\providecommand \Eprint [0]{\href }%
\providecommand \doibase [0]{https://doi.org/}%
\providecommand \selectlanguage [0]{\@gobble}%
\providecommand \bibinfo  [0]{\@secondoftwo}%
\providecommand \bibfield  [0]{\@secondoftwo}%
\providecommand \translation [1]{[#1]}%
\providecommand \BibitemOpen [0]{}%
\providecommand \bibitemStop [0]{}%
\providecommand \bibitemNoStop [0]{.\EOS\space}%
\providecommand \EOS [0]{\spacefactor3000\relax}%
\providecommand \BibitemShut  [1]{\csname bibitem#1\endcsname}%
\let\auto@bib@innerbib\@empty
%</preamble>
\bibitem [{\citenamefont {Poland}\ \emph {et~al.}(2019)\citenamefont {Poland},
  \citenamefont {Rychkov},\ and\ \citenamefont
  {Vichi}}]{PolandRychkovVichi:2019}%
  \BibitemOpen
  \bibfield  {author} {\bibinfo {author} {\bibfnamefont {D.}~\bibnamefont
  {Poland}}, \bibinfo {author} {\bibfnamefont {S.}~\bibnamefont {Rychkov}},\
  and\ \bibinfo {author} {\bibfnamefont {A.}~\bibnamefont {Vichi}},\ }\href
  {https://doi.org/10.1103/RevModPhys.91.015002} {\bibfield  {journal}
  {\bibinfo  {journal} {Rev. Mod. Phys.}\ }\textbf {\bibinfo {volume} {91}},\
  \bibinfo {pages} {015002} (\bibinfo {year} {2019})}\BibitemShut {NoStop}%
\bibitem [{\citenamefont {Castro~Neto}\ \emph {et~al.}(2009)\citenamefont
  {Castro~Neto}, \citenamefont {Guinea}, \citenamefont {Peres}, \citenamefont
  {Novoselov},\ and\ \citenamefont {Geim}}]{Neto:2009}%
  \BibitemOpen
  \bibfield  {author} {\bibinfo {author} {\bibfnamefont {A.~H.}\ \bibnamefont
  {Castro~Neto}}, \bibinfo {author} {\bibfnamefont {F.}~\bibnamefont {Guinea}},
  \bibinfo {author} {\bibfnamefont {N.~M.~R.}\ \bibnamefont {Peres}}, \bibinfo
  {author} {\bibfnamefont {K.~S.}\ \bibnamefont {Novoselov}},\ and\ \bibinfo
  {author} {\bibfnamefont {A.~K.}\ \bibnamefont {Geim}},\ }\href
  {https://doi.org/10.1103/RevModPhys.81.109} {\bibfield  {journal} {\bibinfo
  {journal} {Rev. Mod. Phys.}\ }\textbf {\bibinfo {volume} {81}},\ \bibinfo
  {pages} {109} (\bibinfo {year} {2009})}\BibitemShut {NoStop}%
\bibitem [{\citenamefont {Cayssol}(2013)}]{Cayssol:2013}%
  \BibitemOpen
  \bibfield  {author} {\bibinfo {author} {\bibfnamefont {J.}~\bibnamefont
  {Cayssol}},\ }\href
  {https://doi.org/https://doi.org/10.1016/j.crhy.2013.09.012} {\bibfield
  {journal} {\bibinfo  {journal} {Comptes Rendus Physique}\ }\textbf {\bibinfo
  {volume} {14}},\ \bibinfo {pages} {760} (\bibinfo {year} {2013})}\BibitemShut
  {NoStop}%
\bibitem [{\citenamefont {Armitage}\ \emph {et~al.}(2018)\citenamefont
  {Armitage}, \citenamefont {Mele},\ and\ \citenamefont
  {Vishwanath}}]{Armitage2018}%
  \BibitemOpen
  \bibfield  {author} {\bibinfo {author} {\bibfnamefont {N.~P.}\ \bibnamefont
  {Armitage}}, \bibinfo {author} {\bibfnamefont {E.~J.}\ \bibnamefont {Mele}},\
  and\ \bibinfo {author} {\bibfnamefont {A.}~\bibnamefont {Vishwanath}},\
  }\href {https://doi.org/10.1103/RevModPhys.90.015001} {\bibfield  {journal}
  {\bibinfo  {journal} {Rev. Mod. Phys.}\ }\textbf {\bibinfo {volume} {90}},\
  \bibinfo {pages} {015001} (\bibinfo {year} {2018})}\BibitemShut {NoStop}%
\bibitem [{\citenamefont {Diehl}(1986)}]{Diehl-book:1986}%
  \BibitemOpen
  \bibfield  {author} {\bibinfo {author} {\bibfnamefont {H.~W.}\ \bibnamefont
  {Diehl}},\ }in\ \href@noop {} {\emph {\bibinfo {booktitle} {Phase Transitions
  and Critical Phenomena}}},\ Vol.~\bibinfo {volume} {10},\ \bibinfo {editor}
  {edited by\ \bibinfo {editor} {\bibfnamefont {C.}~\bibnamefont {Domb}}\ and\
  \bibinfo {editor} {\bibfnamefont {J.~L.}\ \bibnamefont {Lebowitz}}}\
  (\bibinfo  {publisher} {Academic London},\ \bibinfo {year}
  {1986})\BibitemShut {NoStop}%
\bibitem [{\citenamefont {Metlitski}(2022)}]{Metlitski:2022}%
  \BibitemOpen
  \bibfield  {author} {\bibinfo {author} {\bibfnamefont {M.~A.}\ \bibnamefont
  {Metlitski}},\ }\href {https://doi.org/10.21468/SciPostPhys.12.4.131}
  {\bibfield  {journal} {\bibinfo  {journal} {SciPost Phys.}\ }\textbf
  {\bibinfo {volume} {12}},\ \bibinfo {pages} {131} (\bibinfo {year}
  {2022})}\BibitemShut {NoStop}%
\bibitem [{\citenamefont {Padayasi}\ \emph {et~al.}(2022)\citenamefont
  {Padayasi}, \citenamefont {Krishnan}, \citenamefont {Metlitski},
  \citenamefont {Gruzberg},\ and\ \citenamefont
  {Meineri}}]{PadayasiKrishnanMetlitskiGruzbergMeineri:2022}%
  \BibitemOpen
  \bibfield  {author} {\bibinfo {author} {\bibfnamefont {J.}~\bibnamefont
  {Padayasi}}, \bibinfo {author} {\bibfnamefont {A.}~\bibnamefont {Krishnan}},
  \bibinfo {author} {\bibfnamefont {M.~A.}\ \bibnamefont {Metlitski}}, \bibinfo
  {author} {\bibfnamefont {I.~A.}\ \bibnamefont {Gruzberg}},\ and\ \bibinfo
  {author} {\bibfnamefont {M.}~\bibnamefont {Meineri}},\ }\href
  {https://doi.org/10.21468/SciPostPhys.12.6.190} {\bibfield  {journal}
  {\bibinfo  {journal} {SciPost Phys.}\ }\textbf {\bibinfo {volume} {12}},\
  \bibinfo {pages} {190} (\bibinfo {year} {2022})}\BibitemShut {NoStop}%
\bibitem [{\citenamefont {Parisen~Toldin}(2021)}]{ParisenToldin:2021}%
  \BibitemOpen
  \bibfield  {author} {\bibinfo {author} {\bibfnamefont {F.}~\bibnamefont
  {Parisen~Toldin}},\ }\href {https://doi.org/10.1103/PhysRevLett.126.135701}
  {\bibfield  {journal} {\bibinfo  {journal} {Phys. Rev. Lett.}\ }\textbf
  {\bibinfo {volume} {126}},\ \bibinfo {pages} {135701} (\bibinfo {year}
  {2021})}\BibitemShut {NoStop}%
\bibitem [{\citenamefont {Parisen~Toldin}\ and\ \citenamefont
  {Metlitski}(2022)}]{ParisenToldinMetlitski:2022}%
  \BibitemOpen
  \bibfield  {author} {\bibinfo {author} {\bibfnamefont {F.}~\bibnamefont
  {Parisen~Toldin}}\ and\ \bibinfo {author} {\bibfnamefont {M.~A.}\
  \bibnamefont {Metlitski}},\ }\href
  {https://doi.org/10.1103/PhysRevLett.128.215701} {\bibfield  {journal}
  {\bibinfo  {journal} {Phys. Rev. Lett.}\ }\textbf {\bibinfo {volume} {128}},\
  \bibinfo {pages} {215701} (\bibinfo {year} {2022})}\BibitemShut {NoStop}%
\bibitem [{\citenamefont {Sun}\ and\ \citenamefont {Lv}(2022)}]{SunLv2022}%
  \BibitemOpen
  \bibfield  {author} {\bibinfo {author} {\bibfnamefont {Y.}~\bibnamefont
  {Sun}}\ and\ \bibinfo {author} {\bibfnamefont {J.-P.}\ \bibnamefont {Lv}},\
  }\href {https://doi.org/10.1103/PhysRevB.106.224502} {\bibfield  {journal}
  {\bibinfo  {journal} {Phys. Rev. B}\ }\textbf {\bibinfo {volume} {106}},\
  \bibinfo {pages} {224502} (\bibinfo {year} {2022})}\BibitemShut {NoStop}%
\bibitem [{\citenamefont {Zhang}\ \emph {et~al.}(2023)\citenamefont {Zhang},
  \citenamefont {Zhu},\ and\ \citenamefont
  {Vishwanath}}]{ZhangZhuVishwanath:2023}%
  \BibitemOpen
  \bibfield  {author} {\bibinfo {author} {\bibfnamefont {Y.-H.}\ \bibnamefont
  {Zhang}}, \bibinfo {author} {\bibfnamefont {Z.}~\bibnamefont {Zhu}},\ and\
  \bibinfo {author} {\bibfnamefont {A.}~\bibnamefont {Vishwanath}},\ }\href
  {https://doi.org/10.1103/PhysRevX.13.031023} {\bibfield  {journal} {\bibinfo
  {journal} {Phys. Rev. X}\ }\textbf {\bibinfo {volume} {13}},\ \bibinfo
  {pages} {031023} (\bibinfo {year} {2023})}\BibitemShut {NoStop}%
\bibitem [{\citenamefont {Abrahams}(2010)}]{Abrahams:2010}%
  \BibitemOpen
  \bibinfo {editor} {\bibfnamefont {E.}~\bibnamefont {Abrahams}},\ ed.,\ \href
  {https://doi.org/10.1142/7663} {\emph {\bibinfo {title} {50 Years of Anderson
  Localization}}}\ (\bibinfo  {publisher} {World Scientific},\ \bibinfo
  {address} {Singapore},\ \bibinfo {year} {2010})\BibitemShut {NoStop}%
\bibitem [{\citenamefont {Evers}\ and\ \citenamefont
  {Mirlin}(2008)}]{Evers:2008}%
  \BibitemOpen
  \bibfield  {author} {\bibinfo {author} {\bibfnamefont {F.}~\bibnamefont
  {Evers}}\ and\ \bibinfo {author} {\bibfnamefont {A.~D.}\ \bibnamefont
  {Mirlin}},\ }\href {https://doi.org/10.1103/RevModPhys.80.1355} {\bibfield
  {journal} {\bibinfo  {journal} {Rev. Mod. Phys.}\ }\textbf {\bibinfo {volume}
  {80}},\ \bibinfo {pages} {1355} (\bibinfo {year} {2008})}\BibitemShut
  {NoStop}%
\bibitem [{\citenamefont {Subramaniam}\ \emph {et~al.}(2006)\citenamefont
  {Subramaniam}, \citenamefont {Gruzberg}, \citenamefont {Ludwig},
  \citenamefont {Evers}, \citenamefont {Mildenberger},\ and\ \citenamefont
  {Mirlin}}]{Subramaniam:2006}%
  \BibitemOpen
  \bibfield  {author} {\bibinfo {author} {\bibfnamefont {A.~R.}\ \bibnamefont
  {Subramaniam}}, \bibinfo {author} {\bibfnamefont {I.~A.}\ \bibnamefont
  {Gruzberg}}, \bibinfo {author} {\bibfnamefont {A.~W.~W.}\ \bibnamefont
  {Ludwig}}, \bibinfo {author} {\bibfnamefont {F.}~\bibnamefont {Evers}},
  \bibinfo {author} {\bibfnamefont {A.}~\bibnamefont {Mildenberger}},\ and\
  \bibinfo {author} {\bibfnamefont {A.~D.}\ \bibnamefont {Mirlin}},\ }\href
  {https://doi.org/10.1103/PhysRevLett.96.126802} {\bibfield  {journal}
  {\bibinfo  {journal} {Phys. Rev. Lett.}\ }\textbf {\bibinfo {volume} {96}},\
  \bibinfo {pages} {126802} (\bibinfo {year} {2006})}\BibitemShut {NoStop}%
\bibitem [{\citenamefont {{Mildenberger}}\ \emph {et~al.}(2007)\citenamefont
  {{Mildenberger}}, \citenamefont {{Subramaniam}}, \citenamefont {{Narayanan}},
  \citenamefont {{Evers}}, \citenamefont {{Gruzberg}},\ and\ \citenamefont
  {{Mirlin}}}]{Mildenberger-Boundary-2007}%
  \BibitemOpen
  \bibfield  {author} {\bibinfo {author} {\bibfnamefont {A.}~\bibnamefont
  {{Mildenberger}}}, \bibinfo {author} {\bibfnamefont {A.~R.}\ \bibnamefont
  {{Subramaniam}}}, \bibinfo {author} {\bibfnamefont {R.}~\bibnamefont
  {{Narayanan}}}, \bibinfo {author} {\bibfnamefont {F.}~\bibnamefont
  {{Evers}}}, \bibinfo {author} {\bibfnamefont {I.~A.}\ \bibnamefont
  {{Gruzberg}}},\ and\ \bibinfo {author} {\bibfnamefont {A.~D.}\ \bibnamefont
  {{Mirlin}}},\ }\href {https://doi.org/10.1103/PhysRevB.75.094204} {\bibfield
  {journal} {\bibinfo  {journal} {\prb}\ }\textbf {\bibinfo {volume} {75}},\
  \bibinfo {eid} {094204} (\bibinfo {year} {2007})}\BibitemShut {NoStop}%
\bibitem [{\citenamefont {{Obuse}}\ \emph {et~al.}(2007)\citenamefont
  {{Obuse}}, \citenamefont {{Subramaniam}}, \citenamefont {{Furusaki}},
  \citenamefont {{Gruzberg}},\ and\ \citenamefont
  {{Ludwig}}}]{Obuse-Multifractality-2007}%
  \BibitemOpen
  \bibfield  {author} {\bibinfo {author} {\bibfnamefont {H.}~\bibnamefont
  {{Obuse}}}, \bibinfo {author} {\bibfnamefont {A.~R.}\ \bibnamefont
  {{Subramaniam}}}, \bibinfo {author} {\bibfnamefont {A.}~\bibnamefont
  {{Furusaki}}}, \bibinfo {author} {\bibfnamefont {I.~A.}\ \bibnamefont
  {{Gruzberg}}},\ and\ \bibinfo {author} {\bibfnamefont {A.~W.~W.}\
  \bibnamefont {{Ludwig}}},\ }\href
  {https://doi.org/10.1103/PhysRevLett.98.156802} {\bibfield  {journal}
  {\bibinfo  {journal} {\prl}\ }\textbf {\bibinfo {volume} {98}},\ \bibinfo
  {eid} {156802} (\bibinfo {year} {2007})}\BibitemShut {NoStop}%
\bibitem [{\citenamefont {{Obuse}}\ \emph
  {et~al.}(2008{\natexlab{a}})\citenamefont {{Obuse}}, \citenamefont
  {{Subramaniam}}, \citenamefont {{Furusaki}}, \citenamefont {{Gruzberg}},\
  and\ \citenamefont {{Ludwig}}}]{Obuse-Corner-2008}%
  \BibitemOpen
  \bibfield  {author} {\bibinfo {author} {\bibfnamefont {H.}~\bibnamefont
  {{Obuse}}}, \bibinfo {author} {\bibfnamefont {A.~R.}\ \bibnamefont
  {{Subramaniam}}}, \bibinfo {author} {\bibfnamefont {A.}~\bibnamefont
  {{Furusaki}}}, \bibinfo {author} {\bibfnamefont {I.~A.}\ \bibnamefont
  {{Gruzberg}}},\ and\ \bibinfo {author} {\bibfnamefont {A.~W.~W.}\
  \bibnamefont {{Ludwig}}},\ }\href
  {https://doi.org/10.1016/j.physe.2007.09.024} {\bibfield  {journal} {\bibinfo
   {journal} {Physica E Low-Dimensional Systems and Nanostructures}\ }\textbf
  {\bibinfo {volume} {40}},\ \bibinfo {pages} {1404} (\bibinfo {year}
  {2008}{\natexlab{a}})}\BibitemShut {NoStop}%
\bibitem [{\citenamefont {{Obuse}}\ \emph
  {et~al.}(2008{\natexlab{b}})\citenamefont {{Obuse}}, \citenamefont
  {{Subramaniam}}, \citenamefont {{Furusaki}}, \citenamefont {{Gruzberg}},\
  and\ \citenamefont {{Ludwig}}}]{Obuse-Boundary-2008}%
  \BibitemOpen
  \bibfield  {author} {\bibinfo {author} {\bibfnamefont {H.}~\bibnamefont
  {{Obuse}}}, \bibinfo {author} {\bibfnamefont {A.~R.}\ \bibnamefont
  {{Subramaniam}}}, \bibinfo {author} {\bibfnamefont {A.}~\bibnamefont
  {{Furusaki}}}, \bibinfo {author} {\bibfnamefont {I.~A.}\ \bibnamefont
  {{Gruzberg}}},\ and\ \bibinfo {author} {\bibfnamefont {A.~W.~W.}\
  \bibnamefont {{Ludwig}}},\ }\href
  {https://doi.org/10.1103/PhysRevLett.101.116802} {\bibfield  {journal}
  {\bibinfo  {journal} {\prl}\ }\textbf {\bibinfo {volume} {101}},\ \bibinfo
  {eid} {116802} (\bibinfo {year} {2008}{\natexlab{b}})}\BibitemShut {NoStop}%
\bibitem [{\citenamefont {{Subramaniam}}\ \emph {et~al.}(2008)\citenamefont
  {{Subramaniam}}, \citenamefont {{Gruzberg}},\ and\ \citenamefont
  {{Ludwig}}}]{Subramaniam-Boundary-2008}%
  \BibitemOpen
  \bibfield  {author} {\bibinfo {author} {\bibfnamefont {A.~R.}\ \bibnamefont
  {{Subramaniam}}}, \bibinfo {author} {\bibfnamefont {I.~A.}\ \bibnamefont
  {{Gruzberg}}},\ and\ \bibinfo {author} {\bibfnamefont {A.~W.~W.}\
  \bibnamefont {{Ludwig}}},\ }\href
  {https://doi.org/10.1103/PhysRevB.78.245105} {\bibfield  {journal} {\bibinfo
  {journal} {\prb}\ }\textbf {\bibinfo {volume} {78}},\ \bibinfo {eid} {245105}
  (\bibinfo {year} {2008})}\BibitemShut {NoStop}%
\bibitem [{\citenamefont {{Babkin}}\ \emph {et~al.}(2023)\citenamefont
  {{Babkin}}, \citenamefont {{Karcher}}, \citenamefont {{Burmistrov}},\ and\
  \citenamefont {{Mirlin}}}]{Babkin-Generalized-2023}%
  \BibitemOpen
  \bibfield  {author} {\bibinfo {author} {\bibfnamefont {S.~S.}\ \bibnamefont
  {{Babkin}}}, \bibinfo {author} {\bibfnamefont {J.~F.}\ \bibnamefont
  {{Karcher}}}, \bibinfo {author} {\bibfnamefont {I.~S.}\ \bibnamefont
  {{Burmistrov}}},\ and\ \bibinfo {author} {\bibfnamefont {A.~D.}\ \bibnamefont
  {{Mirlin}}},\ }\href {https://doi.org/10.1103/PhysRevB.108.104205} {\bibfield
   {journal} {\bibinfo  {journal} {\prb}\ }\textbf {\bibinfo {volume} {108}},\
  \bibinfo {eid} {104205} (\bibinfo {year} {2023})}\BibitemShut {NoStop}%
\bibitem [{\citenamefont {{Babkin}}\ and\ \citenamefont
  {{Burmistrov}}(2023)}]{Babkin-Boundary-2023}%
  \BibitemOpen
  \bibfield  {author} {\bibinfo {author} {\bibfnamefont {S.~S.}\ \bibnamefont
  {{Babkin}}}\ and\ \bibinfo {author} {\bibfnamefont {I.~S.}\ \bibnamefont
  {{Burmistrov}}},\ }\href {https://doi.org/10.1103/PhysRevB.108.205429}
  {\bibfield  {journal} {\bibinfo  {journal} {\prb}\ }\textbf {\bibinfo
  {volume} {108}},\ \bibinfo {eid} {205429} (\bibinfo {year}
  {2023})}\BibitemShut {NoStop}%
\bibitem [{\citenamefont {Syzranov}\ and\ \citenamefont
  {Radzihovsky}(2018)}]{Syzranov:2018}%
  \BibitemOpen
  \bibfield  {author} {\bibinfo {author} {\bibfnamefont {S.~V.}\ \bibnamefont
  {Syzranov}}\ and\ \bibinfo {author} {\bibfnamefont {L.}~\bibnamefont
  {Radzihovsky}},\ }\href
  {https://doi.org/10.1146/annurev-conmatphys-033117-054037} {\bibfield
  {journal} {\bibinfo  {journal} {Ann. Rev. Cond. Mat. Phys.}\ }\textbf
  {\bibinfo {volume} {9}},\ \bibinfo {pages} {35} (\bibinfo {year}
  {2018})}\BibitemShut {NoStop}%
\bibitem [{\citenamefont {Roy}\ and\ \citenamefont
  {Das~Sarma}(2014)}]{Roy:2014}%
  \BibitemOpen
  \bibfield  {author} {\bibinfo {author} {\bibfnamefont {B.}~\bibnamefont
  {Roy}}\ and\ \bibinfo {author} {\bibfnamefont {S.}~\bibnamefont
  {Das~Sarma}},\ }\href {https://doi.org/10.1103/PhysRevB.90.241112} {\bibfield
   {journal} {\bibinfo  {journal} {Phys. Rev. B}\ }\textbf {\bibinfo {volume}
  {90}},\ \bibinfo {pages} {241112(R)} (\bibinfo {year} {2014})},\ \bibinfo
  {note} {\ see also erratum
  \href{https://doi.org/10.1103/PhysRevB.93.119911}{{\bf 93}, 119911
  (2016)}}\BibitemShut {NoStop}%
\bibitem [{\citenamefont {Louvet}\ \emph {et~al.}(2016)\citenamefont {Louvet},
  \citenamefont {Carpentier},\ and\ \citenamefont {Fedorenko}}]{Louvet:2016}%
  \BibitemOpen
  \bibfield  {author} {\bibinfo {author} {\bibfnamefont {T.}~\bibnamefont
  {Louvet}}, \bibinfo {author} {\bibfnamefont {D.}~\bibnamefont {Carpentier}},\
  and\ \bibinfo {author} {\bibfnamefont {A.~A.}\ \bibnamefont {Fedorenko}},\
  }\href {https://doi.org/10.1103/PhysRevB.94.220201} {\bibfield  {journal}
  {\bibinfo  {journal} {Phys. Rev. B}\ }\textbf {\bibinfo {volume} {94}},\
  \bibinfo {pages} {220201(R)} (\bibinfo {year} {2016})}\BibitemShut {NoStop}%
\bibitem [{\citenamefont {Syzranov}\ \emph {et~al.}(2016)\citenamefont
  {Syzranov}, \citenamefont {Ostrovsky}, \citenamefont {Gurarie},\ and\
  \citenamefont {Radzihovsky}}]{Syzranov:2015b}%
  \BibitemOpen
  \bibfield  {author} {\bibinfo {author} {\bibfnamefont {S.~V.}\ \bibnamefont
  {Syzranov}}, \bibinfo {author} {\bibfnamefont {P.~M.}\ \bibnamefont
  {Ostrovsky}}, \bibinfo {author} {\bibfnamefont {V.}~\bibnamefont {Gurarie}},\
  and\ \bibinfo {author} {\bibfnamefont {L.}~\bibnamefont {Radzihovsky}},\
  }\href {https://doi.org/10.1103/PhysRevB.93.155113} {\bibfield  {journal}
  {\bibinfo  {journal} {Phys. Rev. B}\ }\textbf {\bibinfo {volume} {93}},\
  \bibinfo {pages} {155113} (\bibinfo {year} {2016})}\BibitemShut {NoStop}%
\bibitem [{\citenamefont {Sbierski}\ \emph {et~al.}(2014)\citenamefont
  {Sbierski}, \citenamefont {Pohl}, \citenamefont {Bergholtz},\ and\
  \citenamefont {Brouwer}}]{Sbierski:2014}%
  \BibitemOpen
  \bibfield  {author} {\bibinfo {author} {\bibfnamefont {B.}~\bibnamefont
  {Sbierski}}, \bibinfo {author} {\bibfnamefont {G.}~\bibnamefont {Pohl}},
  \bibinfo {author} {\bibfnamefont {E.~J.}\ \bibnamefont {Bergholtz}},\ and\
  \bibinfo {author} {\bibfnamefont {P.~W.}\ \bibnamefont {Brouwer}},\ }\href
  {https://doi.org/10.1103/PhysRevLett.113.026602} {\bibfield  {journal}
  {\bibinfo  {journal} {Phys. Rev. Lett.}\ }\textbf {\bibinfo {volume} {113}},\
  \bibinfo {pages} {026602} (\bibinfo {year} {2014})}\BibitemShut {NoStop}%
\bibitem [{\citenamefont {Sbierski}\ \emph {et~al.}(2016)\citenamefont
  {Sbierski}, \citenamefont {Decker},\ and\ \citenamefont
  {Brouwer}}]{Sbierski:2016}%
  \BibitemOpen
  \bibfield  {author} {\bibinfo {author} {\bibfnamefont {B.}~\bibnamefont
  {Sbierski}}, \bibinfo {author} {\bibfnamefont {K.~S.~C.}\ \bibnamefont
  {Decker}},\ and\ \bibinfo {author} {\bibfnamefont {P.~W.}\ \bibnamefont
  {Brouwer}},\ }\href {https://doi.org/10.1103/PhysRevB.94.220202} {\bibfield
  {journal} {\bibinfo  {journal} {Phys. Rev. B}\ }\textbf {\bibinfo {volume}
  {94}},\ \bibinfo {pages} {220202(R)} (\bibinfo {year} {2016})}\BibitemShut
  {NoStop}%
\bibitem [{\citenamefont {Fradkin}(1986)}]{Fradkin:1986}%
  \BibitemOpen
  \bibfield  {author} {\bibinfo {author} {\bibfnamefont {E.}~\bibnamefont
  {Fradkin}},\ }\href {https://doi.org/10.1103/PhysRevB.33.3263} {\bibfield
  {journal} {\bibinfo  {journal} {Phys. Rev. B}\ }\textbf {\bibinfo {volume}
  {33}},\ \bibinfo {pages} {3263} (\bibinfo {year} {1986})}\BibitemShut
  {NoStop}%
\bibitem [{\citenamefont {Roy}\ \emph {et~al.}(2016)\citenamefont {Roy},
  \citenamefont {Juri\v{c}i\'{c}},\ and\ \citenamefont
  {Das~Sarma}}]{Roy:2016b}%
  \BibitemOpen
  \bibfield  {author} {\bibinfo {author} {\bibfnamefont {B.}~\bibnamefont
  {Roy}}, \bibinfo {author} {\bibfnamefont {V.}~\bibnamefont
  {Juri\v{c}i\'{c}}},\ and\ \bibinfo {author} {\bibfnamefont {S.}~\bibnamefont
  {Das~Sarma}},\ }\href {http://dx.doi.org/10.1038/srep32446} {\bibfield
  {journal} {\bibinfo  {journal} {Sci. Rep.}\ }\textbf {\bibinfo {volume}
  {6}},\ \bibinfo {pages} {32446} (\bibinfo {year} {2016})}\BibitemShut
  {NoStop}%
\bibitem [{\citenamefont {Goswami}\ and\ \citenamefont
  {Chakravarty}(2011)}]{Goswami:2011}%
  \BibitemOpen
  \bibfield  {author} {\bibinfo {author} {\bibfnamefont {P.}~\bibnamefont
  {Goswami}}\ and\ \bibinfo {author} {\bibfnamefont {S.}~\bibnamefont
  {Chakravarty}},\ }\href {https://doi.org/10.1103/PhysRevLett.107.196803}
  {\bibfield  {journal} {\bibinfo  {journal} {Phys. Rev. Lett.}\ }\textbf
  {\bibinfo {volume} {107}},\ \bibinfo {pages} {196803} (\bibinfo {year}
  {2011})}\BibitemShut {NoStop}%
\bibitem [{\citenamefont {Hosur}\ \emph {et~al.}(2012)\citenamefont {Hosur},
  \citenamefont {Parameswaran},\ and\ \citenamefont {Vishwanath}}]{Hosur:2012}%
  \BibitemOpen
  \bibfield  {author} {\bibinfo {author} {\bibfnamefont {P.}~\bibnamefont
  {Hosur}}, \bibinfo {author} {\bibfnamefont {S.~A.}\ \bibnamefont
  {Parameswaran}},\ and\ \bibinfo {author} {\bibfnamefont {A.}~\bibnamefont
  {Vishwanath}},\ }\href {https://doi.org/10.1103/PhysRevLett.108.046602}
  {\bibfield  {journal} {\bibinfo  {journal} {Phys. Rev. Lett.}\ }\textbf
  {\bibinfo {volume} {108}},\ \bibinfo {pages} {046602} (\bibinfo {year}
  {2012})}\BibitemShut {NoStop}%
\bibitem [{\citenamefont {Ominato}\ and\ \citenamefont
  {Koshino}(2014)}]{Ominato:2014}%
  \BibitemOpen
  \bibfield  {author} {\bibinfo {author} {\bibfnamefont {Y.}~\bibnamefont
  {Ominato}}\ and\ \bibinfo {author} {\bibfnamefont {M.}~\bibnamefont
  {Koshino}},\ }\href {https://doi.org/10.1103/PhysRevB.89.054202} {\bibfield
  {journal} {\bibinfo  {journal} {Phys. Rev. B}\ }\textbf {\bibinfo {volume}
  {89}},\ \bibinfo {pages} {054202} (\bibinfo {year} {2014})}\BibitemShut
  {NoStop}%
\bibitem [{\citenamefont {Chen}\ \emph {et~al.}(2015)\citenamefont {Chen},
  \citenamefont {Song}, \citenamefont {Jiang}, \citenamefont {Sun},
  \citenamefont {Wang},\ and\ \citenamefont {Xie}}]{Chen:2015}%
  \BibitemOpen
  \bibfield  {author} {\bibinfo {author} {\bibfnamefont {C.-Z.}\ \bibnamefont
  {Chen}}, \bibinfo {author} {\bibfnamefont {J.}~\bibnamefont {Song}}, \bibinfo
  {author} {\bibfnamefont {H.}~\bibnamefont {Jiang}}, \bibinfo {author}
  {\bibfnamefont {Q.-F.}\ \bibnamefont {Sun}}, \bibinfo {author} {\bibfnamefont
  {Z.}~\bibnamefont {Wang}},\ and\ \bibinfo {author} {\bibfnamefont {X.~C.}\
  \bibnamefont {Xie}},\ }\href {https://doi.org/10.1103/PhysRevLett.115.246603}
  {\bibfield  {journal} {\bibinfo  {journal} {Phys. Rev. Lett.}\ }\textbf
  {\bibinfo {volume} {115}},\ \bibinfo {pages} {246603} (\bibinfo {year}
  {2015})}\BibitemShut {NoStop}%
\bibitem [{\citenamefont {Altland}\ and\ \citenamefont
  {Bagrets}(2015)}]{Altland:2015:2016}%
  \BibitemOpen
  \bibfield  {author} {\bibinfo {author} {\bibfnamefont {A.}~\bibnamefont
  {Altland}}\ and\ \bibinfo {author} {\bibfnamefont {D.}~\bibnamefont
  {Bagrets}},\ }\href {https://doi.org/10.1103/PhysRevLett.114.257201}
  {\bibfield  {journal} {\bibinfo  {journal} {Phys. Rev. Lett.}\ }\textbf
  {\bibinfo {volume} {114}},\ \bibinfo {pages} {257201} (\bibinfo {year}
  {2015})},\ \bibinfo {note} {\
  \href{https://doi.org/10.1103/PhysRevB.93.075113}{Phys. Rev. B {\bf 93},
  075113 (2016)}}\BibitemShut {NoStop}%
\bibitem [{\citenamefont {Szab\'o}\ and\ \citenamefont
  {Roy}(2020)}]{SzaboRoy:2020}%
  \BibitemOpen
  \bibfield  {author} {\bibinfo {author} {\bibfnamefont {A.~L.}\ \bibnamefont
  {Szab\'o}}\ and\ \bibinfo {author} {\bibfnamefont {B.}~\bibnamefont {Roy}},\
  }\href {https://doi.org/10.1103/PhysRevResearch.2.043197} {\bibfield
  {journal} {\bibinfo  {journal} {Phys. Rev. Res.}\ }\textbf {\bibinfo {volume}
  {2}},\ \bibinfo {pages} {043197} (\bibinfo {year} {2020})}\BibitemShut
  {NoStop}%
\bibitem [{\citenamefont {Kobayashi}\ \emph {et~al.}(2014)\citenamefont
  {Kobayashi}, \citenamefont {Ohtsuki}, \citenamefont {Imura},\ and\
  \citenamefont {Herbut}}]{Kobayashi:2014}%
  \BibitemOpen
  \bibfield  {author} {\bibinfo {author} {\bibfnamefont {K.}~\bibnamefont
  {Kobayashi}}, \bibinfo {author} {\bibfnamefont {T.}~\bibnamefont {Ohtsuki}},
  \bibinfo {author} {\bibfnamefont {K.-I.}\ \bibnamefont {Imura}},\ and\
  \bibinfo {author} {\bibfnamefont {I.~F.}\ \bibnamefont {Herbut}},\ }\href
  {https://doi.org/10.1103/PhysRevLett.112.016402} {\bibfield  {journal}
  {\bibinfo  {journal} {Phys. Rev. Lett.}\ }\textbf {\bibinfo {volume} {112}},\
  \bibinfo {pages} {016402} (\bibinfo {year} {2014})}\BibitemShut {NoStop}%
\bibitem [{\citenamefont {Sbierski}\ \emph {et~al.}(2015)\citenamefont
  {Sbierski}, \citenamefont {Bergholtz},\ and\ \citenamefont
  {Brouwer}}]{Sbierski:2015}%
  \BibitemOpen
  \bibfield  {author} {\bibinfo {author} {\bibfnamefont {B.}~\bibnamefont
  {Sbierski}}, \bibinfo {author} {\bibfnamefont {E.~J.}\ \bibnamefont
  {Bergholtz}},\ and\ \bibinfo {author} {\bibfnamefont {P.~W.}\ \bibnamefont
  {Brouwer}},\ }\href {https://doi.org/10.1103/PhysRevB.92.115145} {\bibfield
  {journal} {\bibinfo  {journal} {Phys. Rev. B}\ }\textbf {\bibinfo {volume}
  {92}},\ \bibinfo {pages} {115145} (\bibinfo {year} {2015})}\BibitemShut
  {NoStop}%
\bibitem [{\citenamefont {Liu}\ \emph {et~al.}(2016)\citenamefont {Liu},
  \citenamefont {Ohtsuki},\ and\ \citenamefont {Shindou}}]{Liu:2016}%
  \BibitemOpen
  \bibfield  {author} {\bibinfo {author} {\bibfnamefont {S.}~\bibnamefont
  {Liu}}, \bibinfo {author} {\bibfnamefont {T.}~\bibnamefont {Ohtsuki}},\ and\
  \bibinfo {author} {\bibfnamefont {R.}~\bibnamefont {Shindou}},\ }\href
  {https://doi.org/10.1103/PhysRevLett.116.066401} {\bibfield  {journal}
  {\bibinfo  {journal} {Phys. Rev. Lett.}\ }\textbf {\bibinfo {volume} {116}},\
  \bibinfo {pages} {066401} (\bibinfo {year} {2016})}\BibitemShut {NoStop}%
\bibitem [{\citenamefont {Bera}\ \emph {et~al.}(2016)\citenamefont {Bera},
  \citenamefont {Sau},\ and\ \citenamefont {Roy}}]{Bera:2016}%
  \BibitemOpen
  \bibfield  {author} {\bibinfo {author} {\bibfnamefont {S.}~\bibnamefont
  {Bera}}, \bibinfo {author} {\bibfnamefont {J.~D.}\ \bibnamefont {Sau}},\ and\
  \bibinfo {author} {\bibfnamefont {B.}~\bibnamefont {Roy}},\ }\href
  {https://doi.org/10.1103/PhysRevB.93.201302} {\bibfield  {journal} {\bibinfo
  {journal} {Phys. Rev. B}\ }\textbf {\bibinfo {volume} {93}},\ \bibinfo
  {pages} {201302(R)} (\bibinfo {year} {2016})}\BibitemShut {NoStop}%
\bibitem [{\citenamefont {Fu}\ \emph {et~al.}(2017)\citenamefont {Fu},
  \citenamefont {Zhu}, \citenamefont {Shi}, \citenamefont {Li}, \citenamefont
  {Yang},\ and\ \citenamefont {Zhang}}]{Fu:2017}%
  \BibitemOpen
  \bibfield  {author} {\bibinfo {author} {\bibfnamefont {B.}~\bibnamefont
  {Fu}}, \bibinfo {author} {\bibfnamefont {W.}~\bibnamefont {Zhu}}, \bibinfo
  {author} {\bibfnamefont {Q.}~\bibnamefont {Shi}}, \bibinfo {author}
  {\bibfnamefont {Q.}~\bibnamefont {Li}}, \bibinfo {author} {\bibfnamefont
  {J.}~\bibnamefont {Yang}},\ and\ \bibinfo {author} {\bibfnamefont
  {Z.}~\bibnamefont {Zhang}},\ }\href
  {https://doi.org/10.1103/PhysRevLett.118.146401} {\bibfield  {journal}
  {\bibinfo  {journal} {Phys. Rev. Lett.}\ }\textbf {\bibinfo {volume} {118}},\
  \bibinfo {pages} {146401} (\bibinfo {year} {2017})}\BibitemShut {NoStop}%
\bibitem [{\citenamefont {Sbierski}\ \emph {et~al.}(2017)\citenamefont
  {Sbierski}, \citenamefont {Madsen}, \citenamefont {Brouwer},\ and\
  \citenamefont {Karrasch}}]{Sbierski:2017}%
  \BibitemOpen
  \bibfield  {author} {\bibinfo {author} {\bibfnamefont {B.}~\bibnamefont
  {Sbierski}}, \bibinfo {author} {\bibfnamefont {K.~A.}\ \bibnamefont
  {Madsen}}, \bibinfo {author} {\bibfnamefont {P.~W.}\ \bibnamefont
  {Brouwer}},\ and\ \bibinfo {author} {\bibfnamefont {C.}~\bibnamefont
  {Karrasch}},\ }\href {https://doi.org/10.1103/PhysRevB.96.064203} {\bibfield
  {journal} {\bibinfo  {journal} {Phys. Rev. B}\ }\textbf {\bibinfo {volume}
  {96}},\ \bibinfo {pages} {064203} (\bibinfo {year} {2017})}\BibitemShut
  {NoStop}%
\bibitem [{\citenamefont {Roy}\ \emph {et~al.}(2018)\citenamefont {Roy},
  \citenamefont {Slager},\ and\ \citenamefont {Juri\ifmmode \check{c}\else
  \v{c}\fi{}i\ifmmode~\acute{c}\else \'{c}\fi{}}}]{Roy2016}%
  \BibitemOpen
  \bibfield  {author} {\bibinfo {author} {\bibfnamefont {B.}~\bibnamefont
  {Roy}}, \bibinfo {author} {\bibfnamefont {R.-J.}\ \bibnamefont {Slager}},\
  and\ \bibinfo {author} {\bibfnamefont {V.}~\bibnamefont {Juri\ifmmode
  \check{c}\else \v{c}\fi{}i\ifmmode~\acute{c}\else \'{c}\fi{}}},\ }\href
  {https://doi.org/10.1103/PhysRevX.8.031076} {\bibfield  {journal} {\bibinfo
  {journal} {Phys. Rev. X}\ }\textbf {\bibinfo {volume} {8}},\ \bibinfo {pages}
  {031076} (\bibinfo {year} {2018})}\BibitemShut {NoStop}%
\bibitem [{\citenamefont {Balog}\ \emph {et~al.}(2018)\citenamefont {Balog},
  \citenamefont {Carpentier},\ and\ \citenamefont {Fedorenko}}]{Balog:2018}%
  \BibitemOpen
  \bibfield  {author} {\bibinfo {author} {\bibfnamefont {I.}~\bibnamefont
  {Balog}}, \bibinfo {author} {\bibfnamefont {D.}~\bibnamefont {Carpentier}},\
  and\ \bibinfo {author} {\bibfnamefont {A.~A.}\ \bibnamefont {Fedorenko}},\
  }\href {https://doi.org/10.1103/PhysRevLett.121.166402} {\bibfield  {journal}
  {\bibinfo  {journal} {Phys. Rev. Lett.}\ }\textbf {\bibinfo {volume} {121}},\
  \bibinfo {pages} {166402} (\bibinfo {year} {2018})}\BibitemShut {NoStop}%
\bibitem [{\citenamefont {Louvet}\ \emph {et~al.}(2017)\citenamefont {Louvet},
  \citenamefont {Carpentier},\ and\ \citenamefont {Fedorenko}}]{Louvet:2017}%
  \BibitemOpen
  \bibfield  {author} {\bibinfo {author} {\bibfnamefont {T.}~\bibnamefont
  {Louvet}}, \bibinfo {author} {\bibfnamefont {D.}~\bibnamefont {Carpentier}},\
  and\ \bibinfo {author} {\bibfnamefont {A.~A.}\ \bibnamefont {Fedorenko}},\
  }\href {https://doi.org/10.1103/PhysRevB.95.014204} {\bibfield  {journal}
  {\bibinfo  {journal} {Phys. Rev. B}\ }\textbf {\bibinfo {volume} {95}},\
  \bibinfo {pages} {014204} (\bibinfo {year} {2017})}\BibitemShut {NoStop}%
\bibitem [{\citenamefont {Sbierski}\ and\ \citenamefont
  {Fr\"a\ss{}dorf}(2019)}]{Sbierski:2019}%
  \BibitemOpen
  \bibfield  {author} {\bibinfo {author} {\bibfnamefont {B.}~\bibnamefont
  {Sbierski}}\ and\ \bibinfo {author} {\bibfnamefont {C.}~\bibnamefont
  {Fr\"a\ss{}dorf}},\ }\href {https://doi.org/10.1103/PhysRevB.99.020201}
  {\bibfield  {journal} {\bibinfo  {journal} {Phys. Rev. B}\ }\textbf {\bibinfo
  {volume} {99}},\ \bibinfo {pages} {020201(R)} (\bibinfo {year}
  {2019})}\BibitemShut {NoStop}%
\bibitem [{\citenamefont {Klier}\ \emph {et~al.}(2019)\citenamefont {Klier},
  \citenamefont {Gornyi},\ and\ \citenamefont {Mirlin}}]{Klier:2019}%
  \BibitemOpen
  \bibfield  {author} {\bibinfo {author} {\bibfnamefont {J.}~\bibnamefont
  {Klier}}, \bibinfo {author} {\bibfnamefont {I.~V.}\ \bibnamefont {Gornyi}},\
  and\ \bibinfo {author} {\bibfnamefont {A.~D.}\ \bibnamefont {Mirlin}},\
  }\href {https://doi.org/10.1103/PhysRevB.100.125160} {\bibfield  {journal}
  {\bibinfo  {journal} {Phys. Rev. B}\ }\textbf {\bibinfo {volume} {100}},\
  \bibinfo {pages} {125160} (\bibinfo {year} {2019})}\BibitemShut {NoStop}%
\bibitem [{\citenamefont {Brillaux}\ \emph {et~al.}(2019)\citenamefont
  {Brillaux}, \citenamefont {Carpentier},\ and\ \citenamefont
  {Fedorenko}}]{Brillaux:2020}%
  \BibitemOpen
  \bibfield  {author} {\bibinfo {author} {\bibfnamefont {E.}~\bibnamefont
  {Brillaux}}, \bibinfo {author} {\bibfnamefont {D.}~\bibnamefont
  {Carpentier}},\ and\ \bibinfo {author} {\bibfnamefont {A.~A.}\ \bibnamefont
  {Fedorenko}},\ }\href {https://doi.org/10.1103/PhysRevB.100.134204}
  {\bibfield  {journal} {\bibinfo  {journal} {Phys. Rev. B}\ }\textbf {\bibinfo
  {volume} {100}},\ \bibinfo {pages} {134204} (\bibinfo {year}
  {2019})}\BibitemShut {NoStop}%
\bibitem [{\citenamefont {Brillaux}\ and\ \citenamefont
  {Fedorenko}(2021)}]{Brillaux:2021}%
  \BibitemOpen
  \bibfield  {author} {\bibinfo {author} {\bibfnamefont {E.}~\bibnamefont
  {Brillaux}}\ and\ \bibinfo {author} {\bibfnamefont {A.~A.}\ \bibnamefont
  {Fedorenko}},\ }\href {https://doi.org/10.1103/PhysRevB.103.L081405}
  {\bibfield  {journal} {\bibinfo  {journal} {Phys. Rev. B}\ }\textbf {\bibinfo
  {volume} {103}},\ \bibinfo {pages} {L081405} (\bibinfo {year}
  {2021})}\BibitemShut {NoStop}%
\bibitem [{\citenamefont {Brillaux}\ \emph {et~al.}(2024)\citenamefont
  {Brillaux}, \citenamefont {Fedorenko},\ and\ \citenamefont
  {Gruzberg}}]{Brillaux:2024}%
  \BibitemOpen
  \bibfield  {author} {\bibinfo {author} {\bibfnamefont {E.}~\bibnamefont
  {Brillaux}}, \bibinfo {author} {\bibfnamefont {A.~A.}\ \bibnamefont
  {Fedorenko}},\ and\ \bibinfo {author} {\bibfnamefont {I.~A.}\ \bibnamefont
  {Gruzberg}},\ }\href {https://doi.org/10.1103/PhysRevB.109.174204} {\bibfield
   {journal} {\bibinfo  {journal} {Phys. Rev. B}\ }\textbf {\bibinfo {volume}
  {109}},\ \bibinfo {pages} {174204} (\bibinfo {year} {2024})}\BibitemShut
  {NoStop}%
\bibitem [{\citenamefont {Miao}\ \emph {et~al.}(2017)\citenamefont {Miao},
  \citenamefont {Chu},\ and\ \citenamefont {Guo}}]{MiaoChuGuo:2017}%
  \BibitemOpen
  \bibfield  {author} {\bibinfo {author} {\bibfnamefont {R.-X.}\ \bibnamefont
  {Miao}}, \bibinfo {author} {\bibfnamefont {C.-S.}\ \bibnamefont {Chu}},\ and\
  \bibinfo {author} {\bibfnamefont {W.-Z.}\ \bibnamefont {Guo}},\ }\href
  {https://doi.org/10.1103/PhysRevD.96.046005} {\bibfield  {journal} {\bibinfo
  {journal} {Phys. Rev. D}\ }\textbf {\bibinfo {volume} {96}},\ \bibinfo
  {pages} {046005} (\bibinfo {year} {2017})}\BibitemShut {NoStop}%
\bibitem [{\citenamefont {Herzog}\ and\ \citenamefont
  {Huang}(2017)}]{HerzogHuang:2017}%
  \BibitemOpen
  \bibfield  {author} {\bibinfo {author} {\bibfnamefont {C.~P.}\ \bibnamefont
  {Herzog}}\ and\ \bibinfo {author} {\bibfnamefont {K.-W.}\ \bibnamefont
  {Huang}},\ }\href {https://doi.org/10.1007/JHEP10(2017)189} {\bibfield
  {journal} {\bibinfo  {journal} {J. High Energy Phys.}\ }\textbf {\bibinfo
  {volume} {2017}}\bibinfo  {number} { (10)},\ \bibinfo {pages}
  {189}}\BibitemShut {NoStop}%
\bibitem [{\citenamefont {Shpot}(2021)}]{Shpot:2021}%
  \BibitemOpen
\bibfield  {number} {  }\bibfield  {author} {\bibinfo {author} {\bibfnamefont
  {M.~A.}\ \bibnamefont {Shpot}},\ }\href
  {https://doi.org/10.1007/JHEP01(2021)055} {\bibfield  {journal} {\bibinfo
  {journal} {J. High Energy Phys.}\ }\textbf {\bibinfo {volume} {2021}}\bibinfo
   {number} { (1)},\ \bibinfo {pages} {55}}\BibitemShut {NoStop}%
\bibitem [{\citenamefont {Herzog}\ and\ \citenamefont
  {Schaub}(2024)}]{HerzogSchaub:2024}%
  \BibitemOpen
\bibfield  {number} {  }\bibfield  {author} {\bibinfo {author} {\bibfnamefont
  {C.~P.}\ \bibnamefont {Herzog}}\ and\ \bibinfo {author} {\bibfnamefont
  {V.}~\bibnamefont {Schaub}},\ }\href
  {https://doi.org/10.1103/PhysRevD.109.L061701} {\bibfield  {journal}
  {\bibinfo  {journal} {Phys. Rev. D}\ }\textbf {\bibinfo {volume} {109}},\
  \bibinfo {pages} {L061701} (\bibinfo {year} {2024})}\BibitemShut {NoStop}%
\bibitem [{\citenamefont {Zinn-Justin}(2021)}]{Zinn-Justin:1986}%
  \BibitemOpen
  \bibfield  {author} {\bibinfo {author} {\bibfnamefont {J.}~\bibnamefont
  {Zinn-Justin}},\ }\href {https://doi.org/10.1093/oso/9780198834625.001.0001}
  {\emph {\bibinfo {title} {Quantum field theory and critical phenomena}}},\
  \bibinfo {edition} {5th}\ ed.\ (\bibinfo  {publisher} {Oxford University
  Press},\ \bibinfo {address} {Oxford},\ \bibinfo {year} {2021})\BibitemShut
  {NoStop}%
\bibitem [{\citenamefont {Rosenstein}\ \emph {et~al.}(1993)\citenamefont
  {Rosenstein}, \citenamefont {Yu},\ and\ \citenamefont
  {Kovner}}]{Rosenstein:1993}%
  \BibitemOpen
  \bibfield  {author} {\bibinfo {author} {\bibfnamefont {B.}~\bibnamefont
  {Rosenstein}}, \bibinfo {author} {\bibfnamefont {H.-L.}\ \bibnamefont {Yu}},\
  and\ \bibinfo {author} {\bibfnamefont {A.}~\bibnamefont {Kovner}},\ }\href
  {https://doi.org/https://doi.org/10.1016/0370-2693(93)91253-J} {\bibfield
  {journal} {\bibinfo  {journal} {Phys. Lett. B}\ }\textbf {\bibinfo {volume}
  {314}},\ \bibinfo {pages} {381} (\bibinfo {year} {1993})}\BibitemShut
  {NoStop}%
\bibitem [{\citenamefont {Fei}\ \emph {et~al.}(2016)\citenamefont {Fei},
  \citenamefont {Giombi}, \citenamefont {Klebanov},\ and\ \citenamefont
  {Tarnopolsky}}]{Fei:2016}%
  \BibitemOpen
  \bibfield  {author} {\bibinfo {author} {\bibfnamefont {L.}~\bibnamefont
  {Fei}}, \bibinfo {author} {\bibfnamefont {S.}~\bibnamefont {Giombi}},
  \bibinfo {author} {\bibfnamefont {I.~R.}\ \bibnamefont {Klebanov}},\ and\
  \bibinfo {author} {\bibfnamefont {G.}~\bibnamefont {Tarnopolsky}},\ }\href
  {https://doi.org/10.1093/ptep/ptw120} {\bibfield  {journal} {\bibinfo
  {journal} {Progr. Theor. Exp. Phys.}\ }\textbf {\bibinfo {volume} {2016}},\
  \bibinfo {pages} {12C105} (\bibinfo {year} {2016})}\BibitemShut {NoStop}%
\bibitem [{\citenamefont {Mihaila}\ \emph {et~al.}(2017)\citenamefont
  {Mihaila}, \citenamefont {Zerf}, \citenamefont {Ihrig}, \citenamefont
  {Herbut},\ and\ \citenamefont {Scherer}}]{Mihaila:2017}%
  \BibitemOpen
  \bibfield  {author} {\bibinfo {author} {\bibfnamefont {L.~N.}\ \bibnamefont
  {Mihaila}}, \bibinfo {author} {\bibfnamefont {N.}~\bibnamefont {Zerf}},
  \bibinfo {author} {\bibfnamefont {B.}~\bibnamefont {Ihrig}}, \bibinfo
  {author} {\bibfnamefont {I.~F.}\ \bibnamefont {Herbut}},\ and\ \bibinfo
  {author} {\bibfnamefont {M.~M.}\ \bibnamefont {Scherer}},\ }\href
  {https://doi.org/10.1103/PhysRevB.96.165133} {\bibfield  {journal} {\bibinfo
  {journal} {Phys. Rev. B}\ }\textbf {\bibinfo {volume} {96}},\ \bibinfo
  {pages} {165133} (\bibinfo {year} {2017})}\BibitemShut {NoStop}%
\bibitem [{\citenamefont {Zerf}\ \emph {et~al.}(2017)\citenamefont {Zerf},
  \citenamefont {Mihaila}, \citenamefont {Marquard}, \citenamefont {Herbut},\
  and\ \citenamefont {Scherer}}]{Zerf:2017}%
  \BibitemOpen
  \bibfield  {author} {\bibinfo {author} {\bibfnamefont {N.}~\bibnamefont
  {Zerf}}, \bibinfo {author} {\bibfnamefont {L.~N.}\ \bibnamefont {Mihaila}},
  \bibinfo {author} {\bibfnamefont {P.}~\bibnamefont {Marquard}}, \bibinfo
  {author} {\bibfnamefont {I.~F.}\ \bibnamefont {Herbut}},\ and\ \bibinfo
  {author} {\bibfnamefont {M.~M.}\ \bibnamefont {Scherer}},\ }\href
  {https://doi.org/10.1103/PhysRevD.96.096010} {\bibfield  {journal} {\bibinfo
  {journal} {Phys. Rev. D}\ }\textbf {\bibinfo {volume} {96}},\ \bibinfo
  {pages} {096010} (\bibinfo {year} {2017})}\BibitemShut {NoStop}%
\bibitem [{\citenamefont {Ihrig}\ \emph {et~al.}(2018)\citenamefont {Ihrig},
  \citenamefont {Mihaila},\ and\ \citenamefont {Scherer}}]{Ihrig:2018}%
  \BibitemOpen
  \bibfield  {author} {\bibinfo {author} {\bibfnamefont {B.}~\bibnamefont
  {Ihrig}}, \bibinfo {author} {\bibfnamefont {L.~N.}\ \bibnamefont {Mihaila}},\
  and\ \bibinfo {author} {\bibfnamefont {M.~M.}\ \bibnamefont {Scherer}},\
  }\href {https://doi.org/10.1103/PhysRevB.98.125109} {\bibfield  {journal}
  {\bibinfo  {journal} {Phys. Rev. B}\ }\textbf {\bibinfo {volume} {98}},\
  \bibinfo {pages} {125109} (\bibinfo {year} {2018})}\BibitemShut {NoStop}%
\bibitem [{\citenamefont {Erramilli}\ \emph {et~al.}(2023)\citenamefont
  {Erramilli}, \citenamefont {Iliesiu}, \citenamefont {Kravchuk}, \citenamefont
  {Liu}, \citenamefont {Poland},\ and\ \citenamefont
  {Simmons-Duffin}}]{Erramilli:2023}%
  \BibitemOpen
  \bibfield  {author} {\bibinfo {author} {\bibfnamefont {R.~S.}\ \bibnamefont
  {Erramilli}}, \bibinfo {author} {\bibfnamefont {L.~V.}\ \bibnamefont
  {Iliesiu}}, \bibinfo {author} {\bibfnamefont {P.}~\bibnamefont {Kravchuk}},
  \bibinfo {author} {\bibfnamefont {A.}~\bibnamefont {Liu}}, \bibinfo {author}
  {\bibfnamefont {D.}~\bibnamefont {Poland}},\ and\ \bibinfo {author}
  {\bibfnamefont {D.}~\bibnamefont {Simmons-Duffin}},\ }\href
  {https://doi.org/10.1007/JHEP02(2023)036} {\bibfield  {journal} {\bibinfo
  {journal} {J. High Energy Phys.}\ }\textbf {\bibinfo {volume} {2023}}\bibinfo
   {number} { (2)},\ \bibinfo {pages} {36}}\BibitemShut {NoStop}%
\bibitem [{\citenamefont {Rosa}\ \emph {et~al.}(2001)\citenamefont {Rosa},
  \citenamefont {Vitale},\ and\ \citenamefont
  {Wetterich}}]{RosaVitaleWetterich:2001}%
  \BibitemOpen
\bibfield  {number} {  }\bibfield  {author} {\bibinfo {author} {\bibfnamefont
  {L.}~\bibnamefont {Rosa}}, \bibinfo {author} {\bibfnamefont {P.}~\bibnamefont
  {Vitale}},\ and\ \bibinfo {author} {\bibfnamefont {C.}~\bibnamefont
  {Wetterich}},\ }\href {https://doi.org/10.1103/PhysRevLett.86.958} {\bibfield
   {journal} {\bibinfo  {journal} {Phys. Rev. Lett.}\ }\textbf {\bibinfo
  {volume} {86}},\ \bibinfo {pages} {958} (\bibinfo {year} {2001})}\BibitemShut
  {NoStop}%
\bibitem [{\citenamefont {Knorr}(2016)}]{Knorr:2016}%
  \BibitemOpen
  \bibfield  {author} {\bibinfo {author} {\bibfnamefont {B.}~\bibnamefont
  {Knorr}},\ }\href {https://doi.org/10.1103/PhysRevB.94.245102} {\bibfield
  {journal} {\bibinfo  {journal} {Phys. Rev. B}\ }\textbf {\bibinfo {volume}
  {94}},\ \bibinfo {pages} {245102} (\bibinfo {year} {2016})}\BibitemShut
  {NoStop}%
\bibitem [{\citenamefont {Chandrasekharan}\ and\ \citenamefont
  {Li}(2013)}]{Chandrasekharan:2013}%
  \BibitemOpen
  \bibfield  {author} {\bibinfo {author} {\bibfnamefont {S.}~\bibnamefont
  {Chandrasekharan}}\ and\ \bibinfo {author} {\bibfnamefont {A.}~\bibnamefont
  {Li}},\ }\href {https://doi.org/10.1103/PhysRevD.88.021701} {\bibfield
  {journal} {\bibinfo  {journal} {Phys. Rev. D}\ }\textbf {\bibinfo {volume}
  {88}},\ \bibinfo {pages} {021701} (\bibinfo {year} {2013})}\BibitemShut
  {NoStop}%
\bibitem [{\citenamefont {Giombi}\ \emph {et~al.}(2022)\citenamefont {Giombi},
  \citenamefont {Helfenberger},\ and\ \citenamefont
  {Khanchandani}}]{Giombi:2022}%
  \BibitemOpen
  \bibfield  {author} {\bibinfo {author} {\bibfnamefont {S.}~\bibnamefont
  {Giombi}}, \bibinfo {author} {\bibfnamefont {E.}~\bibnamefont
  {Helfenberger}},\ and\ \bibinfo {author} {\bibfnamefont {H.}~\bibnamefont
  {Khanchandani}},\ }\href {https://doi.org/10.1007/JHEP07(2022)018} {\bibfield
   {journal} {\bibinfo  {journal} {J. High Energy Phys.}\ }\textbf {\bibinfo
  {volume} {2022}},\ \bibinfo {pages} {18}}\BibitemShut {NoStop}%
\bibitem [{\citenamefont {Herzog}\ and\ \citenamefont
  {Schaub}(2023)}]{Herzog:2023}%
  \BibitemOpen
  \bibfield  {author} {\bibinfo {author} {\bibfnamefont {C.~P.}\ \bibnamefont
  {Herzog}}\ and\ \bibinfo {author} {\bibfnamefont {V.}~\bibnamefont
  {Schaub}},\ }\href {https://doi.org/https://doi.org/10.1007/JHEP02(2023)129}
  {\bibfield  {journal} {\bibinfo  {journal} {J. High Energy Phys.}\ }\textbf
  {\bibinfo {volume} {2023}},\ \bibinfo {pages} {129}}\BibitemShut {NoStop}%
\bibitem [{\citenamefont {Jiang}\ \emph {et~al.}(2025)\citenamefont {Jiang},
  \citenamefont {Ge},\ and\ \citenamefont {Jian}}]{Jiang:2025}%
  \BibitemOpen
  \bibfield  {author} {\bibinfo {author} {\bibfnamefont {H.}~\bibnamefont
  {Jiang}}, \bibinfo {author} {\bibfnamefont {Y.}~\bibnamefont {Ge}},\ and\
  \bibinfo {author} {\bibfnamefont {S.-K.}\ \bibnamefont {Jian}},\ }\href
  {https://doi.org/10.1103/tjfk-84f8} {\bibfield  {journal} {\bibinfo
  {journal} {Phys. Rev. Lett.}\ }\textbf {\bibinfo {volume} {135}},\ \bibinfo
  {pages} {141602} (\bibinfo {year} {2025})},\ \Eprint
  {https://arxiv.org/abs/2503.13247} {arXiv:2503.13247 [cond-mat.str-el]}
  \BibitemShut {NoStop}%
\bibitem [{\citenamefont {Shen}\ \emph {et~al.}(2025)\citenamefont {Shen},
  \citenamefont {Wu},\ and\ \citenamefont {Jian}}]{Shen:2024}%
  \BibitemOpen
  \bibfield  {author} {\bibinfo {author} {\bibfnamefont {X.}~\bibnamefont
  {Shen}}, \bibinfo {author} {\bibfnamefont {Z.}~\bibnamefont {Wu}},\ and\
  \bibinfo {author} {\bibfnamefont {S.-K.}\ \bibnamefont {Jian}},\ }\href
  {https://doi.org/10.1103/4lv4-mc81} {\bibfield  {journal} {\bibinfo
  {journal} {Phys. Rev. B}\ }\textbf {\bibinfo {volume} {112}},\ \bibinfo
  {pages} {L041118} (\bibinfo {year} {2025})}\BibitemShut {NoStop}%
\bibitem [{\citenamefont {Ge}\ \emph {et~al.}()\citenamefont {Ge},
  \citenamefont {Jiang}, \citenamefont {Yao},\ and\ \citenamefont
  {Jian}}]{GeJiangYaoJian:2025}%
  \BibitemOpen
  \bibfield  {author} {\bibinfo {author} {\bibfnamefont {Y.}~\bibnamefont
  {Ge}}, \bibinfo {author} {\bibfnamefont {H.}~\bibnamefont {Jiang}}, \bibinfo
  {author} {\bibfnamefont {H.}~\bibnamefont {Yao}},\ and\ \bibinfo {author}
  {\bibfnamefont {S.-K.}\ \bibnamefont {Jian}},\ }\href
  {https://arxiv.org/abs/2510.05230} {}\Eprint
  {https://arxiv.org/abs/2510.05230} {arXiv:2510.05230} \BibitemShut {NoStop}%
\bibitem [{\citenamefont {Redlich}(1984)}]{Redlich:1984}%
  \BibitemOpen
  \bibfield  {author} {\bibinfo {author} {\bibfnamefont {A.~N.}\ \bibnamefont
  {Redlich}},\ }\href {https://doi.org/10.1103/PhysRevD.29.2366} {\bibfield
  {journal} {\bibinfo  {journal} {Phys. Rev. D}\ }\textbf {\bibinfo {volume}
  {29}},\ \bibinfo {pages} {2366} (\bibinfo {year} {1984})}\BibitemShut
  {NoStop}%
\bibitem [{\citenamefont {Biswas}\ and\ \citenamefont
  {Semenoff}(2022)}]{Biswas:2022}%
  \BibitemOpen
  \bibfield  {author} {\bibinfo {author} {\bibfnamefont {S.}~\bibnamefont
  {Biswas}}\ and\ \bibinfo {author} {\bibfnamefont {G.~W.}\ \bibnamefont
  {Semenoff}},\ }\href {https://doi.org/10.1007/JHEP10(2022)045} {\bibfield
  {journal} {\bibinfo  {journal} {J. High Energy Phys.}\ }\textbf {\bibinfo
  {volume} {2022}}\bibinfo  {number} { (45)}}\BibitemShut {NoStop}%
\bibitem [{\citenamefont {McCann}\ and\ \citenamefont
  {Fal'ko}(2004)}]{McCann-Falco:2004}%
  \BibitemOpen
\bibfield  {number} {  }\bibfield  {author} {\bibinfo {author} {\bibfnamefont
  {E.}~\bibnamefont {McCann}}\ and\ \bibinfo {author} {\bibfnamefont {V.~I.}\
  \bibnamefont {Fal'ko}},\ }\href {https://doi.org/10.1088/0953-8984/16/13/016}
  {\bibfield  {journal} {\bibinfo  {journal} {Journal of Physics: Condensed
  Matter}\ }\textbf {\bibinfo {volume} {16}},\ \bibinfo {pages} {2371}
  (\bibinfo {year} {2004})}\BibitemShut {NoStop}%
\bibitem [{\citenamefont {{Akhmerov}}\ and\ \citenamefont
  {{Beenakker}}(2008)}]{Akhmerov-Boundary-2008}%
  \BibitemOpen
  \bibfield  {author} {\bibinfo {author} {\bibfnamefont {A.~R.}\ \bibnamefont
  {{Akhmerov}}}\ and\ \bibinfo {author} {\bibfnamefont {C.~W.~J.}\ \bibnamefont
  {{Beenakker}}},\ }\href {https://doi.org/10.1103/PhysRevB.77.085423}
  {\bibfield  {journal} {\bibinfo  {journal} {\prb}\ }\textbf {\bibinfo
  {volume} {77}},\ \bibinfo {eid} {085423} (\bibinfo {year}
  {2008})}\BibitemShut {NoStop}%
\bibitem [{\citenamefont {Shtanko}\ and\ \citenamefont
  {Levitov}(2018)}]{ShtankoLevitov2018}%
  \BibitemOpen
  \bibfield  {author} {\bibinfo {author} {\bibfnamefont {O.}~\bibnamefont
  {Shtanko}}\ and\ \bibinfo {author} {\bibfnamefont {L.}~\bibnamefont
  {Levitov}},\ }\href {https://doi.org/10.1073/pnas.1722663115} {\bibfield
  {journal} {\bibinfo  {journal} {PNAS}\ }\textbf {\bibinfo {volume} {115}},\
  \bibinfo {pages} {5908} (\bibinfo {year} {2018})}\BibitemShut {NoStop}%
\bibitem [{\citenamefont {Hashimoto}\ \emph {et~al.}(2017)\citenamefont
  {Hashimoto}, \citenamefont {Kimura},\ and\ \citenamefont
  {Wu}}]{hashimoto_boundary_2017}%
  \BibitemOpen
  \bibfield  {author} {\bibinfo {author} {\bibfnamefont {K.}~\bibnamefont
  {Hashimoto}}, \bibinfo {author} {\bibfnamefont {T.}~\bibnamefont {Kimura}},\
  and\ \bibinfo {author} {\bibfnamefont {X.}~\bibnamefont {Wu}},\ }\href
  {https://doi.org/10.1093/ptep/ptx053} {\bibfield  {journal} {\bibinfo
  {journal} {Prog. Theor. Exp. Phys.}\ }\textbf {\bibinfo {volume} {2017}},\
  \bibinfo {pages} {053I01} (\bibinfo {year} {2017})}\BibitemShut {NoStop}%
\bibitem [{\citenamefont {Faraei}\ \emph {et~al.}(2018)\citenamefont {Faraei},
  \citenamefont {Farajollahpour},\ and\ \citenamefont
  {Jafari}}]{faraei_greens_2018}%
  \BibitemOpen
  \bibfield  {author} {\bibinfo {author} {\bibfnamefont {Z.}~\bibnamefont
  {Faraei}}, \bibinfo {author} {\bibfnamefont {T.}~\bibnamefont
  {Farajollahpour}},\ and\ \bibinfo {author} {\bibfnamefont {S.~A.}\
  \bibnamefont {Jafari}},\ }\href {https://doi.org/10.1103/PhysRevB.98.195402}
  {\bibfield  {journal} {\bibinfo  {journal} {Phys. Rev. B}\ }\textbf {\bibinfo
  {volume} {98}},\ \bibinfo {pages} {195402} (\bibinfo {year} {2018})},\
  \bibinfo {note} {arXiv: 1807.01139}\BibitemShut {NoStop}%
\bibitem [{\citenamefont {Rohim}\ and\ \citenamefont
  {Yamamoto}(2021)}]{Rohim:2021}%
  \BibitemOpen
  \bibfield  {author} {\bibinfo {author} {\bibfnamefont {A.}~\bibnamefont
  {Rohim}}\ and\ \bibinfo {author} {\bibfnamefont {K.}~\bibnamefont
  {Yamamoto}},\ }\href {https://doi.org/10.1093/ptep/ptab122} {\bibfield
  {journal} {\bibinfo  {journal} {Prog. Theor. Exp. Phys.}\ }\textbf {\bibinfo
  {volume} {2021}},\ \bibinfo {pages} {113B01} (\bibinfo {year}
  {2021})}\BibitemShut {NoStop}%
\bibitem [{\citenamefont {Diatlyk}\ \emph {et~al.}()\citenamefont {Diatlyk},
  \citenamefont {Giombi},\ and\ \citenamefont {Sun}}]{Diatlyk:2026}%
  \BibitemOpen
  \bibfield  {author} {\bibinfo {author} {\bibfnamefont {O.}~\bibnamefont
  {Diatlyk}}, \bibinfo {author} {\bibfnamefont {S.}~\bibnamefont {Giombi}},\
  and\ \bibinfo {author} {\bibfnamefont {Z.}~\bibnamefont {Sun}},\ }\href
  {https://doi.org/10.48550/arXiv.2606.07510} {}\Eprint
  {https://arxiv.org/abs/2606.07510} {arXiv:2606.07510} \BibitemShut {NoStop}%
\bibitem [{\citenamefont {Brey}\ and\ \citenamefont
  {Fertig}(2006)}]{BreyFertig:2006}%
  \BibitemOpen
  \bibfield  {author} {\bibinfo {author} {\bibfnamefont {L.}~\bibnamefont
  {Brey}}\ and\ \bibinfo {author} {\bibfnamefont {H.~A.}\ \bibnamefont
  {Fertig}},\ }\href {https://doi.org/10.1103/PhysRevB.73.235411} {\bibfield
  {journal} {\bibinfo  {journal} {Phys. Rev. B}\ }\textbf {\bibinfo {volume}
  {73}},\ \bibinfo {pages} {235411} (\bibinfo {year} {2006})}\BibitemShut
  {NoStop}%
\end{thebibliography}

\end{document}